\documentclass[aps,prd,letter,superscriptaddress,amsmath,twocolumn,floatfix,showpacs]{revtex4-1}
\newcommand{\beq}{\begin{equation}}
\newcommand{\eeq}{\end{equation}}
\newcommand{\beqn}{\begin{eqnarray}}
\newcommand{\eeqn}{\end{eqnarray}}

\newcommand{\apjl}{Astrophys.\ J.\ Lett.}
\newcommand{\apjs}{Astrophys.\ J.\ Supp.}
\newcommand{\mnras}{Mon. Not. Roy. Astron. Soc.}

\newcommand \ttau {\tilde{\tau}}

\newcommand \tS {\tilde{S}}
\newcommand \rhos {\rho_*}
\newcommand \sg {\alpha \sqrt{\gamma}}
\newcommand \sgam {\sqrt{\gamma}}

\newcommand{\bB}{\bar{B}}

\usepackage{graphicx}
\usepackage{epsf}
\usepackage{graphics,epsfig,placeins,subfigure,wrapfig}
\begin{document}
\title{General relativistic simulations of black hole-neutron
  star mergers: \\ Effects of magnetic fields}

\author{Zachariah B. Etienne}
\email{zetienne@illinois.edu}
\author{Yuk Tung Liu}
\author{Vasileios Paschalidis}
\author{Stuart~L.~Shapiro}
\altaffiliation{Also at Department of Astronomy and NCSA, University of
  Illinois at Urbana-Champaign, Urbana, IL 61801}
\affiliation{Department of Physics, University of Illinois at
  Urbana-Champaign, Urbana, IL 61801}

\begin{abstract}
As a neutron star (NS) is tidally disrupted by a black hole (BH) companion 
at the end of a BH--NS binary inspiral, its magnetic fields will be
stretched and amplified.  If sufficiently strong, these magnetic
fields may impact the gravitational waveforms, merger evolution and mass of the remnant
disk.  Formation of highly-collimated magnetic field lines in the 
disk+spinning BH remnant may launch relativistic jets, providing the engine
for a short-hard GRB.  We analyze this scenario
through fully general relativistic, magnetohydrodynamic (GRMHD) BHNS
simulations from inspiral through merger and disk formation.
Different initial magnetic field configurations and strengths are
chosen for the NS interior
for both nonspinning and moderately spinning
($a_{\rm BH}/M_{\rm BH}$=0.75) BHs aligned with the orbital angular momentum.
Only strong interior ($B_{\max} \sim10^{17}$G) initial magnetic
fields in the NS significantly influence merger dynamics,
enhancing the remnant disk mass by 100\% and 40\% in the nonspinning
and spinning BH cases, respectively.  However, detecting the
imprint of even a strong magnetic field may be challenging for Advanced
LIGO.  Though there is no evidence of mass outflows or magnetic field 
collimation
during the preliminary simulations we
have performed, higher
resolution, coupled with longer disk evolutions and different initial
magnetic field configurations, may be required to definitively assess
the possibility of BHNS binaries as short-hard GRB progenitors.
\end{abstract}


\maketitle

\section{Introduction}

With the first direct detection of gravitational waves (GWs) expected in the next few years, 
numerical relativity simulations will be crucial for
distinguishing different GW sources from one
another.  Mergers of black hole-neutron star (BHNS) binaries 
are among the most promising sources of gravitational waves detectable 
by ground-based laser interferometers
like LIGO~\cite{LIGO1,LIGO2}, VIRGO~\cite{VIRGO1,VIRGO2},
GEO~\cite{GEO}, LCGT~\cite{LCGT}, and AIGO~\cite{aigo}, as well as by
the proposed space-based LISA-like interferometers~\cite{LISA} and
DECIGO~\cite{DECIGO} and third-generation ground-based detectors such as 
the Einstein telescope~\cite{ET1,ET2}.  Analysis of
gravitational waveforms from BHNS mergers may spark new insights into
the behavior of matter at nuclear densities.

Theoretical models indicate that a neutron
star-neutron star (NSNS)~\cite{HMNS1,HMNS2,ShiUIUC,STU1,STU2,ST,SST,2008PhRvD..78h4033B}
or BHNS~\cite{FBSTR,FBST,SU1,SU2,SST,ST07,eflstb08,elsb09} binary merger may result in
a hot, massive disk around a black hole, whose 
temperatures and densities could be sufficient to trigger a short-hard
gamma-ray burst (SGRB).  Indeed, SGRBs have been associated
with galaxies with extremely low star formation rates (see~\cite{SGRB_local}
and references therein for a review), indicating
that the source is likely to involve an evolved population, rather
than main sequence stars. 
The number of detectable BHNS mergers in the observable universe is still an
open question, due to uncertainties in population synthesis calculations.
The estimated event rate of BHNS mergers observable by an
Advanced LIGO detector typically falls in the range ${\cal R}\sim$
0.2--300~${\rm yr}^{-1}$~\cite{KBKOW,aetal10}.

Motivated by the significance of BHNS binaries both as detectable GW
sources and SGRB candidates, many simulations of BHNS systems
have been performed in past years in a Newtonian or
post-Newtonian framework (see, e.g.,
\cite{Lee00,RSW,Rosswog,Koba,RKLRasio}) and in conformally-flat
relativistic gravitation~\cite{FBSTR,FBST}. Recently, several groups have performed
dynamical simulations of BHNS binary inspirals and
mergers in full GR~\cite{LRA,SU1,SU2,ST07,yst08,skyt09,kst10,eflstb08,elsb09,dfkpst08,dfkot10,fdkt11,cabllmn10,sep11,eps11,kost11,st11,fdksst11}.

Over the past few years, we have studied BHNS mergers beginning with the 
construction of quasiequilibrium circular orbit 
initial data~\cite{BSS,TBFS05,TBFS06,TBFS07a,TBFS07b}
and following up with full GR dynamical simulations~\cite{eflstb08,elsb09}. 
Our GR simulations, and those by other groups, suggest that for
initially nonspinning BHs, the remnant disk mass is substantial 
for $q \equiv M_{\rm BH}/M_{\rm NS} \lesssim 3$ and tends to increase
with decreasing $q$.
For a fixed $q$, the disk mass also increases for smaller NS compaction and
for more rapidly spinning BHs aligned with the orbital angular
momentum of the binary.  For sufficiently high spins, small mass
ratios, and/or lower NS compactions, a substantial disk can form following
the merger, favoring BHNS mergers as plausible central engines
for SGRBs.  However, these simulations have yet to account for magnetic
field effects -- a crucial component in many SGRB models involving a disk
around a spinning BH (see, e.g.~\cite{vk03a,vk03b,SGRB_review}).

For NS surface field strengths $B \lesssim 10^{16}$G, 
magnetic fields are unlikely to affect the dynamics of the BHNS
inspiral and merger~\footnote{However, it has been suggested
that during the BHNS inspiral the interaction of BH and NS magnetic
fields could drive a Poynting flux, which might give rise to observable
electromagnetic signals~\cite{ml11}.}. This was shown to be the case for NSNS binary 
inspirals in~\cite{IT00,vkmg03,mvg05,2009MNRAS.399L.164G}. Despite this conclusion, magnetic fields
may significantly influence the post-merger dynamics, as
the fields are likely to be amplified during and after merger. Magnetic 
fields could stir turbulence in the remnant disk, resulting in angular 
momentum transport and accretion onto the BH. They could also lead to 
matter outflow and jets along the BH remnant spin axis \cite{2011ApJ...732L...6R}, 
another ingredient required by most SGRB models~\cite{SGRB_review,vk03a,vk03b}.

In this paper, we present a new set of fully relativistic 
BHNS simulations that probes
how magnetic fields influence the dynamics and outcome of the merger
using our new adaptive mesh refinement (AMR)
GRMHD code~\cite{els10,APPENDIXPAPER}.
Fixing the BH:NS mass ratio at $q=3$, we consider the cases where the BH
possesses no spin (A cases) and moderate spin $a_{\rm BH}/M_{\rm BH}=0.75$ 
aligned with the orbital angular momentum (B cases).
Since the {\em internal} magnetic field
strength and configuration in a NS is not known, we vary the strength
and geometry of the internal fields to study their effects. We find that for
low and moderate field strengths $\lesssim 10^{16}\rm G$, magnetic fields do not
significantly alter the inspiral and merger dynamics, which is consistent
with the result reported in~\cite{cabllmn10}. Here, by low and moderate magnetic fields we mean those with field strengths small or moderate when 
compared to the virial value of about $10^{18}$G; well below this value
the field is dynamically unimportant. However, 
when the central field strength approaches $\sim 10^{17}$G, corresponding 
to a magnetic to gas pressure ratio of $\sim 0.5$\%, 
the merger dynamics and remnant disk mass are affected
significantly. Yet even with such strong internal magnetic fields,
the emitted GWs are not appreciably different
from the unmagnetized case, at least for the preliminary set of models
considered here. In the late inspiral and merger phases,
tidal deformation and disruption of the NS play key roles in
distinguishing GWs from BHNSs and BHBHs.

During merger, most of the magnetized NS matter is captured by
the BH.  Only when the NS interior is seeded with strong magnetic
fields ($B_{\rm max} \sim 10^{17}$G, near the center of the NS) is a
significant impact on the dynamics observed, resulting in a disk that has up to
twice the rest mass as the corresponding unmagnetized case.  
In all cases the disk accretion rate onto the BH decreases
with time immediately after merger, before settling down to a quasistationary
state.  Most of the magnetic field lines are
tightly wound within the remnant disk, and no evidence of magnetic
field collimation around the final spinning BH is observed by the time 
we terminate our simulations. The remnant disk is 
hot ($T\sim 1$MeV) and massive ($M_{\rm disk} \sim
0.02M_\odot$ and $\sim 0.1M_\odot$ for cases A and B, respectively). 

The magnetic fields threading the remnant disk may be amplified 
and tangled on a longer timescale than we simulate, stirring up MHD turbulence. 
Based on extrapolation of the accretion rates near the end of
our simulations, the lifetime of the disk is roughly 0.3($M_0/1.4M_\odot$)s. 
While there is no evidence of outflows during these preliminary simulations, 
longer disk evolutions, higher
resolution and different B-field geometries may be required 
to definitively assess the possibility of BHNS binaries 
as short-hard GRB progenitors.

The following sections are organized as
follows. Sections~\ref{sec:basic_eqns} and~\ref{sec:numerical}
review the basic equations, including our initial data, gauge
conditions, matter evolution equations, and diagnostics, as well as
their implementation in our GRMHD code. Section~\ref{sec:results}
presents the results of our magnetized BHNS merger
simulations. Finally, we summarize our findings and comment on future
directions in Sec.~\ref{sec:summaryandfuturework}.

\section{Basic Equations}
\label{sec:basic_eqns}

This section introduces our notation, summarizes our method, and
points out the latest changes to our numerical technique as summarized
in~\cite{eflstb08,elsb09,els10,APPENDIXPAPER}.  Geometrized units ($G
= c = 1$) are adopted, except when stated otherwise.  Greek indices
denote all four spacetime dimensions (0, 1, 2, and 3), and Latin
indices label spatial parts only (1, 2, and 3).

We use the 3+1 formulation of general relativity and decompose
the metric into the following form:
\beq
  ds^2 = -\alpha^2 dt^2
+ \gamma_{ij} (dx^i + \beta^i dt) (dx^j + \beta^j dt) \ .
\eeq
The fundamental variables for metric evolution are the spatial
three-metric $\gamma_{ij}$ and extrinsic curvature $K_{ij}$. We adopt
the Baumgarte-Shapiro-Shibata-Nakamura (BSSN) 
formalism~\cite{SN,BS} in which 
the evolution variables are the conformal exponent $\phi
\equiv \ln (\gamma)/12$, the conformal 3-metric $\tilde
\gamma_{ij}=e^{-4\phi}\gamma_{ij}$, three auxiliary functions
$\tilde{\Gamma}^i \equiv -\tilde \gamma^{ij}{}_{,j}$, the trace of
the extrinsic curvature $K$, and the trace-free part of the conformal extrinsic
curvature $\tilde A_{ij} \equiv e^{-4\phi}(K_{ij}-\gamma_{ij} K/3)$.
Here, $\gamma={\rm det}(\gamma_{ij})$. The full spacetime metric $g_{\mu \nu}$
is related to the three-metric $\gamma_{\mu \nu}$ by $\gamma_{\mu \nu}
= g_{\mu \nu} + n_{\mu} n_{\nu}$, where the future-directed, timelike
unit vector $n^{\mu}$ normal to the time slice can be written in terms
of the lapse $\alpha$ and shift $\beta^i$ as $n^{\mu} = \alpha^{-1}
(1,-\beta^i)$. Evolution equations for these BSSN variables are 
given by Eqs.~(9)--(13) in~\cite{eflstb08}. 
We adopt standard puncture gauge conditions: an advective
``1+log'' slicing condition for the lapse and a
``$\Gamma$-freezing'' condition for the shift~\cite{GodGauge}. The
evolution equations for $\alpha$ and $\beta^i$ are given by
Eqs.~(2)--(4) in~\cite{elsb09}, with the
$\eta$ parameter set to $2.2/M$ for the initially nonspinning BH cases 
and $3.3/M$ for the spinning BH cases,
where $M$ is the ADM mass of the BHNS binary.
We add a fifth-order Kreiss-Oliger dissipation term to all evolved
BSSN, lapse and shift variables to reduce high-frequency numerical
noise associated with AMR refinement interfaces 
(see \cite{BSBook} for a review and references).

The fundamental MHD variables are the rest-mass density 
$\rho_0$, specific internal energy $\epsilon$, pressure $P$, 
four-velocity $u^{\mu}$, and magnetic field $B^\mu=n_{\nu} F^{* \nu \mu}$. 
Here $F^{* \mu \nu}$ is the dual of the Faraday tensor $F^{\mu \nu}$. 
Note that $B^\mu$ is purely spatial ($B^0 =-n_\mu B^\mu/\alpha=0$). 
We adopt a $\Gamma$-law equation of state (EOS)
$P=(\Gamma-1)\rho_0 \epsilon$ with $\Gamma=2$, which reduces to
an $n=1$ polytropic law for the initial (cold) neutron star matter.
The stress-energy tensor is given by
\beq
  T_{\mu \nu} = (\rho_0 h +b^2) u_\mu u_\nu 
+ \left(P +\frac{b^2}{2}\right) g_{\mu \nu} - b_\mu b_\nu \ ,
\eeq
where $h=1+\epsilon+P/\rho_0$ is the specific enthalpy and 
\beq
  b_\mu = -\frac{P_{\mu \nu} B^\nu}{\sqrt{4\pi}\, n_{\nu} u^\nu}
\label{eq:bmu}
\eeq
is the magnetic field measured in fluid's comoving frame, modulo 
a factor of $1/\sqrt{4\pi}$. Here $P_{\mu \nu}=g_{\mu \nu} + u_\mu u_\nu$ 
and $b^2=b^\mu b_\mu$. The comoving magnetic energy density is
$u_\mu u_\nu T^{\mu\nu}_{EM}=b^2/2$, where 
$T^{\mu \nu}_{\rm EM} = b^2 u^\mu u^\nu + (b^2/2) g^{\mu \nu} -b^\mu b^\nu$
is the stress-energy tensor associated with the magnetic field.
 In the ideal MHD limit, in which the plasma is 
assumed to have perfect conductivity, the Faraday tensor can be written 
as $F^{\mu \nu} = \sqrt{4\pi}\, u_{\gamma} \epsilon^{\gamma \mu \nu \delta} b_\delta$.

In the standard numerical implementation of the MHD
equations using a conservative scheme, it is useful to introduce the 
``conservative'' variables 
$\rho_*$, $\tilde{S}_i$, $\tilde{\tau}$ and $\tilde{B}^i$. They are 
defined as 
\beqn
&&\rho_* \equiv - \sqrt{\gamma}\, \rho_0 n_{\mu} u^{\mu} \ ,
\label{eq:rhos} \\
&& \tilde{S}_i \equiv -  \sqrt{\gamma}\, T_{\mu \nu}n^{\mu} \gamma^{\nu}_{~i}
\ , \\
&& \tilde{\tau} \equiv  \sqrt{\gamma}\, T_{\mu \nu}n^{\mu} n^{\nu} - \rho_* \ , 
\label{eq:S0} \\
&& \tilde{B}^i \equiv \sqrt{\gamma}\, B^i .
\label{eq:Btilde}
\eeqn
The evolution equations for $\rho_*$, $\tilde{S}_i$ and $\tilde{\tau}$ can 
be derived from the conservation of rest mass $\nabla_\mu (\rho_* u^\mu)=0$ 
and the conservation of energy-momentum 
$\nabla_\mu T^{\mu \nu}=0$, giving rise to 
Eqs.~(27)--(30) in~\cite{els10}. 

In the ideal MHD limit, the Maxwell equation 
$\nabla_\nu F^{* \mu \nu}=0$ yields the magnetic constraint
$\partial_j \tilde{B}^j=0$ and induction equation $\partial_t
\tilde{B}^i + \partial_j (v^j \tilde{B}^i - v^i \tilde{B}^j)=0$. As
shown in~\cite{bs03} and \cite{els10}, these equations can be
rewritten by introducing the electromagnetic 4-vector potential
${\cal A}_\mu = \Phi n_\mu+ A_\mu$, with $n^\mu A_\mu=0$. 
The magnetic constraint and induction equations become 
\beqn
  B^i &=& \epsilon^{ijk} \partial_j A_k \ , \label{eq:BfromA} \\ 
  \partial_t A_i &=& \epsilon_{ijk} v^j B^k 
- \partial_i (\alpha \Phi -\beta^j A_j) \ ,
\label{eq:Aevol}
\eeqn
where $\epsilon^{ijk}=n_\mu \epsilon^{\mu ijk}$ is the 3-dimensional 
Levi-Civita tensor. In~\cite{els10}, we evolve the vector potential and
choose the algebraic EM gauge $\Phi=\beta^j A_j/\alpha=-n^j A_j$. We have found
that for BHNS simulations, we can achieve better numerical results by imposing the
Lorenz gauge $\nabla_\mu {\cal A}^\mu = 0$, which gives the evolution equation 
\beq
  \partial_t (\sqrt{\gamma}\, \Phi) + \partial_j (\alpha \sqrt{\gamma}\, A^j 
- \sqrt{\gamma}\, \beta^j \Phi) = 0 
\label{eq:Phievol}
\eeq
for $\Phi$ in place of the algebraic gauge
condition~\cite{APPENDIXPAPER}. Our numerical implementation of
Eqs.~(\ref{eq:BfromA}) and (\ref{eq:Aevol}) guarantees numerically
identical $B^i$ (see~\cite{els10}) regardless of EM gauge
in simulations with a uniform-resolution grid.  However,
interpolations performed on $A_i$ at refinement boundaries on AMR
grids will modify $A_i$, resulting in different $B^i$ near these
boundaries.  We have shown that in the algebraic gauge $\Phi=-n^j
A_j$, 
there exists a zero-speed mode, which in BHNS simulations manifests itself
as a trail of nonzero $A_i$ left behind the orbiting NS~\cite{APPENDIXPAPER}.
When AMR refinement boxes
tracking the motion of the NS cross this ``trail'', spurious, strong
magnetic fields appear on the refinement boundaries.  On the other
hand, the Lorenz gauge exhibits no zero-speed modes. As a
result, the behavior of the $B^i$ fields on refinement boundaries is 
drastically improved~\cite{APPENDIXPAPER}.  We therefore adopt the
Lorenz gauge for all simulations in this paper.

\section{Numerical Methods}
\label{sec:numerical}

\subsection{Initial data}
\label{sec:initialdata}

Our initial data are constructed by solving Einstein's constraint
equations in the conformal thin-sandwich (CTS) formalism, which allows
us to impose an approximate helical Killing vector by setting the time
derivatives of the conformally related metric $\tilde{\gamma}_{ij}$ 
to zero in the frame corotating with the binary. We model the NS
as an irrotational $n=1$ polytrope, and impose the black hole
equilibrium boundary conditions of Cook and Pfeiffer~\cite{CP04}
on the black hole horizon. The CTS initial data correspond to a 
binary in circular quasiequilibrium with a separation chosen to 
be outside the tidal disruption radius.
Details of this method can be found in~\cite{TBFS07b,elsb09}.  
The initial data used in this paper are the same as case~A (for 
an initially nonspinning BH) and case~B (for an initial BH spin $J_{\rm BH}/M_{\rm BH}^2=0.75$) 
described in~\cite{elsb09}.

The initial data are calculated using the {\tt Lorene} spectral methods numerical
libraries \cite{web:Lorene}. 
The
excised BH region is filled with constraint-violating initial data, using the ``smooth junk''
technique we developed and validated in~\cite{EFLSB} (see also 
\cite{Turducken,Turducken2}).   
In particular, we extrapolate all initial data quantities from the BH
exterior into the interior with a 7$^{\rm th}$ order polynomial, using a uniform
stencil spacing of $\Delta r \approx 0.3 r_{\rm AH}$. 

All of the NSs considered in this paper have a compaction of
${\cal C}=M_{\rm NS}/R_{\rm NS} = 0.145$, where
$M_{\rm NS}$ is the ADM mass and $R_{\rm NS}$ is the
(circumferential) radius of the NS in isolation. Since we model
the NS with an $n=1$
($\Gamma=2$) polytropic EOS, the rest mass of the NS, $M_0$, scales with the
polytropic constant $\kappa$ as $M_0 \propto \kappa^{1/2}$.
For a NS with compaction ${\cal C}=0.145$, we find the ADM mass
for the isolated NS to be $M_{\rm NS}=1.30M_\odot (M_0/1.4M_\odot)$,
with an isotropic radius $R_{\rm iso}=11.2{\rm km} (M_0/1.4M_\odot)$
and circumferential (Schwarzschild) radius of
$R_{\rm NS}=13.2{\rm km} (M_0/1.4M_\odot)$. The maximum
rest-mass density of this NS is $\rho_{0,\rm max} = 9\times
10^{14} \mbox{g cm}^{-3} (1.4M_\odot/M_0)^2$.

\begin{figure}
\epsfxsize=3.4in
\leavevmode
\epsffile{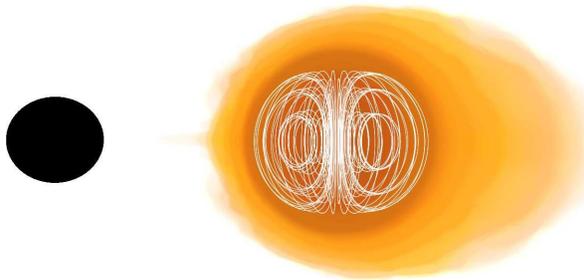}
\caption{Magnetic field lines (white) at the time when they are seeded 
into the NS for case A4. The NS density profile is shown such that darker 
colors indicate higher densities.
The BH apparent horizon is shown in
black, and the BH-NS 
coordinate separation is $6.12M$}
\label{fig:binitial}
\end{figure}

We add a small, poloidal magnetic field via the vector potential of the form 
\beqn
  A_i &=& \left( -\frac{y-y_c}{\varpi_c^2}\delta^x{}_i 
+ \frac{x-x_c}{\varpi^2_c}\delta^y{}_i\right) A_\varphi \label{ini:Ai} \\ 
  A_\varphi &=& A_b \varpi^2_c \max (P- P_{\rm cut},0)^{n_b} 
\label{ini:Aphi}
\eeqn
where $(x_c,y_c,0)$ is the coordinate location of the center of mass of the NS,
$\varpi^2_c=(x-x_c)^2+(y-y_c)^2$, and $A_b$, $n_p$ and $P_{\rm cut}$ are free
parameters. The cutoff pressure parameter $P_{\rm cut}$ confines the B-field
inside the neutron star to reside
within $P>P_{\rm cut}$. The parameter $n_b$ determines the degree of 
central condensation of the magnetic field. Similar profiles of 
initial magnetic fields are also used in numerical 
simulations of magnetized accretion disks (see, e.g.~\cite{dhk03,mg04}) and
magnetized compact binaries (see, e.g.~\cite{ahllmnpt08,grb11,cabllmn10,lset08}).
We set $P_{\rm cut}$
to be 4\% of the maximum pressure and $n_b$=1 or 2. The parameter $A_b$ controls 
the strength of the initial magnetic field, which can be characterized by 
the maximum magnetic field inside the NS, as well as the magnetic energy 
${\cal M}$ defined as 
\beq
  {\cal M} = \int n_{\mu} n_{\nu} T^{\mu \nu}_{\rm EM} dV ,
\label{eq:Eem}
\eeq
where 
$dV=\psi^6 d^3x$ is the proper volume element on a $t=constant$ spatial
slice. ${\cal M}$ is the EM energy measured by a normal observer.

Since the magnetic field is expected to remain frozen into the NS during 
the inspiral phase, we add the fields immediately 
before tidal disruption to minimize numerical error. 
Magnetic fields added to the NS at $t=0$ maintain their original 
profile within the star for much of the first orbit 
($t\lesssim$ milliseconds).  
Table~\ref{table:id} summarizes the initial data used in our simulations.
Figure~\ref{fig:binitial} shows the magnetic field configuration for one of the 
cases (A4) at $t=448.5M$, the time at which the NS is seeded with
magnetic fields.
The seed magnetic fields are too weak
to significantly perturb the quasiequilibrium NS,
leading to virtually no change in
gravitational field constraint violations.

\begin{table*}
\caption{Initial data for the BHNS simulations. Here $a_{BH}/M_{BH}$ is the BH spin, $\Omega$
is the orbital angular frequency, $B_{\rm max}$ 
is the maximum value of magnetic field inside the NS, assuming the rest mass 
of the NS is $1.4 M_\odot$; ${\cal M}$ is the energy of magnetic field defined in Eq.~(\ref{eq:Eem}) at the
time when B-field is added; $W$ is the gravitational potential energy of the NS
in isolation defined in Eq.~(65) of~\cite{CookShapTeuk}. All BHNS systems considered 
here have the BH:NS mass ratio of 3:1.}
\begin{tabular}{cccccccc}
  \hline
  Case & $a_{BH}/M_{BH}$ & $M\Omega$ & $B_{\rm max}$ (G) & ${\cal M}/|W|$
& $P_c$ & $n_b$ & Time $B$ is added \\
  \hline
  A0 & 0.0 & 0.0330 & 0 & 0 & -- & -- & -- \\
  A1 &  &  & $1.3\times 10^{16}$ & $3.1 \times 10^{-6}$ & 0.04 & 2 & 448.5$M$ \\
  A2 &  &  & $1.3\times 10^{16}$ & $1.1 \times 10^{-5}$ & 0.04 & 1 &  \\
  A3  &  &  & $1.4\times 10^{16}$ & $1.2 \times 10^{-5}$ & 0.001 & 1 & \\
  A4  &  &  & $9.7\times 10^{16}$ & $5.9 \times10^{-4}$ & 0.001 & 1 & \\ 
   & & & & & & & \\
  B0 & 0.75  & 0.0328 & 0 & 0 & -- & -- & -- \\
\hspace{0.05cm}  B1a &    &  & $1.3\times 10^{16}$ & $3.1 \times 10^{-6}$ & 0.04 & 2 & 633.6$M$ \\
\hspace{0.05cm}  B1b &    &  & $1.2\times 10^{16}$ & $2.9 \times 10^{-6}$ & 0.04 & 2 & 752.8$M$ \\
  B2 &    &  & $1.3\times 10^{16}$ & $1.1 \times 10^{-5}$ & 0.04 & 1 &  \\
  B3 &    &  & $1.4\times 10^{16}$ & $1.2 \times 10^{-5}$ & 0.001 & 1 &  \\
  B4  &    &  & $9.7\times 10^{16}$ & $5.9 \times10^{-4}$ & 0.001 & 1 &  \\
\hline
\end{tabular}
\label{table:id}
\end{table*}

\subsection{Evolution of the metric and MHD}
\label{sec:num_metric_hydro}

We evolve the BSSN equations
with fourth-order accurate, centered finite-differencing stencils,
except on shift advection terms, where we use fourth-order accurate
upwind stencils.  We apply Sommerfeld outgoing wave boundary
conditions to all BSSN fields.  Our code is embedded in
the Cactus parallelization framework~\cite{Cactus}, and our
fourth-order Runge-Kutta timestepping is managed by the {\tt MoL}
(Method of Lines) thorn, with a Courant-Friedrichs-Lewy (CFL) factor
set between 0.0625 (in the coarsest refinement level) and 0.5 (in the innermost 
4 refinement levels) in all simulations. We decrease the CFL factor 
in the coarse refinement levels so that we can use a larger value for 
the parameter $\eta$ in the shift equation [Eq.~(4) in~\cite{elsb09}].
We use the
Carpet~\cite{Carpet} infrastructure to implement the moving-box
adaptive mesh refinement. In all AMR simulations presented here, we
use second-order temporal prolongation, coupled with fifth-order
spatial prolongation. The apparent horizon (AH) of the
BH is computed with the {\tt AHFinderDirect} Cactus
thorn~\cite{ahfinderdirect}.

The GRMHD equations are evolved by a high-resolution
shock-capturing (HRSC) technique~\cite{DLSS} that employs 
PPM~\cite{PPM} coupled to
the Harten, Lax, and van Leer (HLL) approximate Riemann solver~\cite{HLL}.
The adopted MHD scheme is second-order accurate for smooth
flows, and first-order accurate when discontinuities (e.g.\ shocks)
arise. To stabilize our scheme in regions where there is no
matter, we maintain a tenuous atmosphere on our grid, with a density
floor $\rho_{\rm atm}$ set equal to $10^{-10}$ times the initial
maximum density on our grid. The initial atmospheric pressure
$P_{\rm atm}$ is set equal to the cold
polytropic value $P_{\rm atm} = \kappa \rho_{\rm atm}^{\Gamma}$.
Throughout the
evolution, we impose limits on the atmospheric pressure to prevent
spurious heating and negative values of the internal energy
$\epsilon$ due to numerical errors. Specifically, we require $P_{\rm min}\leq P \leq P_{\rm max}$, 
where $P_{\rm max}=10 \kappa \rho_0^\Gamma$ and $P_{\rm min}=\kappa 
\rho_0^\Gamma/2$. 
Whenever $P$ exceeds $P_{\rm max}$ or drops below $P_{\rm min}$, we 
reset $P$ to $P_{\rm max}$ or $P_{\rm min}$, respectively.  Applying
these limits everywhere on our grid would artificially
sap the angular momentum in the tidally disrupted NS, allowing
matter to fall spuriously into the BH and thereby suppressing 
disk formation~\cite{elsb09}. To effectively 
eliminate this spurious
angular momentum loss, we impose these pressure limits only 
in regions where the rest-mass density remains very low ($\rho_0 < 100
\rho_{\rm atm}$) or deep inside the AH, where $\psi^6 > \psi^6_{\rm
  thr}$ as in~\cite{elsb09}. Here $\psi=e^\phi$ and we set 
$\psi^6_{\rm thr}$ between 10 and 30.

\subsection{Evolution of magnetic field}

We evolve the magnetic induction equation via the 4-vector
potential using Eqs.~(\ref{eq:Aevol}) and (\ref{eq:Phievol}).
We stagger the $A_i$ and $B^i$ as in~\cite{els10}.
We store $\Phi$ on a staggered grid $(i^+,j^+,k^+)$ [all the other
hydrodynamic, BSSN, lapse and shift variables are stored at $(i,j,k)$],
where $i^+=i+1/2$ and similarly for $j^+$ and $k^+$.
We treat the term $-\partial_j (\beta^j \sqrt{\gamma}\, \Phi)$ in
Eq.~(\ref{eq:Phievol}) using a second-order upwind scheme. We evolve
Eq.~(\ref{eq:Aevol}) using the finite-volume equations similar to
Eqs.~(63)--(65) in~\cite{els10},
modified to take into account the second term in Eq.~(\ref{eq:Aevol}),
which does not vanish in the Lorenz gauge. The detailed implementation is described
in~\cite{APPENDIXPAPER}.

The particular staggering of the $A_i$ and $\Phi$ variables coupled with the 
particular implementation of the HRSC scheme 
are designed to ensure that the resulting 
$B$ field obtained by taking the curl operator on $A_i$ [Eqs.~(60)--(62)
in~\cite{els10}] is numerically identical to the standard constrained
transport scheme based on a staggered algorithm~\cite{eh88}.
We have carefully designed an algorithm for the extra term in Eq.~(\ref{eq:Aevol})
so that the additional terms in the $A_i$ evolution equations
cancel exactly after taking the
curl operator. The resulting numerical values of $B^i$ are thus
gauge-invariant in unigrid simulations. We have confirmed numerically
that this is indeed the case. However, in simulations with an AMR grid,
since we perform interpolations on $A_i$ between refinement levels,
values of $A_i$ are not the same in different EM gauges at the refinement
boundaries. The resulting $B$ field at the refinement boundaries is
also different in general but should converge to a unique, true solution
with increasing resolution in any gauge.

As in other numerical relativity
simulations, some gauges are better behaved than others. 
We have demonstrated in~\cite{APPENDIXPAPER} that the 
Lorenz gauge is superior to the algebraic gauge 
in magnetized BHNS simulations. We therefore adopt the Lorenz gauge 
in all of the magnetized BHNS simulations presented here.

\subsection{Recovery of primitive variables}
\label{sec:inversion}

At each timestep,
we need to recover the ``primitive variables''
$\rho_0$, $P$, and $v^i$ from the ``conservative'' variables
$\rho_*$, $\tilde{\tau}$, and $\tilde{S}_i$.
We perform the inversion by numerically solving two nonlinear
equations via the Newton-Raphson method as described in~\cite{ngmz06},
using the code developed by Noble et al~\cite{harmsolver}.

Sometimes the ``conservative'' variables may assume values which are 
out of physical range, resulting in unphysical primitive
variables after inversion (e.g.\ negative pressure or even
complex solutions). This usually happens in the low-density 
``atmosphere'' or deep inside the BH interior where high-accuracy 
evolution is difficult to maintain. Various techniques have been 
suggested to handle the inversion failure (see, e.g.~\cite{bs2011}). 
Our approach is mainly to impose constraints on the conservative 
variables to reduce the inversion failure. 

One reason for the inversion failure comes from $\gamma_{ij}$ 
losing positive-definiteness during the BSSN evolution due to 
numerical inaccuracy, which occurs only near the ``puncture'' deep inside 
the BH. Before performing 
the inversion, we check if $\gamma_{ij}$ is positive-definite by finding its eigenvalues. 
If $\gamma_{ij}$ is not positive definite, we reset 
$\gamma_{ij} \rightarrow \psi^4 f_{ij}$ during the inversion, where 
$f_{ij}$ is the 3D flat metric tensor.

In the absence of magnetic fields, the inversion failure 
can be avoided completely by enforcing the constraints (see~\cite{eflstb08} 
and Appendix~\ref{sec:inequalities})
\beqn
 \tilde{S}^2 \equiv \gamma^{ij} \tilde{S}_i \tilde{S}_j &\leq & \tilde{\tau}
(\tilde{\tau}+2\rho_*), {\rm and} \label{eq:S2con} \\
  \tilde{\tau} &\geq& 0 \ , \label{eq:taucon}
\eeqn
which are the necessary and sufficient conditions for the inversion to
produce the primitive variables in the physical range for the $\Gamma$-law
EOS with $1 < \Gamma \leq 2$ (see Appendix~\ref{sec:inequalities}).
We enforce these constraints in regions where there are no magnetic fields.
When the second condition is not met, we reset 
$\ttau=\ttau_{\rm atm}=10^{-10} \ttau_{\rm 0max}$, where $\ttau_{\rm 0max}$ 
is the maximum value of $\ttau$ initially.  
When the first condition is violated we rescale
$\tilde{S}_i$ so that its
new magnitude is $\tilde{S}^2=
\tilde{\tau}(\tilde{\tau}+2\rho_*)$. 

In the presence of 
magnetic fields, no simple analogous formulae are available. 
However, one can prove that (see Appendix~\ref{sec:inequalities})
\beq
  \tilde{\tau} \geq \psi^{-6} \frac{ \tilde{B}^2}{8\pi} 
\label{eq:tau_ineq1}
\eeq
for any value of primitive variables in the physical range. 
We therefore impose inequality (\ref{eq:tau_ineq1}) everywhere: 
if it is violated, we reset $\tilde{\tau}=\ttau_{\rm atm} +
\psi^{-6} \tilde{B}^2/8\pi$. However, this does 
not guarantee that the inversion always produce physically acceptable 
primitive variables. If failures occur outside the BH after imposing \eqref{eq:tau_ineq1}, we apply a fix, which consists of
replacing the energy equation~(\ref{eq:S0}) by the cold EOS,
$P =P_{\rm cold}(\rho_0)= \kappa \rho_0^\Gamma$ when solving the system of equations, where $\kappa$ is
the polytropic constant. One can show that this procedure always produces
physically valid primitive variables (see Sec.~\ref{sec:fontfix}). However, we find that this
fix gives rise to discontinuous data deep inside the BH and these
data will eventually propagate out of the BH horizon. Note that
the success of the ``smooth junk'' technique~\cite{EFLSB} requires
that the constraint violating initial data filling the interior 
of the BH horizon be {\it smooth}. If discontinuous 
data are used, then information can leak out of the BH horizon. 
To avoid information leakage outside the BH, we apply the following conditions
deep inside the BH:
\beq
  \tilde{\tau} - \frac{\psi^6}{2} \bar{B}^2
 - \frac{ \bar{B}^2 \tilde{S}^2 - (\bar{B}^i \tilde{S}_i)^2}
{2\psi^6 (W_{m}+\bar{B}^2)^2} \equiv \tilde{\tau}_m \geq 0 ,
\label{eq:tau_ineqs}
\eeq
and
\beq
  \tilde{\tau}_m (\tilde{\tau}_m+2\rho_*) \geq \tilde{S}^2  ,
\label{eq:tau_stilde_ineqs}
\eeq
where
$\bar{B}^i = B^i/\sqrt{4\pi}$, $\bar{B}^2 = \gamma_{ij} \bar{B}^i \bar{B}^j$, and
$W_m$ satisfies the quartic equation 
\begin{eqnarray}
\nonumber  (\psi^{12} W_m^2-\rho_*^2)(W_m+\bB^2)^2&  &-W_m^2\tS^2  \\
\nonumber & & - (\bB^i \tS_i)^2(\bB^2+2W_m)=0 .
\end{eqnarray}
It can be shown that the inequalities
(\ref{eq:tau_ineqs}) and (\ref{eq:tau_stilde_ineqs}) are {\it sufficient}
(but not necessary) conditions for the inversion to yield physically
valid primitive variables (see Appendix~\ref{sec:inequalities}). 
We therefore only use them deep inside the 
BH where $\psi^6 > \psi^6_{\rm thr}$. We choose the parameter 
$\psi^6_{\rm thr}$ between 10 and 30. 
Since the inequalities~(\ref{eq:tau_ineqs}) 
and (\ref{eq:tau_stilde_ineqs}) are sufficient but not necessary conditions, 
we do not impose them strictly, but adopt the procedures described at the end of 
Sec.~\ref{app:taust_mhd}. We find that this technique gives rise to smoother
data in the BH interior preventing contraint-violating information from leaking
out of the BH horizon.

\subsection{Diagnostics}
\label{sec:diagnostics}

During the evolution, we monitor the Hamiltonian and momentum
constraints calculated by Eqs.~(40)--(43) of~\cite{eflstb08}.
We also monitor the interior mass $M_{\rm int}$ and ($z$-component of) 
the interior
 angular momentum $J_{\rm int}$ 
of the system contained in the simulation domain. These quantities are defined 
as integrals over the surface of the outer boundary $\partial V)$ of the computational domain:
\beqn
  M_{\rm int} &=&\frac{1}{2\pi}\oint_{\partial V} \left(\frac{1}{8}\tilde{\Gamma}^i -
\tilde{\gamma}^{ij}\partial_j\psi \right)d\Sigma_i,\label{eq:mintsurf}\\
J_{\rm int}&=&\frac{1}{8\pi}{\tilde{\epsilon}_{zj}}^k\oint_{\partial V} x^j(K^m_k-\delta^m_k
K)d\Sigma_m,\label{eq:jintsurf}
\eeqn 
where $\tilde{\epsilon}_{ijk}$ is the flat-space Levi-Civita tensor.
As pointed out in~\cite{elsb09}, the integrals can be evaluated more accurately 
by alternative expressions via Gauss's law~\cite{BSBook}: 
\beqn
  M_{\rm int}&=& \int_V d^3x \left(\psi^5\rho + {1\over16\pi}\psi^5 \tilde{A}_{ij}
\tilde{A}^{ij} - {1\over16\pi}\tilde{\Gamma}^{ijk}\tilde{\Gamma}_{jik} \right.
\ \  \cr
&& \left. + {1-\psi\over16\pi}\tilde{R} - {1\over24\pi}\psi^5K^2\right) \cr
 && + {1\over 2\pi} \oint_S
\left( \frac{1}{8}\tilde{\Gamma}^i-\tilde{\gamma}^{ij} \partial_j \psi
\right) d\Sigma_i \ , \label{eq:Mint_sur_vol}
\eeqn
\beqn
  J_{\rm int} &=& {1\over8\pi} \tilde{\epsilon}_{zj}{}^n\int_V d^3x
           \psi^6(\tilde{A}^j{}_n + {2\over3}x^j\partial_nK \cr
         && - {1\over2} x^j\tilde{A}_{km}\partial_n\tilde{\gamma}^{km}
         + 8\pi x^j \tilde{S}_n) \cr
           && + {1\over8\pi} \tilde{\epsilon}_{zj}{}^n \oint_S
           \psi^6 x^j \tilde{A}^m{}_n d\Sigma_m \ , \label{eq:Jint_sur_vol}
\eeqn
where $S$ is a surface surrounding the BH horizon, $V$ is the volume between 
$S$ and the outer boundary, $\rho=n_\mu n_\nu T^{\mu \nu}$, and $\tilde{R}$ is 
the Ricci scalar associated with the conformal 3-metric $\tilde{\gamma}_{ij}$.
If our outer boundary were extended to spatial infinity, these integrals
would yield the ADM mass and angular momentum of the system and would be
constant in time. While our outer boundary $\partial V$ resides in the nearly 
Minkowski asymptotic regime, it is at a finite distance from the BHNS
system. Thus the integrals are only approximately equal to the ADM $M$ and $J$
at $t=0$ 
and decreases with time, due to GWs carrying away energy and angular momentum
through $\partial V$.

When hydrodynamic matter is evolved on a fixed uniform grid, our
hydrodynamic scheme guarantees that the rest mass $M_0$ is conserved
to machine roundoff error.  This strict conservation is no longer maintained
in an AMR grid, where spatial and temporal prolongation is performed
at the refinement boundaries.  Hence we also monitor the
rest mass
\beq
  M_0 = \int \rho_* d^3x
\label{eq:m0}
\eeq
during the evolution. Rest-mass conservation is also violated whenever 
$\rho_0$ is reset to the atmosphere value. This usually happens only in the 
very low-density atmosphere or deep inside the AH where high accuracy
is difficult to maintain.

We measure the thermal energy generated by shocks via the entropy 
parameter $K\equiv P/P_{\rm cold}$, where $P_{\rm cold}=\kappa \rho_0^\Gamma$ is
the pressure associated with the cold EOS. 
The specific internal energy can be decomposed into a
cold part and a thermal part: $\epsilon = \epsilon_{\rm cold} +
\epsilon_{\rm th}$ with 
\begin{equation}
\epsilon_{\rm cold} = -\int P_{\rm cold}
d(1/\rho_0) = \frac{\kappa}{\Gamma - 1} \rho_0^{\Gamma-1}\ .
\end{equation}
Hence the relationship between $K$ and $\epsilon_{\rm th}$ is 
\beqn
\epsilon_{\rm th} & = & \epsilon - \epsilon_{\rm cold} =
\frac{1}{\Gamma - 1} \frac{P}{\rho_0} - \frac{\kappa}{\Gamma - 1}
\rho_0^{\Gamma-1} \nonumber \\
& = & (K - 1) \epsilon_{\rm cold} \ .
\eeqn
For shock-heated gas, we always have $K>1$ (see 
Appendix~B of~\cite{elsb09}.

Finally, we monitor the mass and spin of the BH during the evolution. 
They are computed using the isolated and dynamical horizon formalism~\cite{IH_DH},
with the approximate axial Killing vector on the horizon computed as in~\cite{dkss03}.

\subsection{Gravitational wave extraction}

Gravitational
waves are extracted using the Newman-Penrose Weyl scalar $\psi_4$ at
various extraction radii between $50M$ and $130M$. We decompose
$\psi_4$ into $s=-2$ spin-weighted spherical harmonics up to and including
$l=4$ modes. At each extraction radius, the retarded time is computed
using the technique described in Sec.~IIB of~\cite{bm09} to reduce the near-field effect.
The wavetrain $h_+$ and $h_\times$ for each mode are computed
by integrating the corresponding mode of $\psi_4$ twice with time using
the fixed frequency integration technique described in~\cite{rp10}.

We compute the radiated energy $\Delta E_{GW}$, $z$-component of
angular momentum $\Delta J_{GW}$ and linear momentum $\Delta P^i_{GW}$
using expressions equivalent to Eqs.~(33), (39), (40) and (49) of~\cite{rant08}.
To check the violation of energy and angular momentum conservation, we monitor the quantities
\beqn
  \delta E &=& [M-M_{\rm int}(t)-\Delta E_{\rm GW}(t)]/M \ , \label{eq:deltae} \\
  \delta J &=& [J-J_{\rm int}(t)-\Delta J_{\rm GW}(t)]/J \ , \label{eq:deltaj}
\eeqn
where $J$ is the ADM angular momentum of the initial binary,
$M_{\rm int}(t)$ and $J_{\rm int}(t)$ are the interior mass and angular momentum
of the system at time $t$ as calculated by Eqs.~\eqref{eq:Mint_sur_vol}
and \eqref{eq:Jint_sur_vol}.

\section{Results}
\label{sec:results}

\begin{table*}
\caption{Grid configurations. Here, $N_{\rm
    AH}$ denotes the number of grid points covering the diameter of
    the (spherical) AH initially, and $N_{\rm NS}$
    denotes the number of grid points covering the smallest diameter
    of the neutron star initially.}
\begin{tabular}{ccccc}
  \hline
  Case & Grid Hierarchy (in units of $M$)$^{(a)}$
  & Max.~resolution & $N_{\rm AH}$ & $N_{\rm NS}$ \\
  \hline 
  A0--A4 & (196.7, 98.35, 49.18, 24.59, 12.29, 6.147, 3.073,
  1.414 [1.660]) & $M/32.5$ & 41 & 85 \\
  B0--B4 & (210.2, 92.49, 46.24, 23.12, 11.56, 5.780, 2.890,
  1.445 [1.642], 0.7554 [N/A]) & $M/60.9$ & 56 & 80 \\
\hline
\end{tabular}
\begin{flushleft}
$^{(a)}$ There are two sets of nested refinement boxes: one centered
  on the NS and one on the BH.  This column specifies the half side length
  of the refinement boxes centered on both the BH and NS. When
  the side length around the NS is different, we specify the NS half side
  length in square brackets.  If there is no corresponding NS
  refinement box (as is the case when the NS is significantly larger
  than the BH), we write [N/A] for that box.
\end{flushleft}
\label{table:GridStructure}
\end{table*}

\begin{table*}
\caption{Magnetized BHNS simulation results. Here $M_{\rm disk}$ is the
  rest mass of the material outside the AH at the end of the simulation,
  $\tilde{a}_{\rm f}=a_{\rm BH}/M_{\rm BH}$ is the spin parameter of the BH at late times determined by
  the isolated horizon formalism.  The total energy and angular
  momentum carried off by the gravitational radiation are given by $\Delta E_{\rm
  GW}$ and $\Delta J_{\rm GW}$, respectively. $v_{\rm kick}$ is the kick velocity
  due to recoil.
  $N_{\rm orbits}$ specifies the
  number of orbits in the inspiral phase before merger, defined as the time 
  at which the (2,2) mode of the GW amplitude reaches maximum.
  $\delta E$ and $\delta J$ measure the violation of energy and
  angular momentum conservation, as defined in Eqs.~(\ref{eq:deltae})
  and (\ref{eq:deltaj}), respectively, at the end of the simulation. The error in
  $v_{\rm kick}$ is estimated by comparing the
  results obtained by several GW extraction radii between 50$M$--100$M$.
}
\begin{tabular}{ccccccccc}
\hline
Case & $M_{\rm disk}/M_0$ & $\tilde{a}_{\rm f}$ & $\Delta
E_{\rm{GW}}/M$ & $\Delta J_{\rm{GW}}/J$ & $v_{\rm kick}$ (km/s) &
$N_{\rm{orbits}}$ & $\delta E$ & $\delta J$ \\
\hline
A0 & 0.019 & 0.55 & 0.011 & 0.20 &$40\pm 2$ & 4.8 & 0.2\% & 2\% \\
A1 & 0.017 & 0.55 & 0.011 & 0.20 & $40\pm 2$ & 4.8 & 0.2\% & 2\% \\
A2 & 0.017 & 0.55 & 0.011 & 0.20 & $40\pm 2$ & 4.8 & 0.2\% & 2\% \\
A3 & 0.015  & 0.55 & 0.011 & 0.20 & $40\pm 2$ & 4.8 & 0.2\% & 3\% \\
A4 & 0.028 & 0.55 & 0.011 & 0.20 & $40\pm 2$ & 4.8 & 0.1\% & 1\% \\
B0 & 0.098 & 0.84 & 0.011 & 0.15 & $67\pm 6$ & 6.9 & 0.6\% & 8\% \\
\hspace{0.05cm} B1a & 0.090  & 0.84 & 0.011 & 0.15 &$67\pm 6$ & 6.9 & 0.6\% & 8\%  \\
\hspace{0.05cm} B1b & 0.090 & 0.84 & 0.011 & 0.15 & $67\pm 6$ & 6.9 & 0.6\% & 8\% \\
B2 & 0.090 & 0.84 & 0.011 & 0.15 & $67\pm 6$ & 6.9 & 0.6\% & 7\% \\
B3 & 0.090 & 0.84 & 0.011 & 0.15 & $67\pm 6$ & 6.9 & 0.6\% & 7\% \\
B4  & 0.117 & 0.85 & 0.010 & 0.14 & $54 \pm 4$ & 6.8 & 0.4\% & 7\% \\
\hline
\end{tabular}
\label{table:results}
\end{table*}

We have performed magnetized simulations of BHNS binaries with 
BH:NS mass ratio 3:1 including both initially nonspinning BHs (the ``A''
cases) and BHs with spin parameter set to 0.75 initially (the ``B''
cases).  Table~\ref{table:GridStructure} specifies the AMR grid
structure used in the simulations and Table~\ref{table:results}
summarizes the quantitative results.  For readers interested only
in a brief summary of the most interesting results, please skip to
Sec.~\ref{sec:summaryandfuturework}.  Detailed simulation
results are described below.

\subsection{Magnetic Field Study: Nonspinning Black Hole}

\begin{figure*}
\vspace{-4mm}
\begin{center}
\epsfxsize=2.3in
\leavevmode
\epsffile{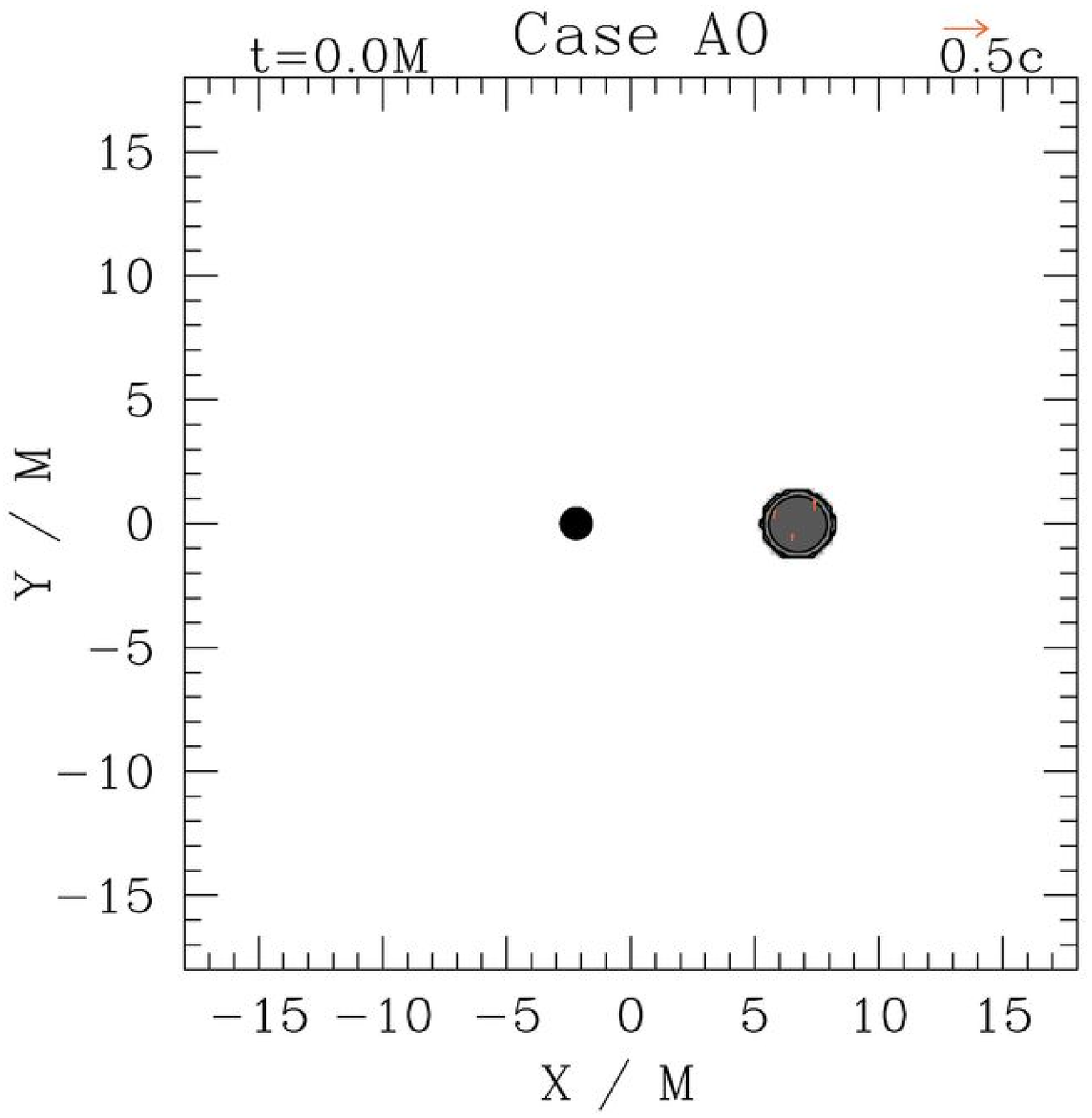}
\epsfxsize=2.3in
\leavevmode
\epsffile{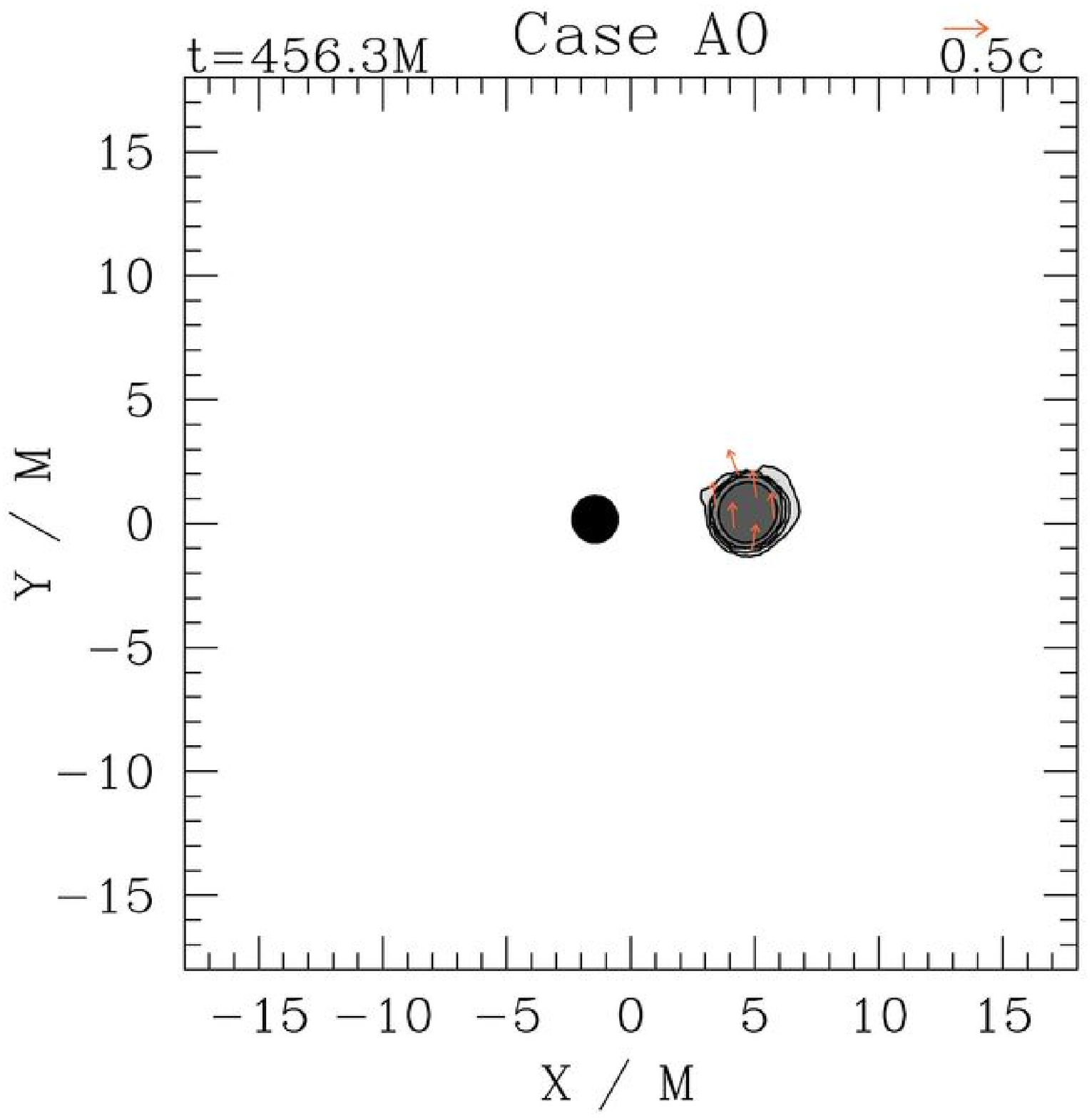}
\epsfxsize=2.3in
\leavevmode
\epsffile{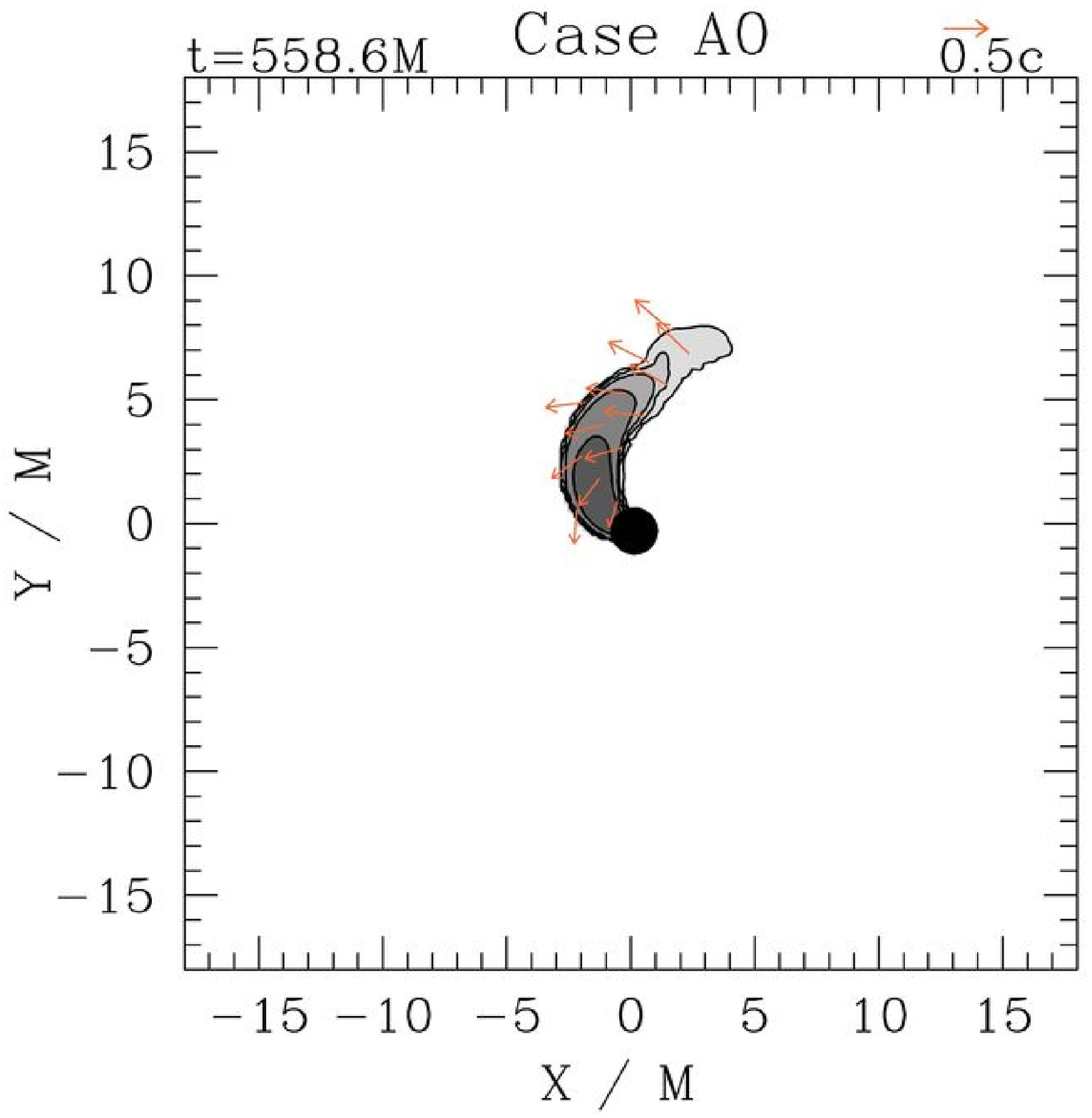}
\epsfxsize=2.3in
\leavevmode
\epsffile{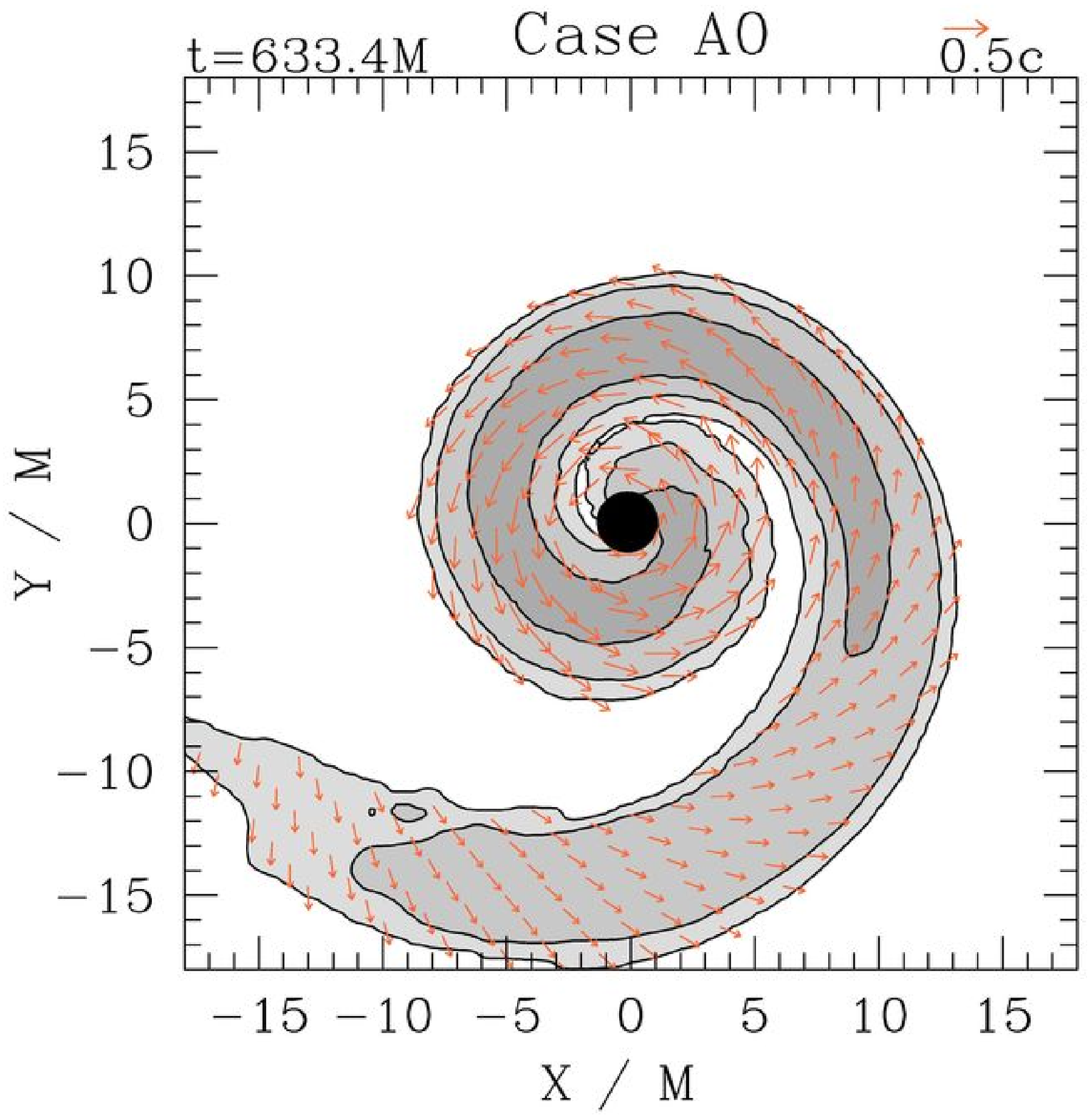}
\epsfxsize=2.3in
\leavevmode
\epsffile{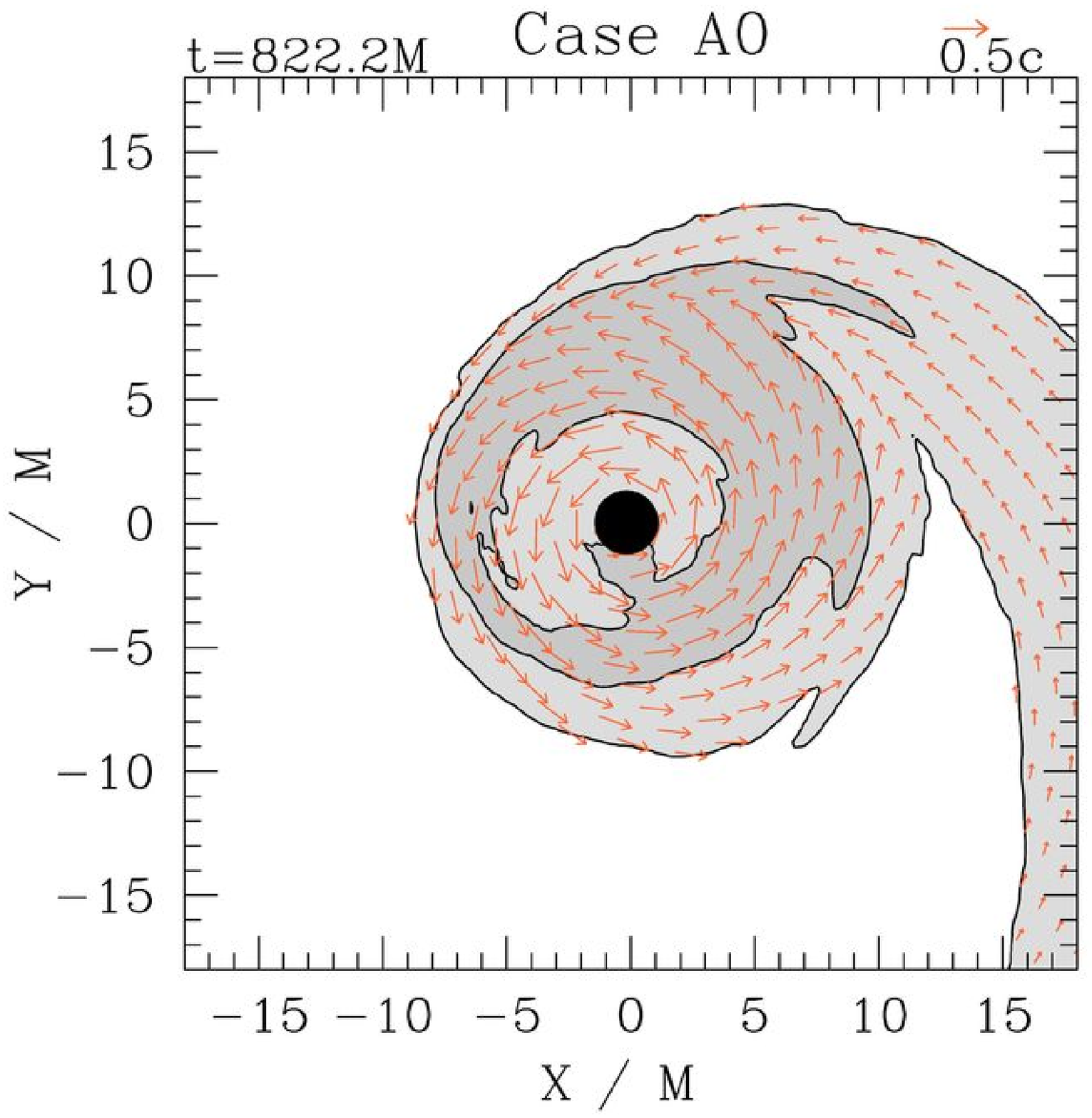}
\epsfxsize=2.3in
\leavevmode
\epsffile{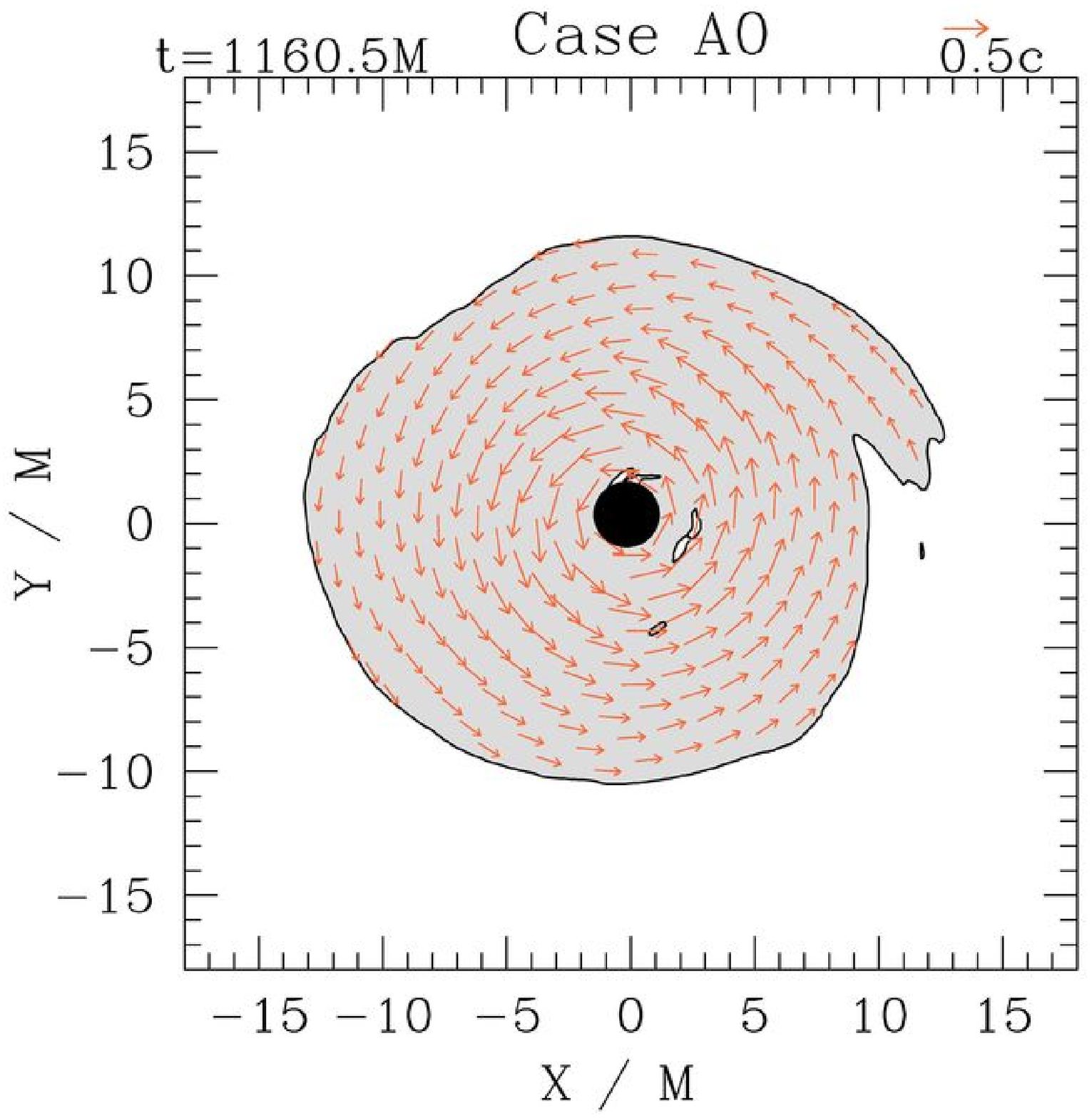}
\caption{Orbital-plane rest mass
  density contours at selected times
  for case A0. Contours are plotted according to $\rho_0 =
  \rho_{0,{\rm max}} (10^{-0.92j})$,
  ($j$=0, 1, ... 5), with darker greyscaling for higher density.  The maximum initial NS density is
 $\kappa \rho_{0,{\rm max}} = 0.126$, or $\rho_{0,{\rm
      max}}=9\times 10^{14}\mbox{g cm}^{-3}(1.4M_\odot/M_0)^2$.
  Arrows represent the velocity field in the
  orbital plane. The black hole AH interior is marked by a filled
  black circle.  The ADM mass for this case is
  $M=2.5\times 10^{-5}(M_0/1.4M_\odot)$s$=7.6(M_0/1.4M_\odot)$km.}
\label{A:evolution_story}
\end{center}
\end{figure*}

\begin{figure*}
\vspace{-4mm}
\begin{center}
\epsfxsize=2.2in
\leavevmode
\epsffile{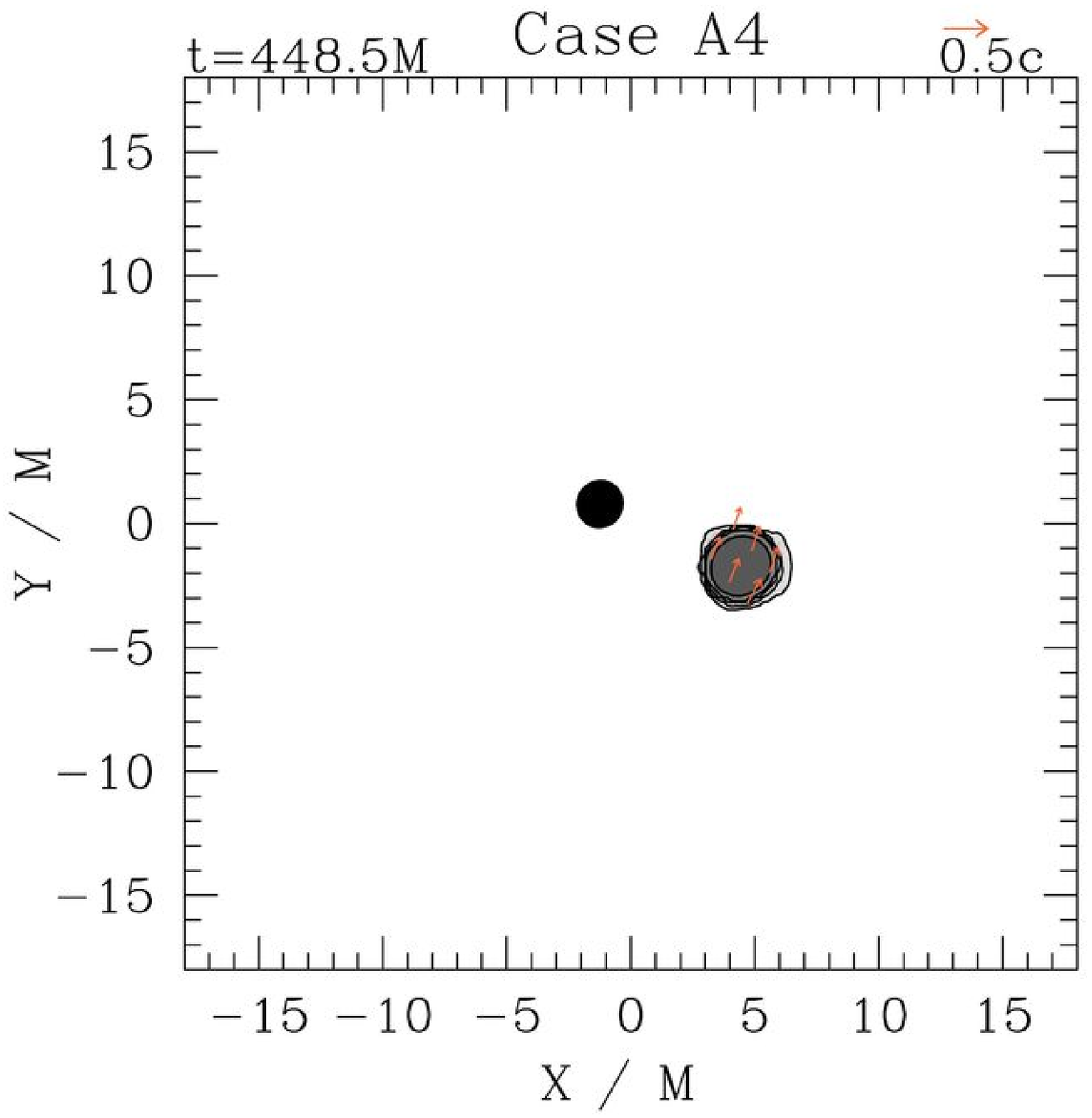}
\epsfxsize=2.3in
\leavevmode
\epsffile[20 -60 575 437]{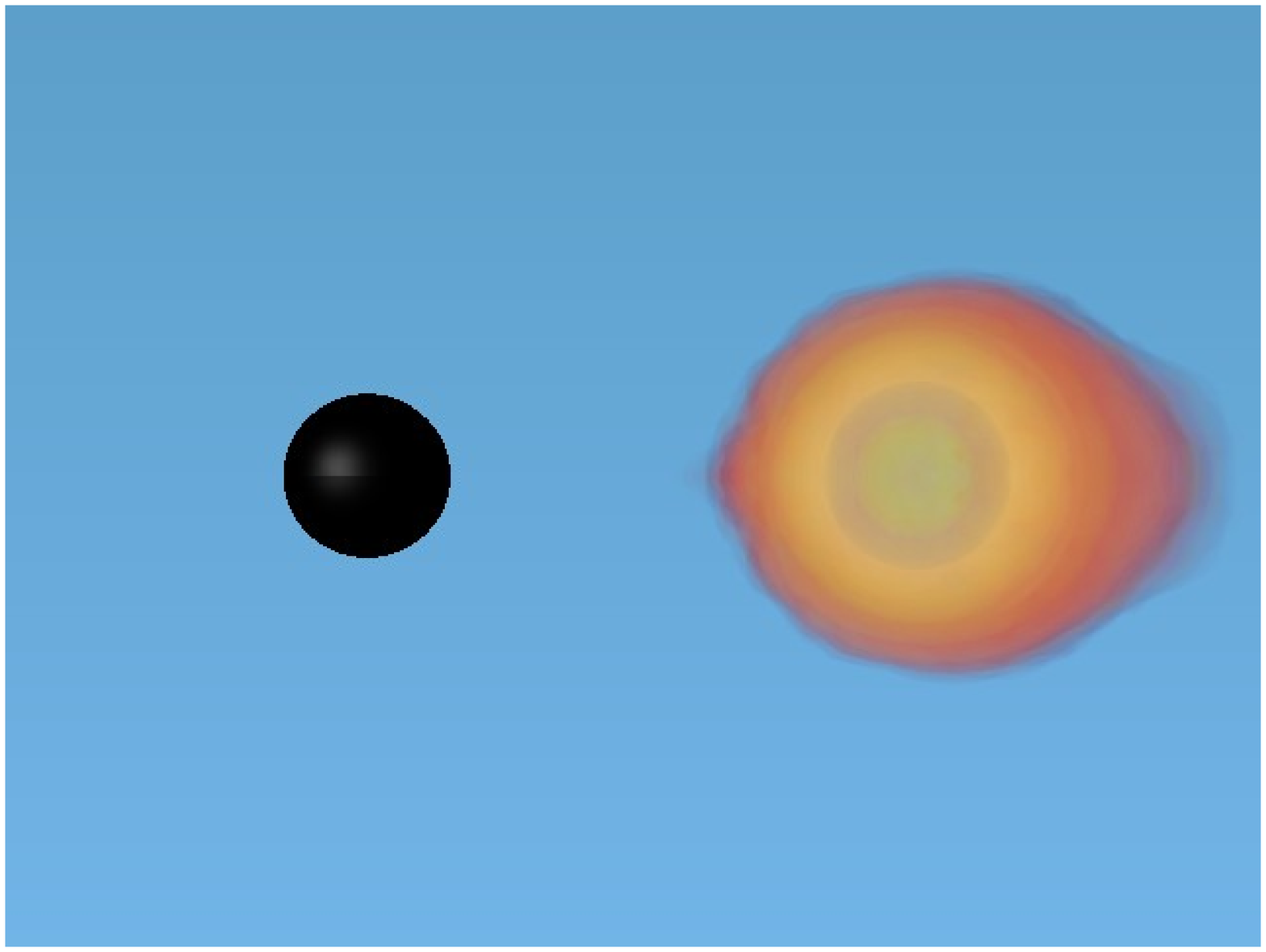}
\epsfxsize=2.3in
\leavevmode
\epsffile[20 -60 575 437]{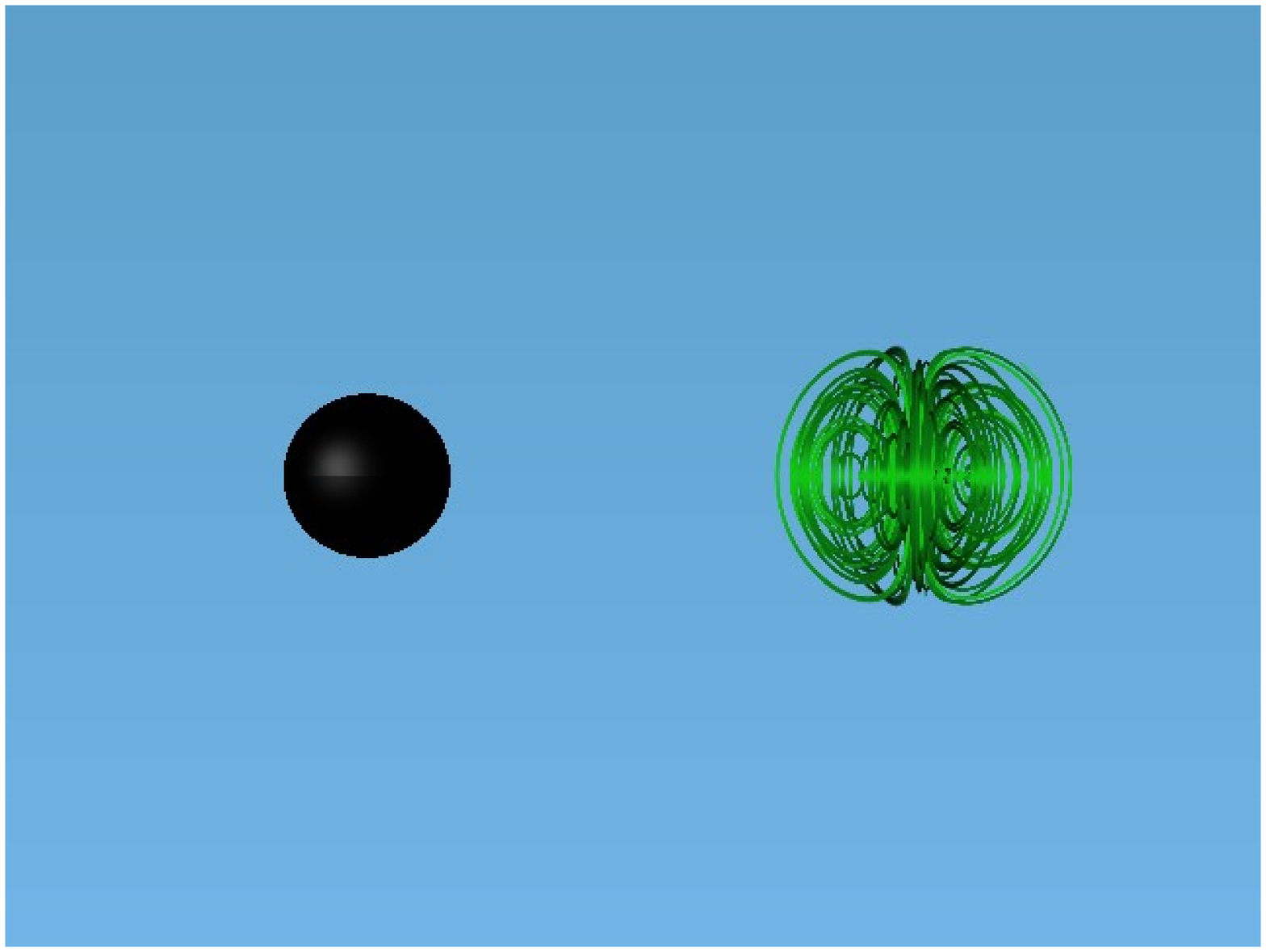}\\
\epsfxsize=2.2in
\leavevmode
\epsffile{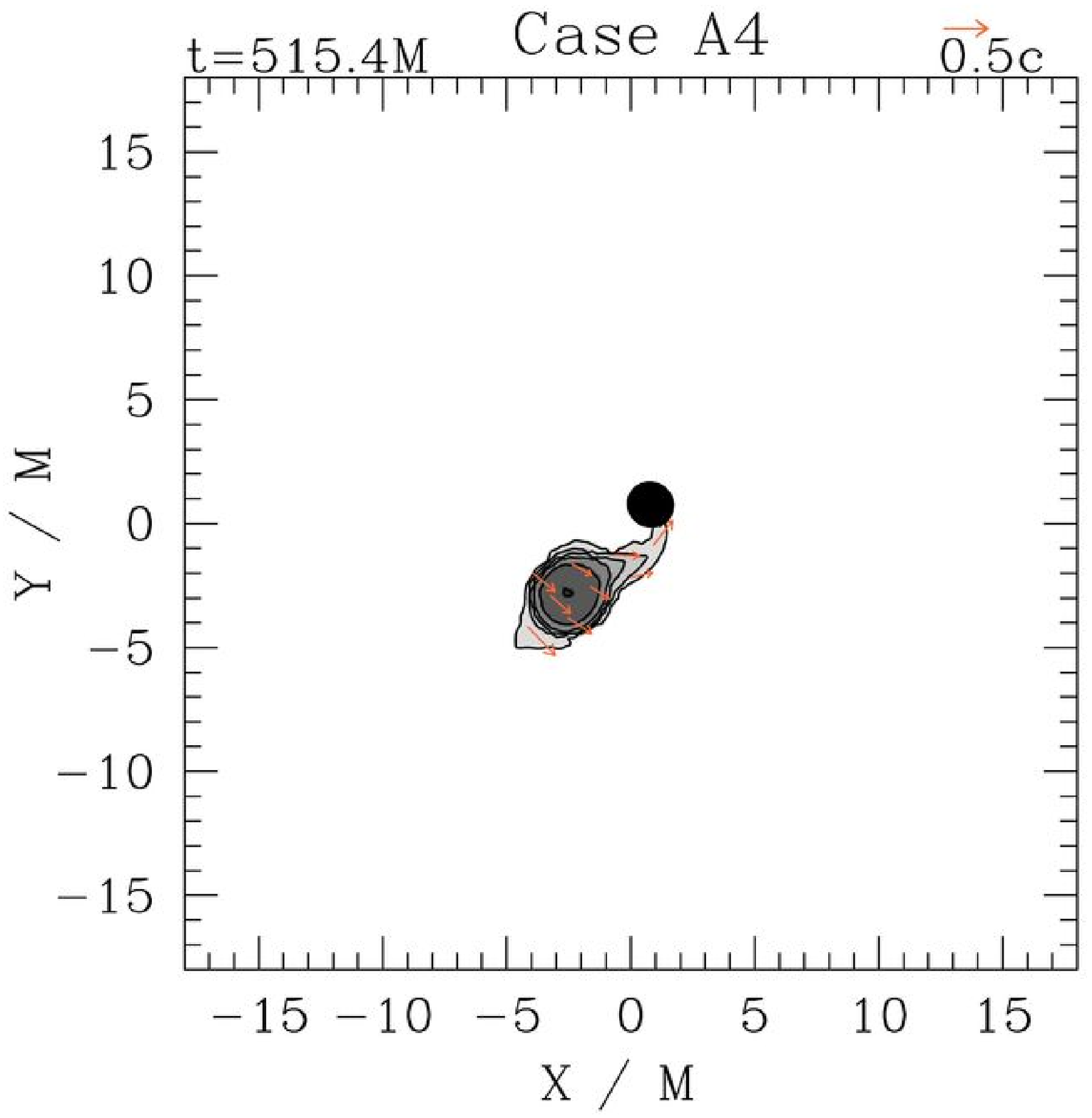}
\epsfxsize=2.3in
\leavevmode
\epsffile[20 -60 575 437]{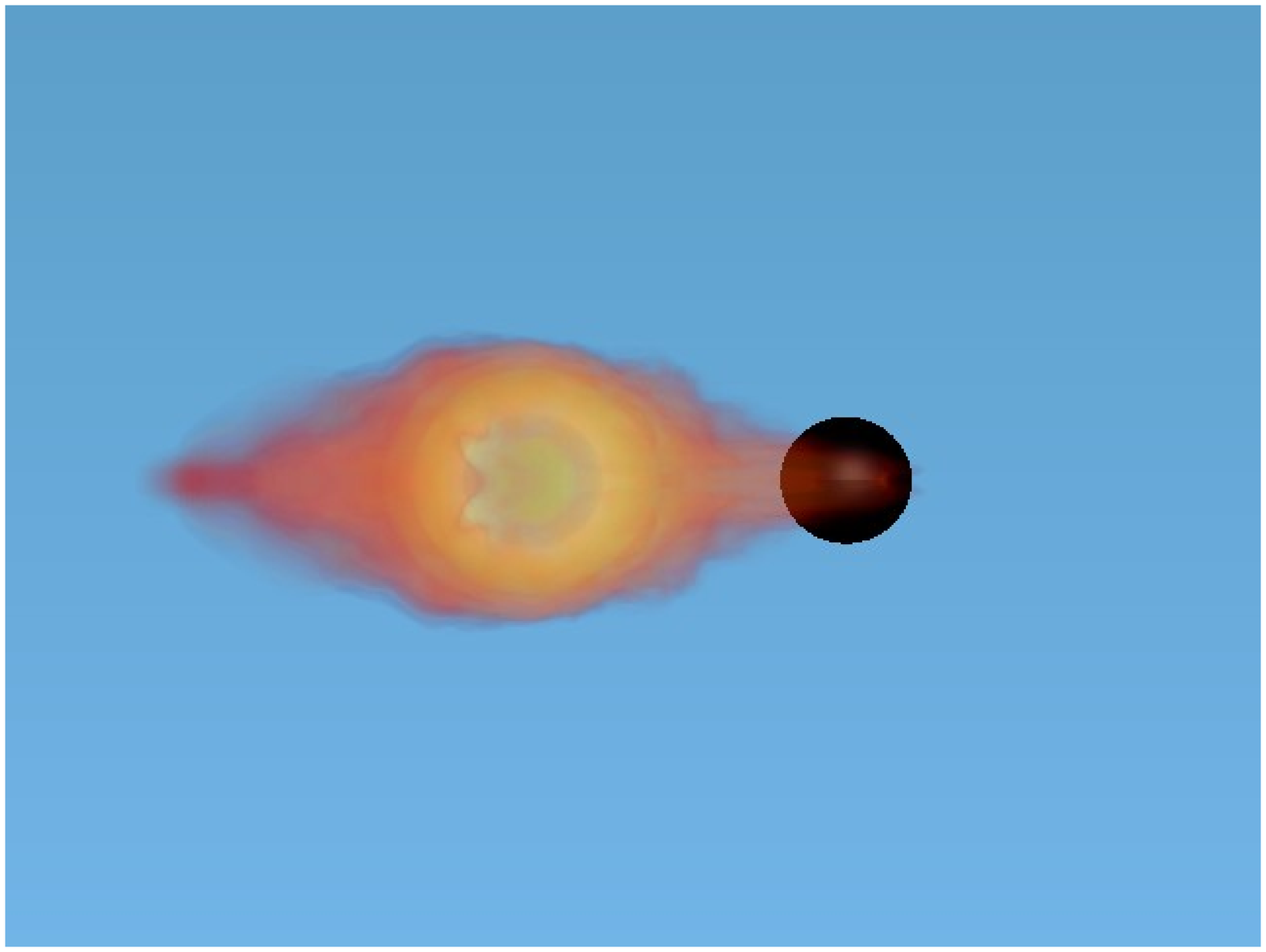}
\epsfxsize=2.3in
\leavevmode
\epsffile[20 -60 575 437]{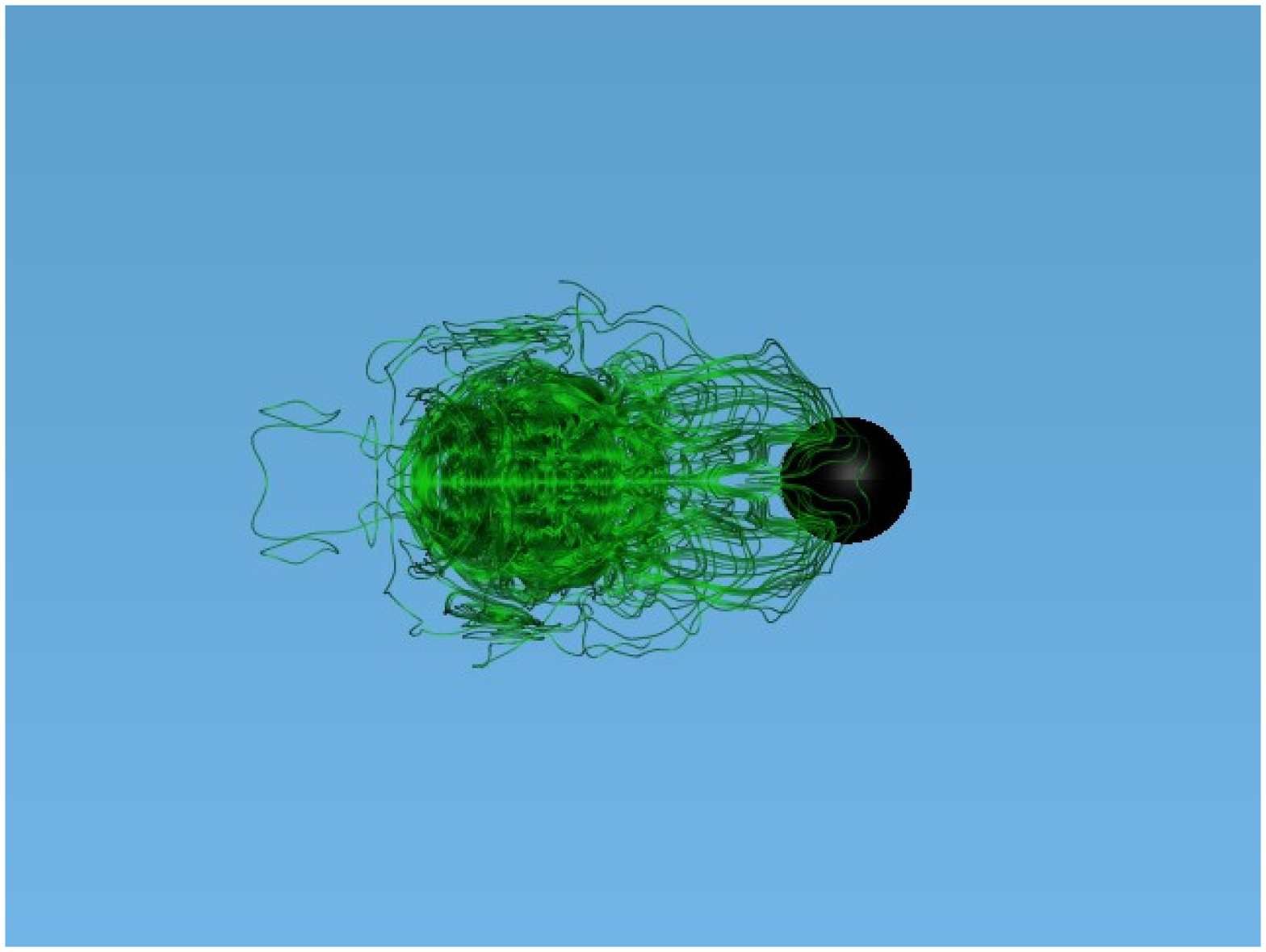}\\
\epsfxsize=2.2in
\leavevmode
\epsffile{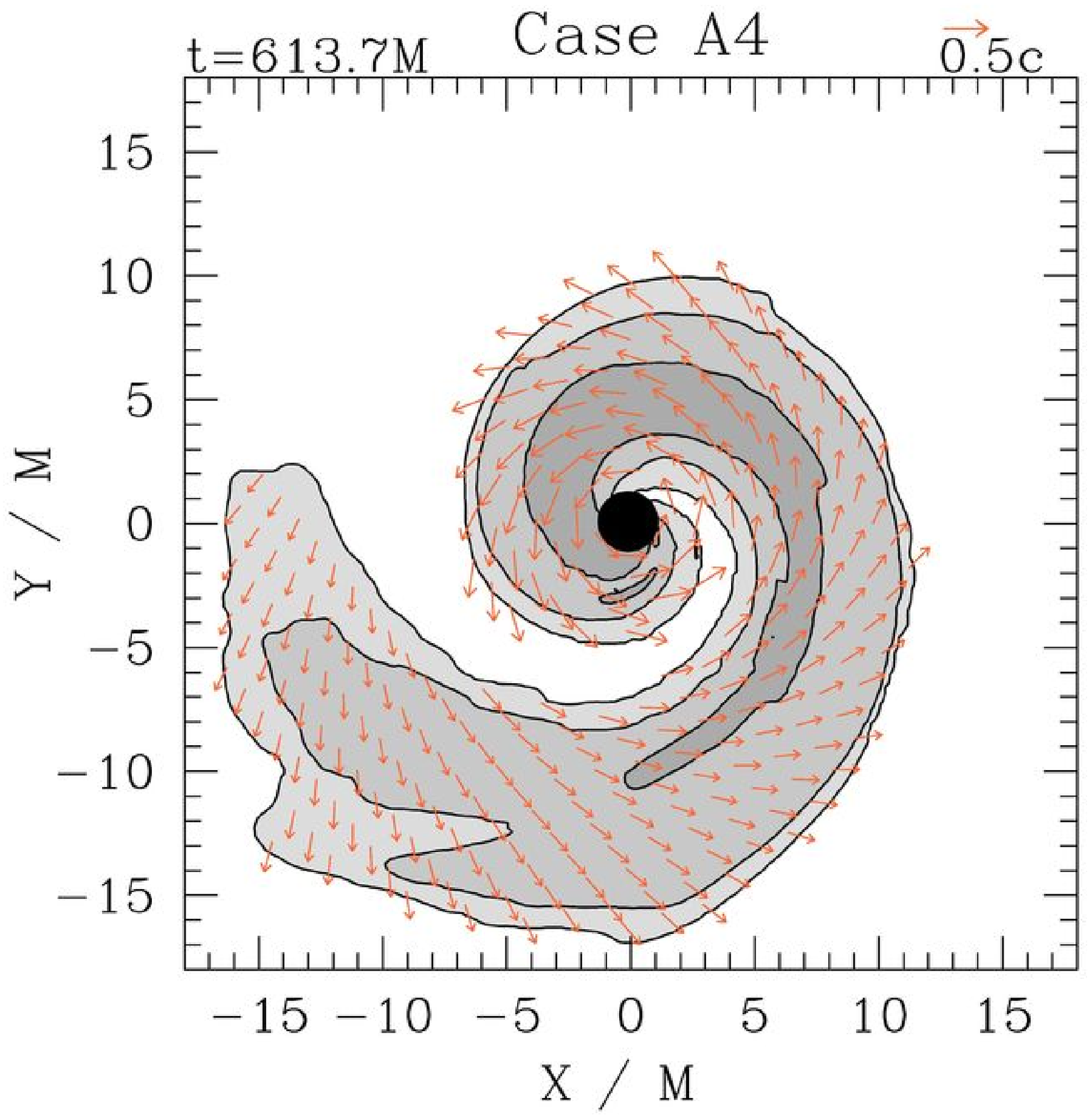}
\epsfxsize=2.3in
\leavevmode
\epsffile[20 -60 575 437]{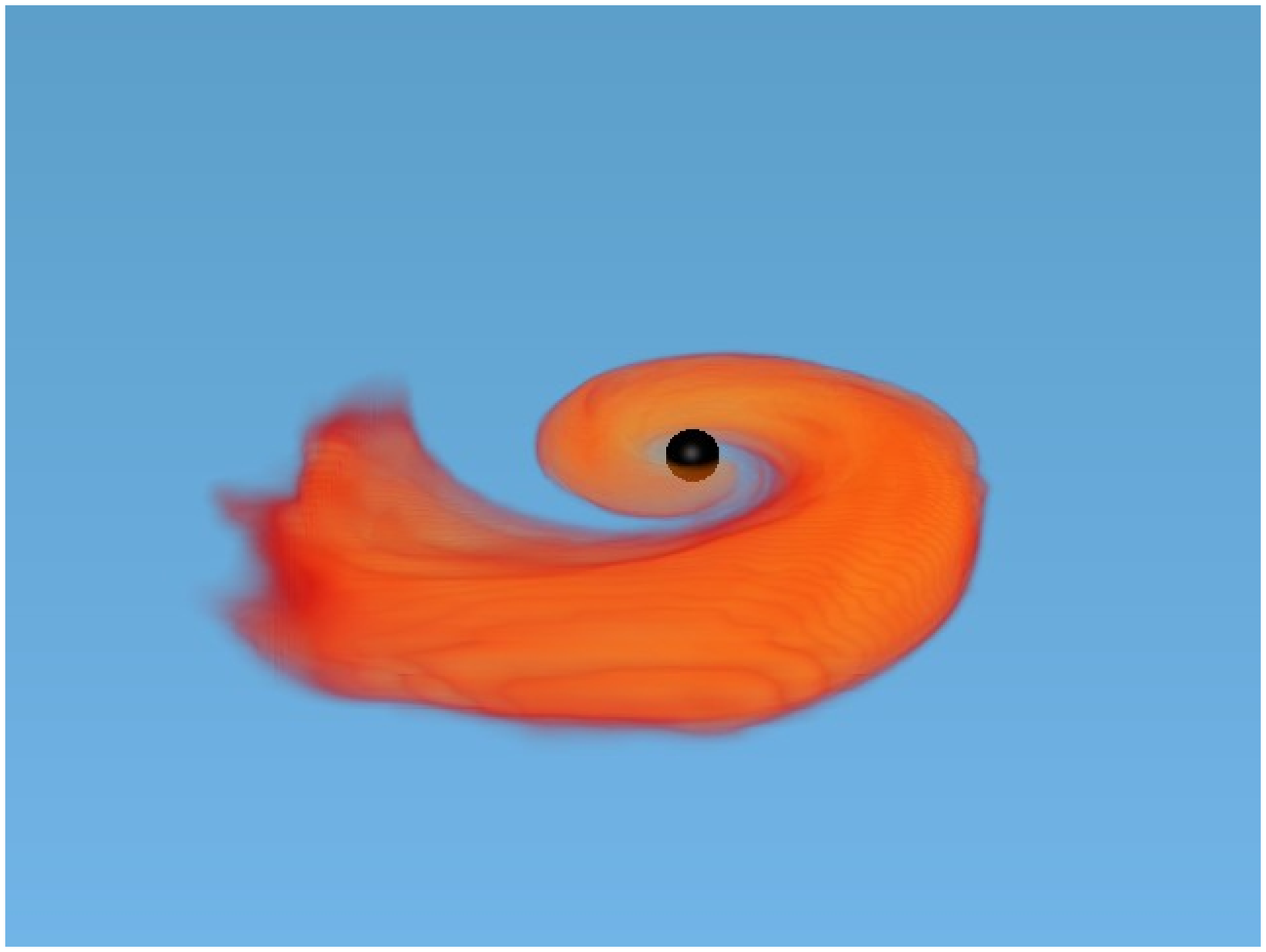}
\epsfxsize=2.3in
\leavevmode
\epsffile[20 -60 575 437]{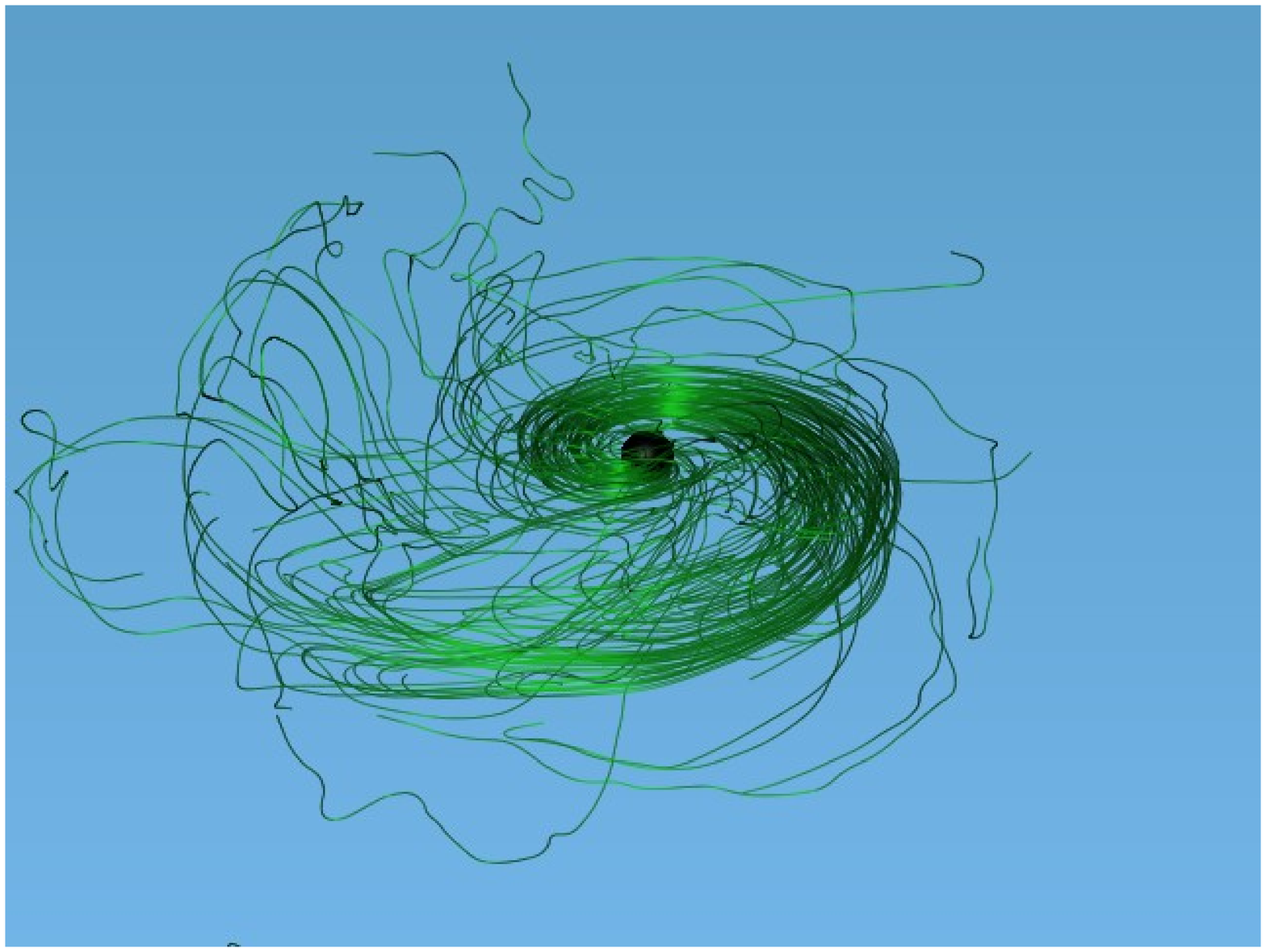}\\
\epsfxsize=2.2in
\leavevmode
\epsffile{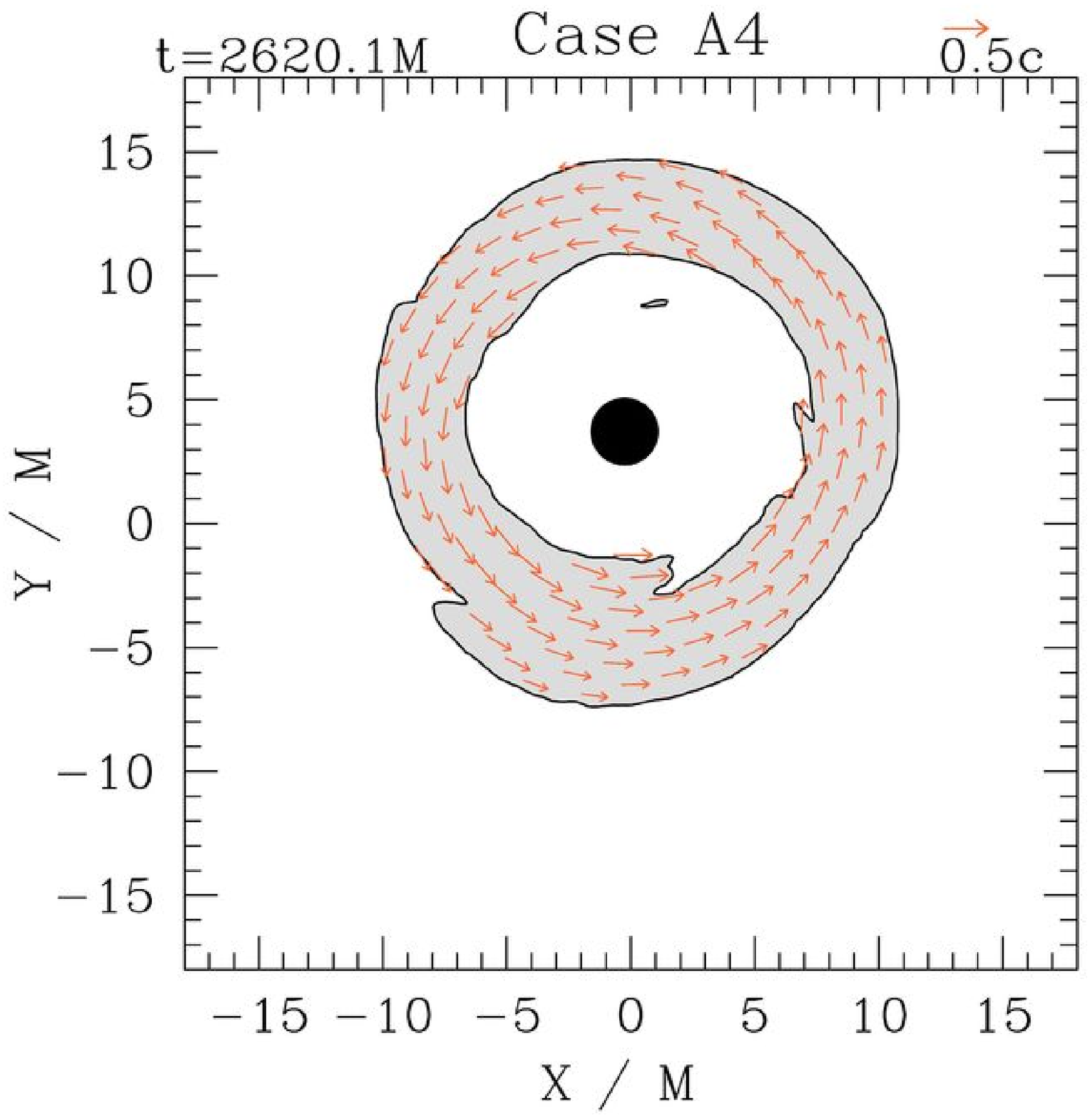}
\epsfxsize=2.3in
\leavevmode
\epsffile[20 -60 575 437]{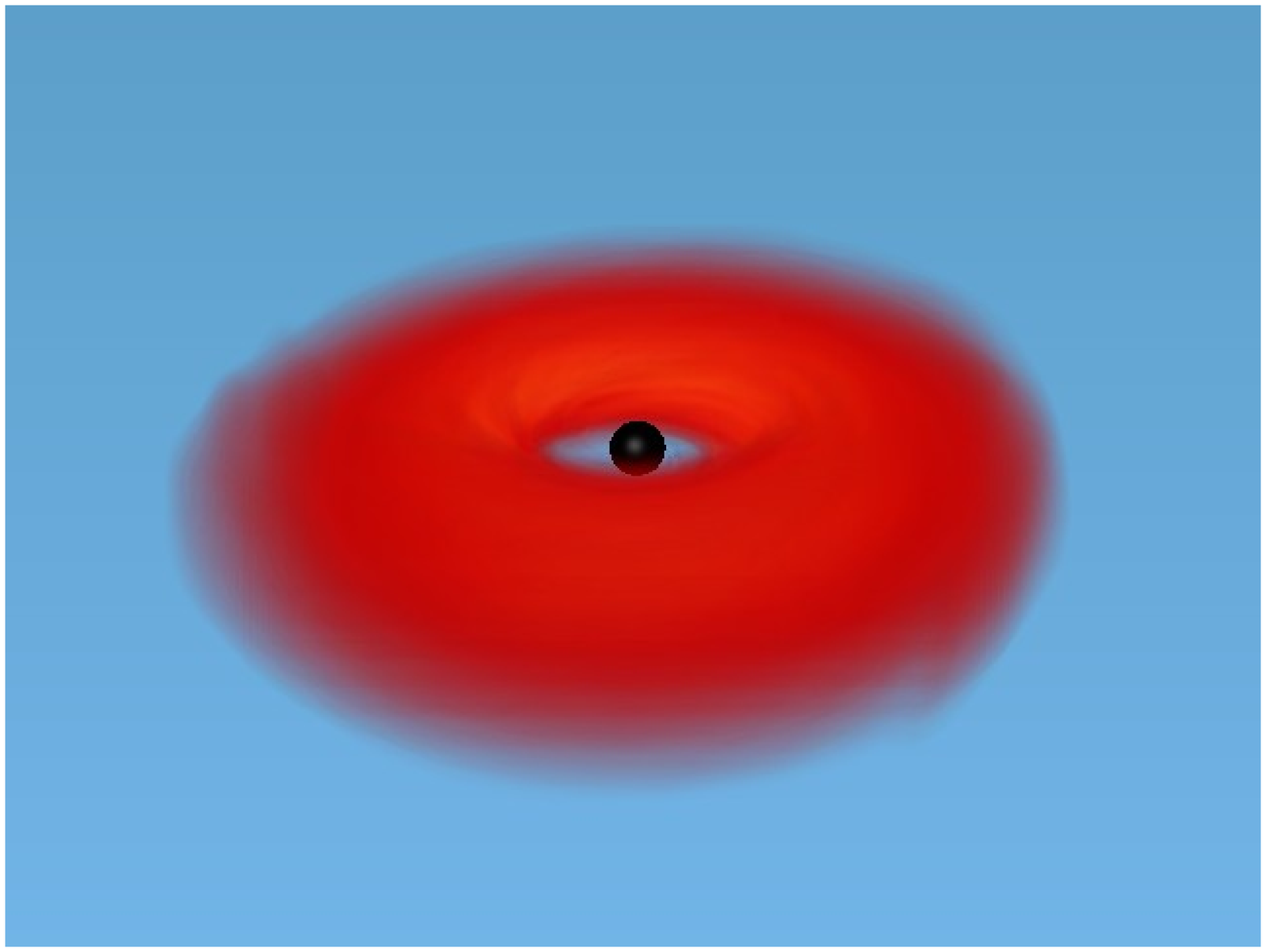}
\epsfxsize=2.3in
\leavevmode
\epsffile[20 -60 575 437]{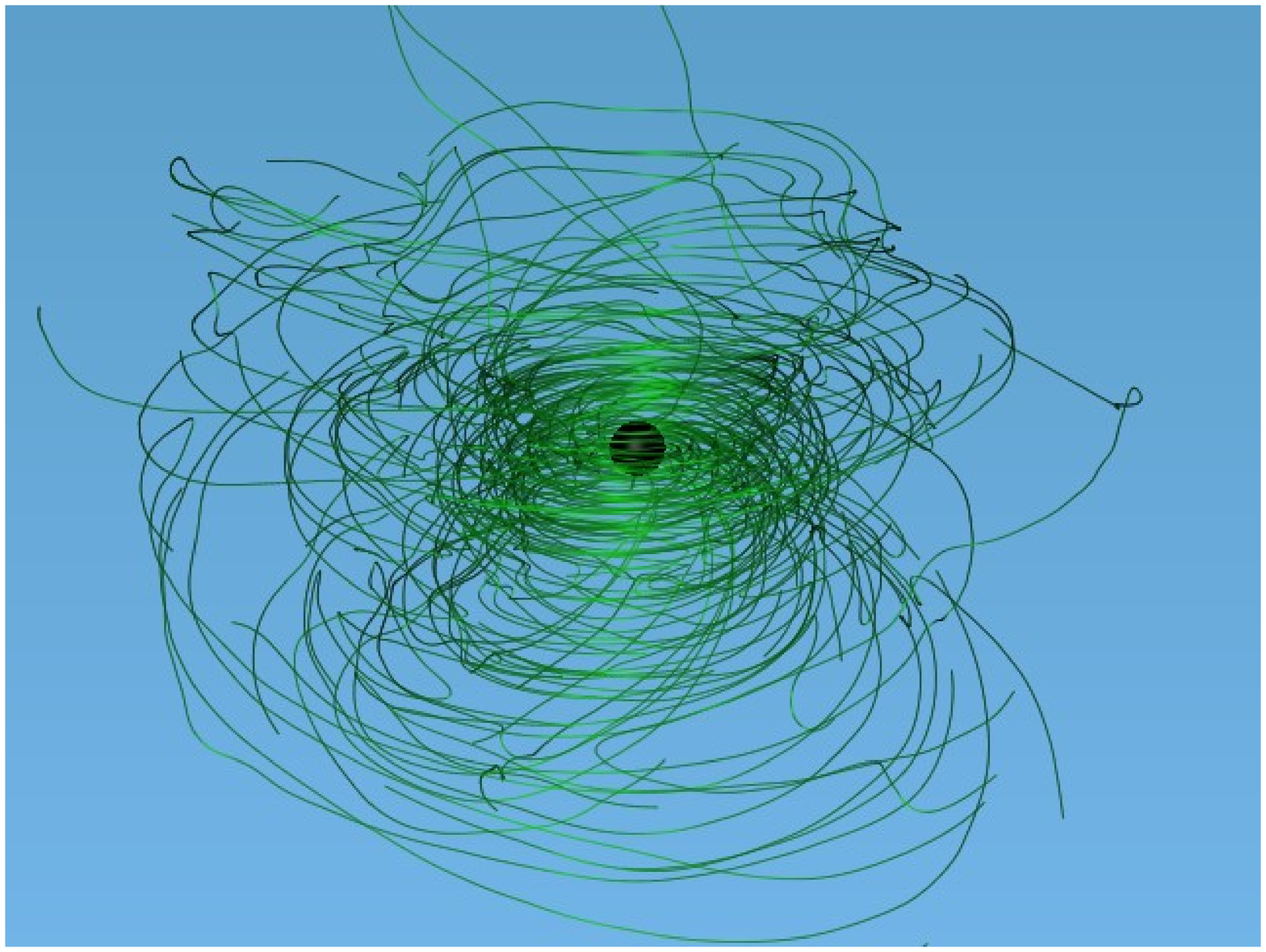}
\caption{Orbital-plane rest-mass density contours (left column), 3D density 
  profiles (middle column) and
  3D magnetic field lines (right column) at four selected times
  for case A4.  The times in the rows (top to bottom) are 
  $t/M=$448.5, 515.4, 613.7, and 2620.   Density
  contours in the orbital plane (left column) are plotted according to
  $\rho_0 = \rho_{0,{\rm max}} (10^{-0.92j})$,  ($j$=0, 1, ... 5),
  with darker greyscaling for higher density.  The maximum initial NS
  density is $\kappa \rho_{0,{\rm max}} = 0.126$, or $\rho_{0,{\rm
      max}}=9\times 10^{14}\mbox{g cm}^{-3}(1.4M_\odot/M_0)^2$.
  Arrows in density contour plots represent the velocity field in the
  orbital plane, and the black hole AH interior is marked by a filled
  black circle.  
  Magnetic fields are plotted as streamlines of the magnetic field
  vector $B^i$,
  distributed in proportion to $|B^i|$.
  The ADM mass for this case is
  $M=2.5\times 10^{-5}(M_0/1.4M_\odot)$s$=7.6(M_0/1.4M_\odot)$km.
The 3D visualizations were produced using the ZIBamira software 
system \cite{Stalling:AmiraVDA-2005}.}
\label{A:A4_magnetic_geometry}
\end{center}
\end{figure*}

Figure~\ref{A:evolution_story} shows density contours on the orbital
plane at selected times for the unmagnetized, zero BH spin case, A0.
Notice that the NS density contours in the top-left plot are 
nearly unchanged after three orbits (top-center plot), confirming that the
initial data are consistent with quasiequilibrium.  After about
3.5 orbits the NS tidally disrupts (top-right 
plot), and about 95\% of the NS matter promptly falls into the BH.
Matter in the low-density NS outer layers far from the accreting
funnel of matter forms a tidal tail that wraps around the black hole
and smashes into itself near the BH, generating a large amount of
shock heating (bottom-left plot). Meanwhile, accretion slows
considerably.  After intersecting itself, the inner regions of the
tidal tail forms a disk that orbits the BH, while the fluid velocity
distribution in the outer tail indicates slow
accretion onto the disk (bottom-middle
plot).  Shortly after disk formation,
only about 2\% of the NS matter remains outside
the BH (bottom-right plot), and the density directly outside the AH begins to plummet,
ultimately forming a low-density cavity around the BH similar to the one shown more prominently
in the bottom left frame of
Fig.~\ref{A:A4_magnetic_geometry}. This cavity indicates the presence of an innermost stable circular
orbit (ISCO).

\begin{figure}
\epsfxsize=3.4in
\leavevmode
\epsffile{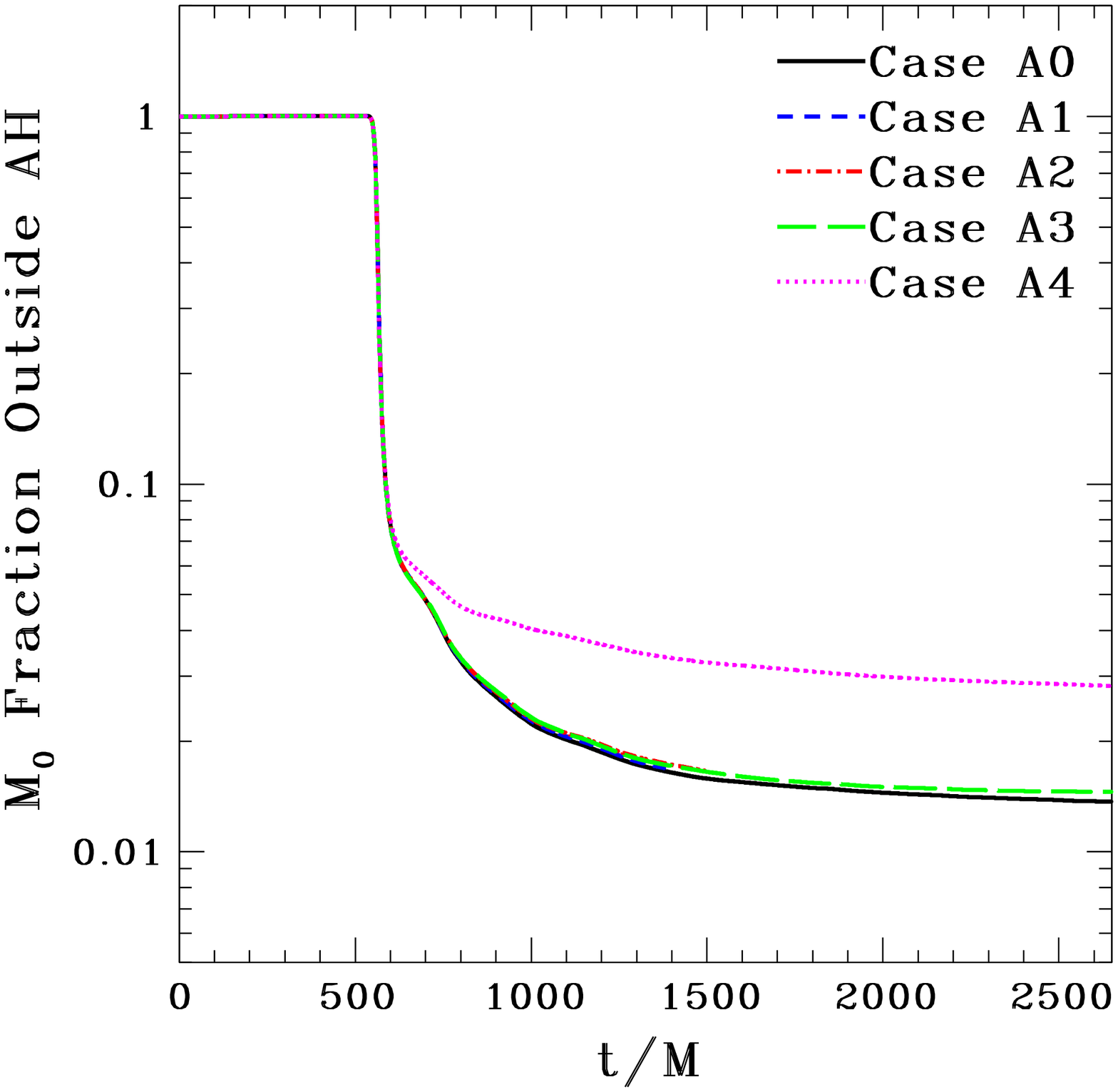}
\caption{Rest-mass fraction outside the AH for all cases in which the
  BH is initially nonspinning.}
\label{A:accretion_history}
\end{figure}

Figure~\ref{A:accretion_history} shows the accretion history for all
cases in which the BH has zero spin initially.  Regardless of magnetic field
configuration or strength, by $t\approx570M$ about 95\% of the NS
matter -- including the most strongly magnetized matter in the star --
has been accreted by the BH.  After this violent merger, the only case that
noticeably deviates from the magnetic-free case (A0) is case A4, the case
in which the initial seed magnetic fields are both strong ($|B|_{\rm
  max}\sim10^{17}$G) and pushed to the NS surface
($P_c=0.001$).  The disk mass in case A4 is two times larger than any
other case, but the final disk is only about 2.5\% of the initial
NS rest mass -- much smaller than in similar BHNS simulations with a
moderate aligned BH spin, in which disk masses of $\sim 10\%$ are
common. Case A3 is identical to case A4, except for the fact that 
its seed magnetic fields are about an order of magnitude weaker
($|B|_{\rm max}\sim10^{16}$G), and its accretion history is virtually
indistinguishable from that of case A0.  Thus it is the magnetic
field strength -- and not the different geometries explored here 
-- that significantly
influences the dynamics.  The final accretion rate implies a disk
half-life of between 3500--5500$M$ or 100--150($M_0/1.4M_\odot$)ms 
(depending on what points are chosen to calculate the final slope).

\begin{figure}
\epsfxsize=3.4in
\leavevmode
\epsffile{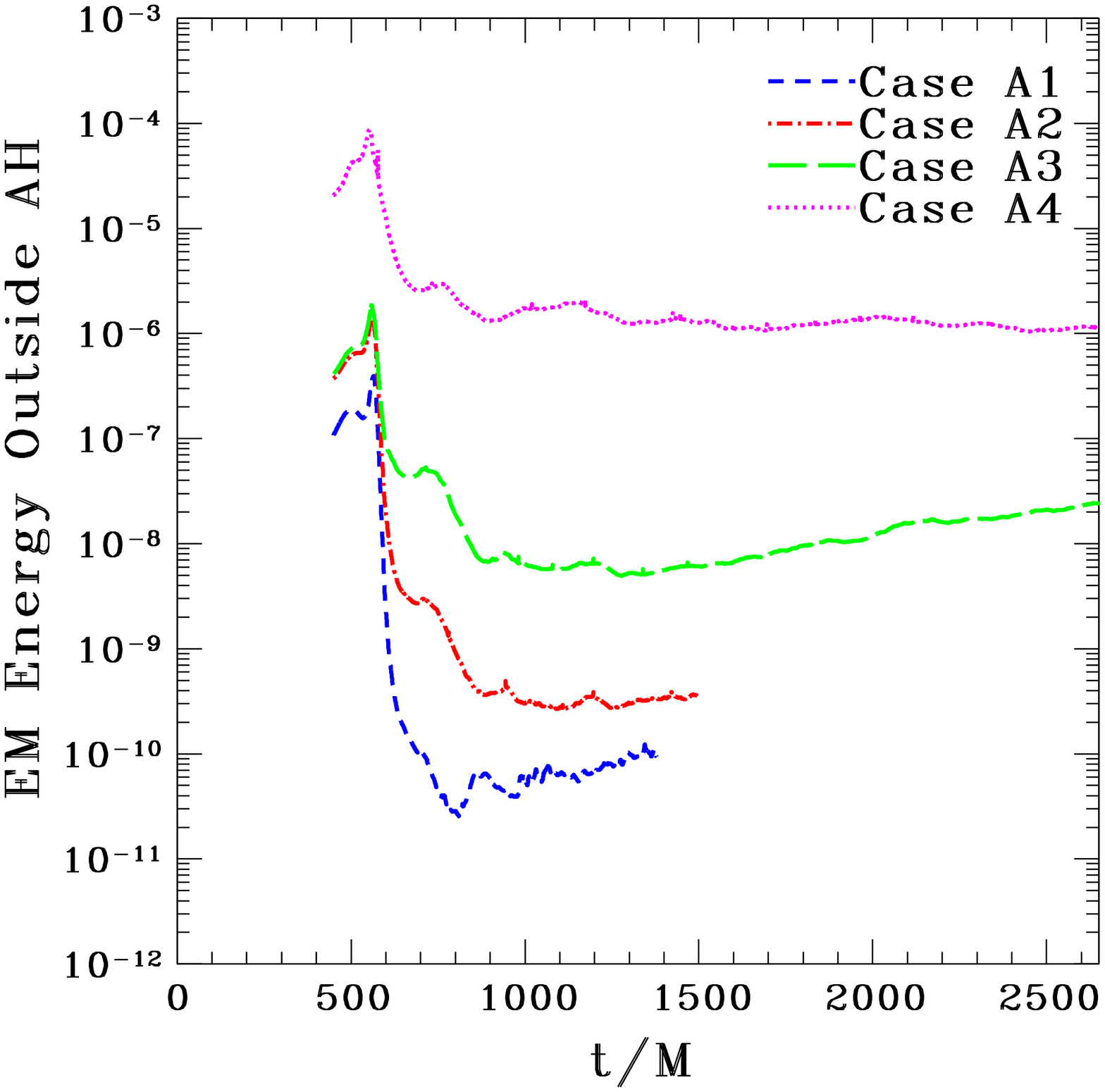}
\caption{Total magnetic energy ${\cal M}$ outside the AH for all magnetized cases
  in which the initial BH possesses zero spin, normalized by 
the ADM mass 
$M = 9.3 \times 10^{54} (M_0/1.4M_\odot)$erg.} 
\label{A:magnetic_energy_history}
\end{figure}

Figure~\ref{A:magnetic_energy_history} plots magnetic energy outside
the AH versus time, for all magnetized nonspinning BH cases studied.
Magnetic fields were added shortly before tidal disruption.
The magnetic fields do
not change significantly in the NS prior to disruption, but at the
point of disruption, there are two competing effects
that influence the magnetic energy. 
For one, the NS is being tidally disrupted, stretching
the magnetic field lines, amplifying the magnetic field strength and energy.
On the other hand, the magnetized fluids comprising the NS are being
rapidly accreted into the BH.  Our results show that there is a
slight amplification of magnetic energy during tidal disruption, but
after tidal disruption the magnetic energy is always less than the
magnetic energy in the seed magnetic fields.  The evolution is followed 
for $\sim 50(M_0/1.4M_\odot)$ms (assuming NS rest mass of 1.4$M_\odot$) after disk
formation, but ultimately no large magnetic energy amplification
is observed. This is likely due to the fact that the magnetic fields
in the disk are mostly toroidal (see
Fig.~\ref{A:A4_magnetic_geometry}) and once the disks have formed magnetic winding 
saturates. Amplification of magnetic fields by instabilities such
as the magnetorational instability (MRI) may occur, but the resolution in our simulations may not 
be high enough to resolve the small-scale turbulence associated with these instabilities. 

Notice that the magnetic energy in case A2 is about three times that
of A1, both initially and when the simulation was terminated.  These cases
differ only in the degree of central condensation of the initial magnetic 
field (see Table~\ref{table:id}). 

Cases A2 and A3 are identical except for the pressure cutoff of the seed magnetic
fields.  In case A2 magnetic fields are set to zero for pressures $P<P_c=0.04P_{\rm max}$, where
$P_{\rm max}$ is the maximum pressure of the NS. However, in case A3
$P_c$ is set to 0.001, so the seed magnetic fields are pushed much
closer to the NS surface.  This results in about a 16\% amplification
of initial magnetic energy in case A3
(Fig.~\ref{A:magnetic_energy_history}), {\it but increases the final
  magnetic energy by about a factor of 15}.  This is
consistent with the fact that the core of the NS is invariably
accreted into the BH during merger in these BHNS simulations,
so the disk is comprised of what were the outer layers
of the NS.
Thus, the stronger the seed magnetic field in the outer
layers of the NS, the stronger and more dynamically relevant the
magnetic fields in the disk.

Cases A3 and A4 differ only in initial seed magnetic field
strength; the seed magnetic fields are uniformly about an order of magnitude
stronger in case A4.  Figure~\ref{A:accretion_history} demonstrates
that the physical extent of the disk is very strongly influenced by the strong
seed magnetic fields of case A4.
Figure~\ref{A:magnetic_energy_history} reinforces that observation;
though only about 1\% of the NS rest mass exists in the disk of A3,
less than 0.5\% of the seed magnetic energy remains in the disk.
Compare this to case A4, where the disk mass is about twice as large,
but where more than an order of magnitude more magnetic energy remains in the
disk.

Magnetic fields play an important dynamical role in only one
nonspinning case, A4, amplifying the disk mass by a factor of two.
Figure~\ref{A:A4_magnetic_geometry} shows how the magnetic field configuration
evolves in this case, from $t=448.5M$ when the magnetic fields are
first seeded into the NS (top-right), until the simulation is stopped at $t=2620M$
(bottom-right). 
Magnetic
field lines are greatly stretched during disk formation (third row 
on the right),
resulting in a strongly-magnetized disk.  At late times, the magnetic
fields that remain in the disk are very tightly wound.
The bottom left and center frames of the figure show that a cavity has formed around the BH 
near the end of the simulation. This hollow region is indicative of the presence of an ISCO, 
which is interesting because it has been suggested that stresses in magnetized disks may 
suppress the presence of an ISCO \cite{kh02,bhk08}. However, it may be that longer 
simulations are required for the disk cavity to be filled. 

\begin{figure*}
\vspace{-4mm}
\begin{center}
\epsfxsize=3.5in
\leavevmode
\epsffile{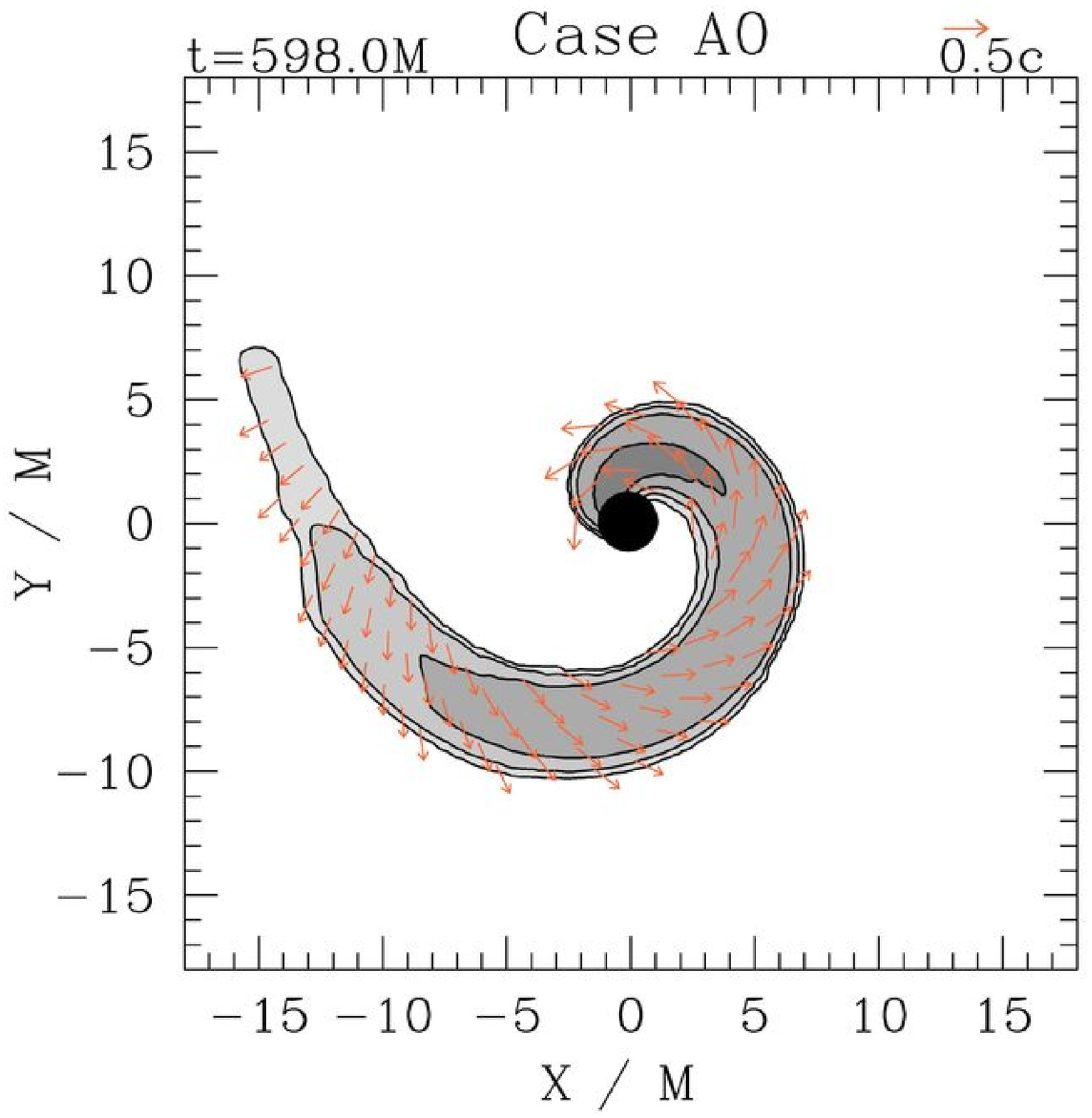}
\epsfxsize=3.5in
\leavevmode
\epsffile{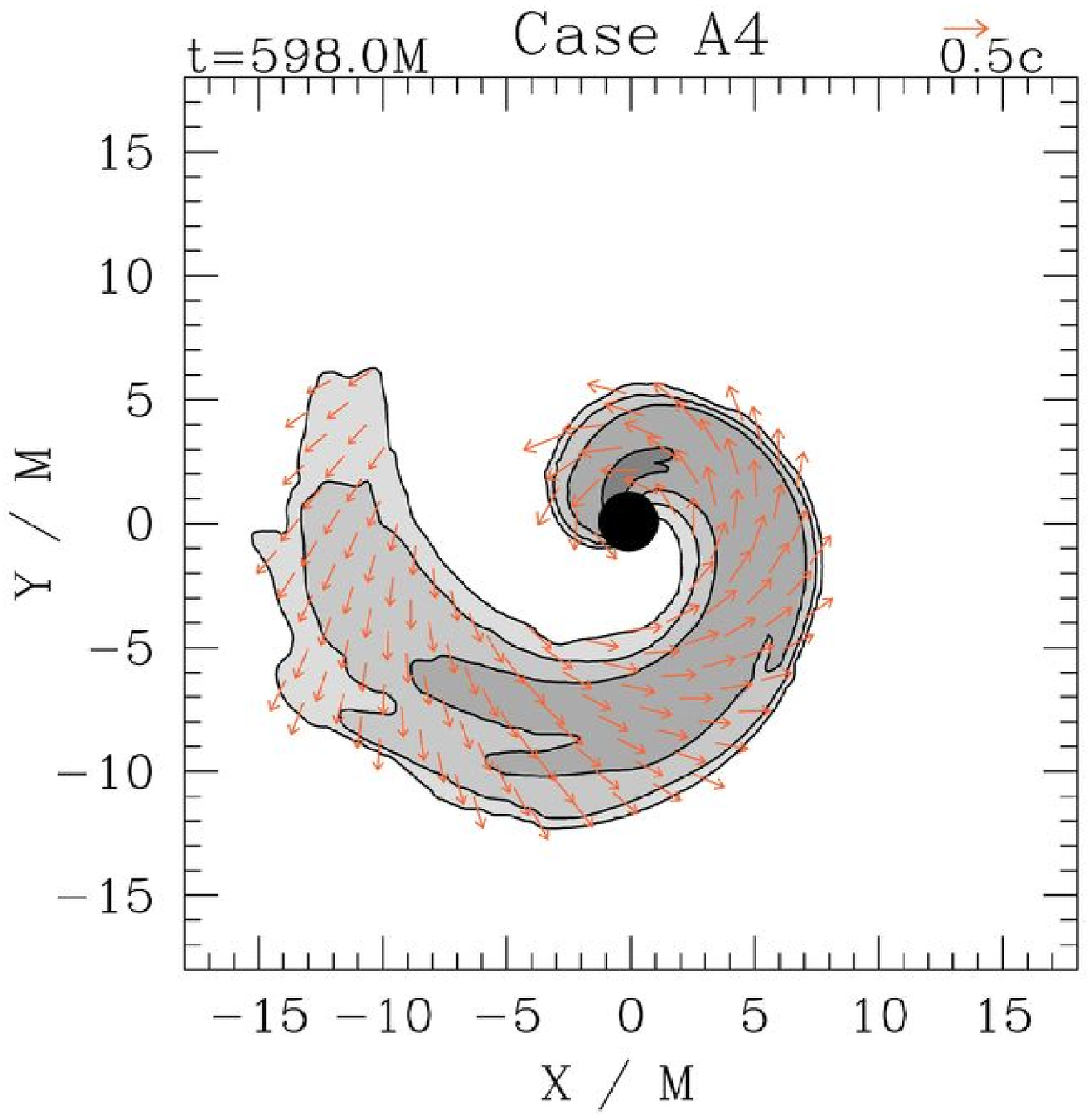}
\caption{Rest-mass density and velocity profile snapshots during NS tidal disruption
  for cases~A0 (left) and A4 (right). Density contours are
  plotted in the orbital plane according to $\rho_0 = \rho_{0,{\rm
      max}} (10^{-0.92j})$,  ($j$=0, 1, ... 5), with darker
  greyscaling for higher density. 
  The maximum initial NS density is $\kappa \rho_{0,{\rm max}} =
  0.126$, 
  or $\rho_{0,{\rm max}}=9\times 10^{14}\mbox{g
    cm}^{-3}(1.4M_\odot/M_0)^2$.  
  Arrows represent the velocity field
  in the orbital plane, and the black hole AH interior is marked 
  by a filled black circle.  The
  ADM mass for this case is $M=2.5\times 10^{-5}(M_0/1.4M_\odot)$
  s$=7.6(M_0/1.4M_\odot)$km.}
\label{A:rho_disruption}
\end{center}
\end{figure*}

\begin{figure*}
\vspace{-4mm}
\begin{center}
\epsfxsize=3.5in
\leavevmode
\epsffile{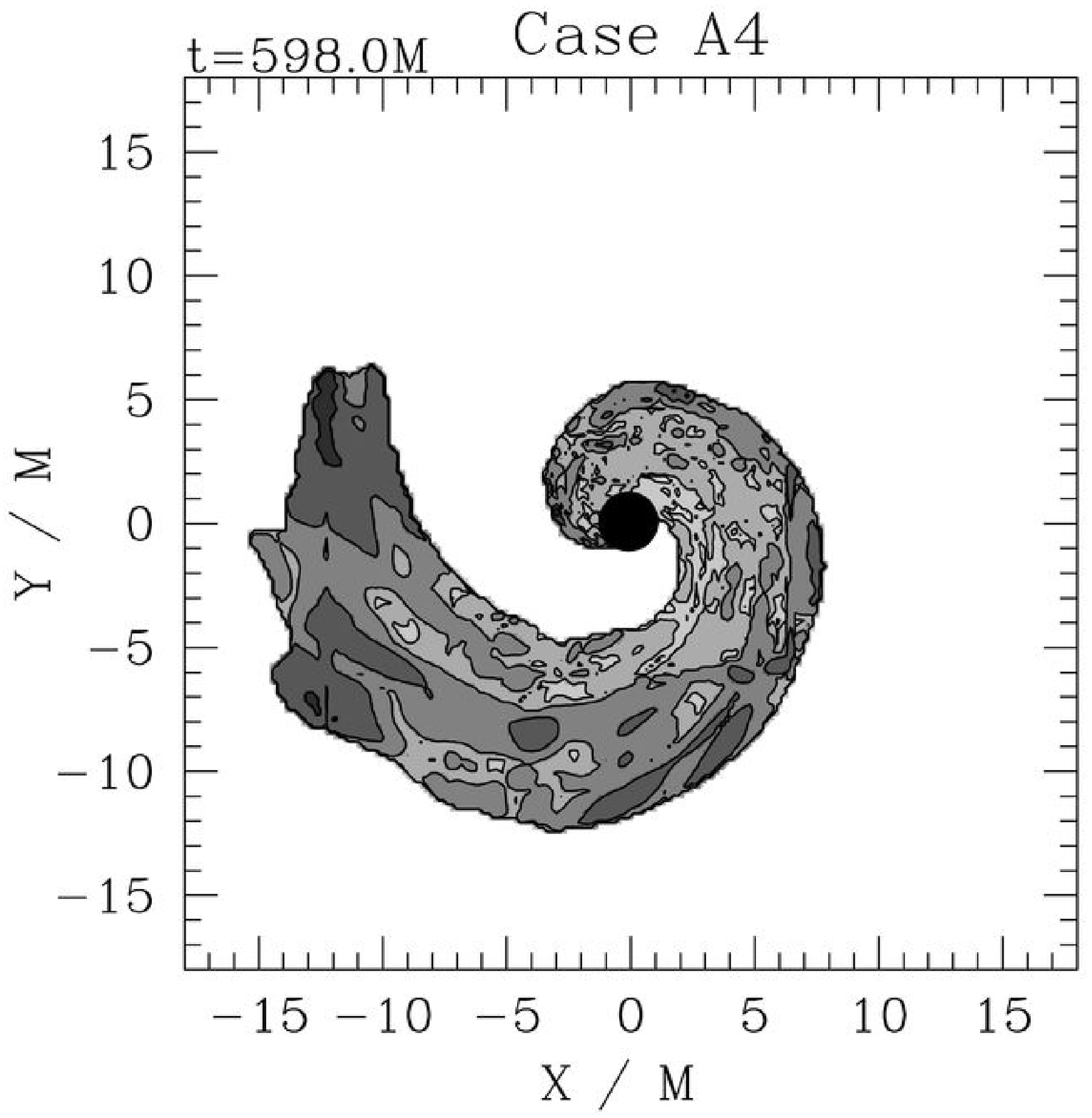}
\epsfxsize=3.5in
\leavevmode
\epsffile{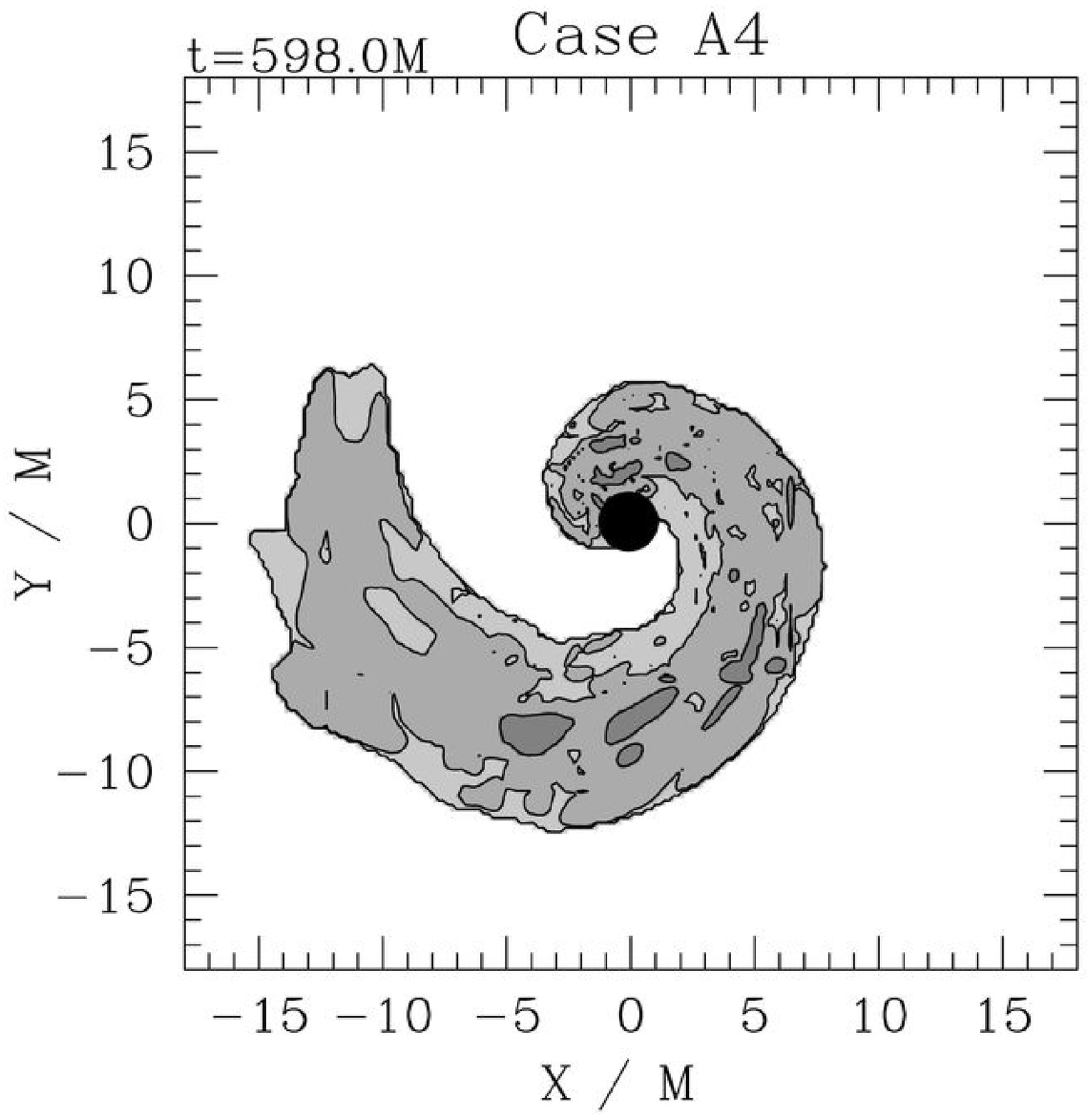}
\caption{Pressure ratio $b^2/(2P)$ (left) and magnetic pressure $b^2/2$ (right) contours during NS tidal disruption, at the same
  time as Fig.~\ref{A:rho_disruption}, plotted according to
  $b^2/(2P)=10^{-1.5}(10^{-1.3j})$, ($j$=0, 1, ... 5), and
  $\kappa b^2 =10^{-5}(10^{-2.2j})$, ($j$=0, 1, ... 5). 
 Darker greyscaling denotes higher values. Contours are only plotted for regions with densities
  higher than the lowest-density $\rho_0$ contours in Fig.~\ref{A:rho_disruption}. 
  In cgs units,
  $\kappa^{-1}=6\times 10^{36}{\rm dyn\ cm}^{-2} (1.4 M_\odot/M_0)^2$.}
\label{A:b2andb2overP_disruption}
\end{center}
\end{figure*}

The significant boost in disk mass in case A4 indicates that with sufficiently strong NS
magnetic fields, merger dynamics may be significantly affected.
Figure~\ref{A:rho_disruption} compares orbital-plane
density contours in cases A0 (unmagnetized) and A4 during the late
stages of tidal disruption.  Notice that the stronger magnetic pressures in case A4
push out the outer layers of the NS during tidal disruption.  
Nevertheless, Fig.~\ref{A:b2andb2overP_disruption} shows that the magnetic pressure in this
matter distribution does not exceed 3\% the gas pressure at the same
time as Fig.~\ref{A:rho_disruption}.  Therefore, large
changes in remnant disk mass do not require huge magnetic-to-gas
pressure ratios.

\begin{figure}
\epsfxsize=3.4in
\leavevmode
\epsffile{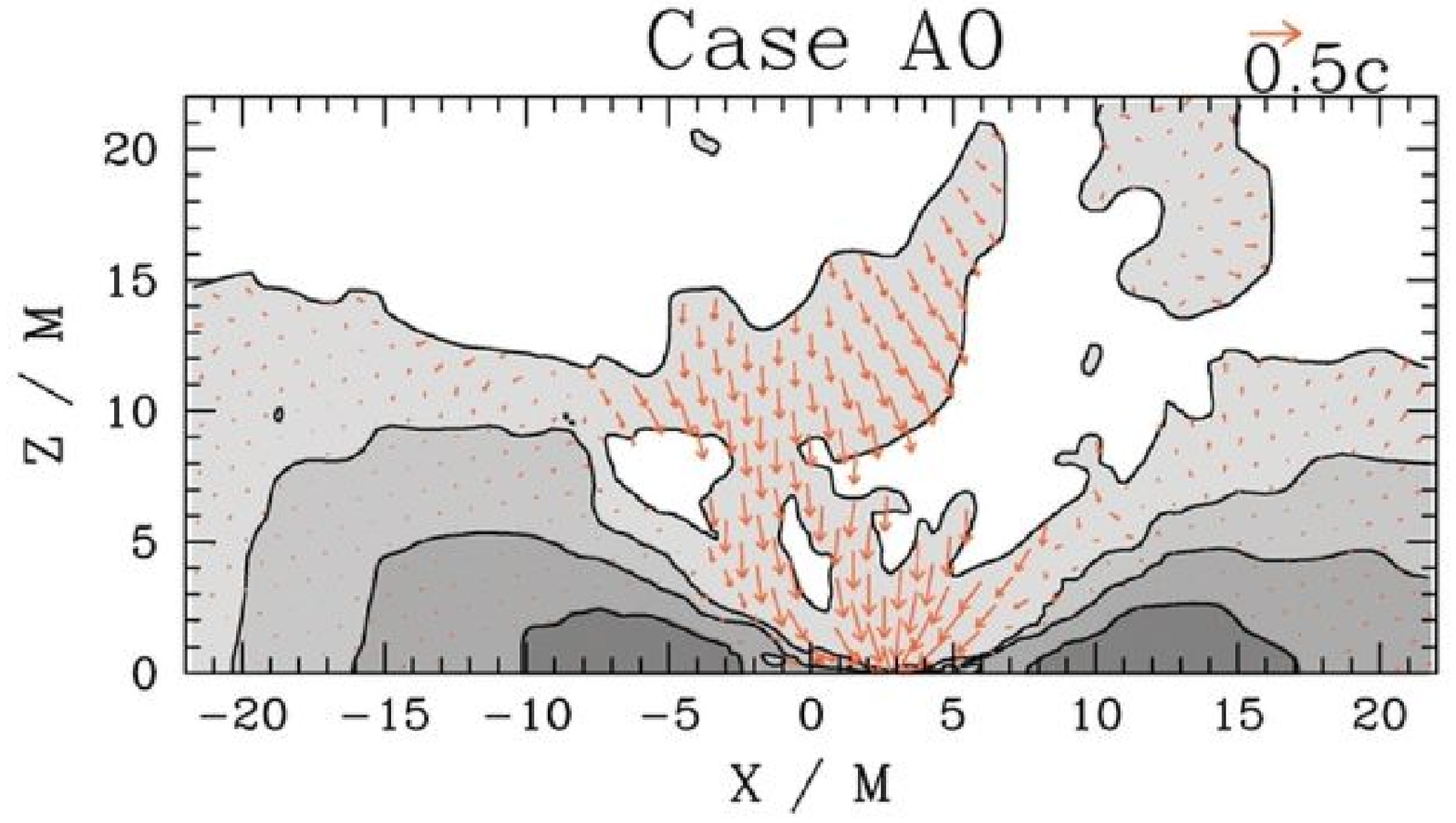}\\
\epsfxsize=3.4in
\leavevmode
\epsffile{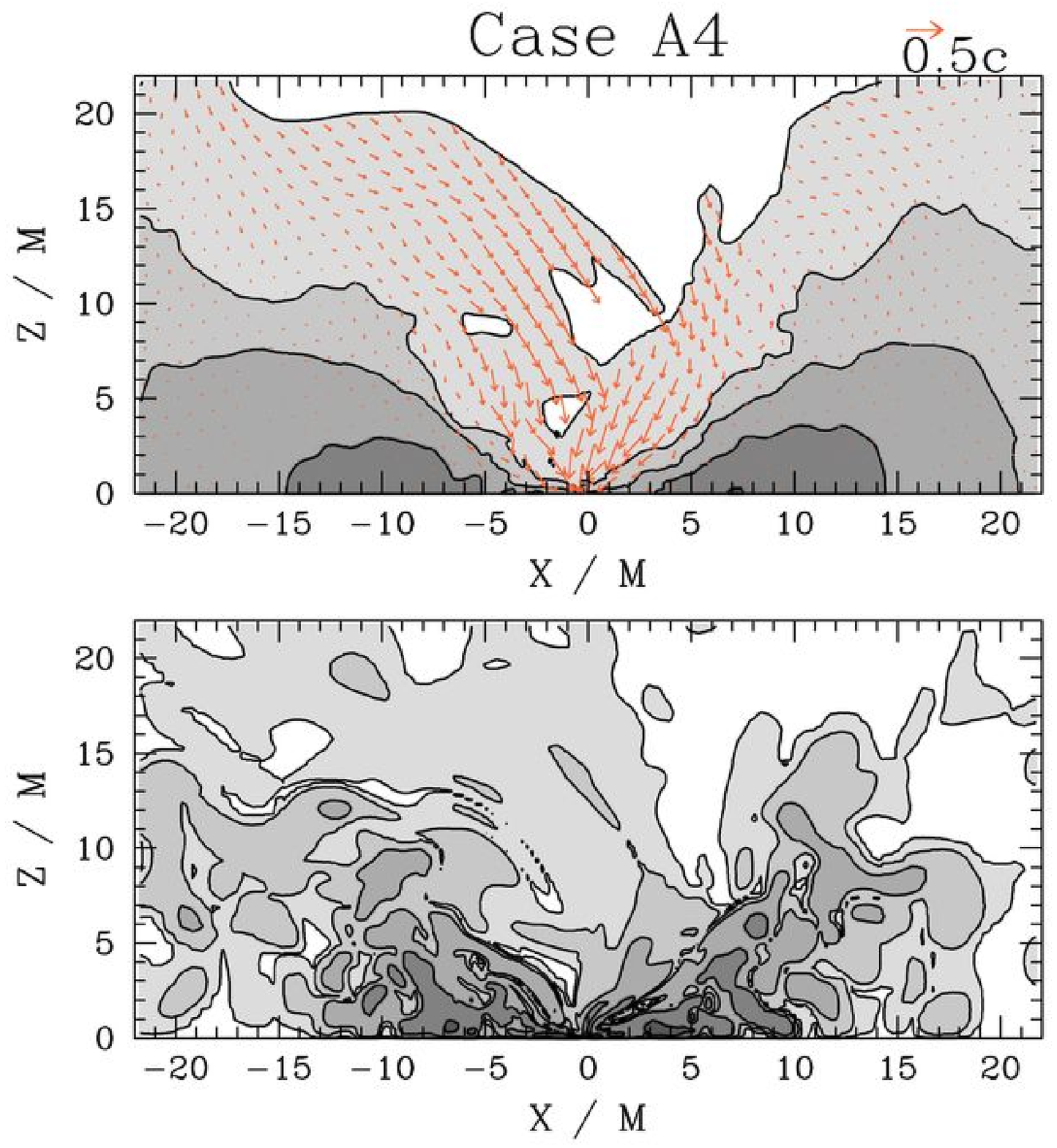}
\caption{Rest-mass density contours for case A0 (top), A4 (middle), 
and magnetic energy density $b^2$ (bottom) 
profile for A4 in the meridional plane at the end of simulations  
($t=2620M$). Density contours are plotted according to 
$\rho_0 = \rho_{0,{\rm max}} 10^{-7.6+0.717j}$, ($j=0,...,6$), 
with darker greyscaling for higher density. $b^2$ contours are plotted 
according to $\kappa b^2 =10^{-12+0.833j}$, ($j=0,...,6$).
  The maximum initial NS density is $\kappa \rho_{0,{\rm max}} =
  0.126$, 
  or $\rho_{0,{\rm max}}=9\times 10^{14}\mbox{g
    cm}^{-3}(1.4M_\odot/M_0)^2$.  
  Arrows represent the velocity field
  in the meridional plane.  
In cgs units, the 
  ADM mass for this case is $M=2.5\times 10^{-5}(M_0/1.4M_\odot)$
  s$=7.6(M_0/1.4M_\odot)$km, and 
$\kappa^{-1}=6\times 10^{36}{\rm erg\ cm}^{-3} (1.4 M_\odot/M_0)^2 
= 7\times 10^{15}\mbox{g cm}^{-3}(1.4M_\odot/M_0)^2$.}
\label{A:meridional}
\end{figure}

Figure~\ref{A:meridional} compares A0 and A4 rest-mass density
profiles when these simulations were stopped at $t=2620M$.
The density profiles in these two cases 
are similar; at the end of the simulation, low-density matter still   
flows {\it into} the BH from the poles.  For A4,
it is clear from the $b^2$ contour plot (bottom graph) that most of
the magnetic fields are confined inside the remnant disk near the equatorial
plane, consistent with the magnetic field-line plots in Fig.~\ref{A:A4_magnetic_geometry}.

\begin{figure*}
\vspace{-4mm}
\begin{center}
\epsfxsize=3.5in
\leavevmode
\epsffile{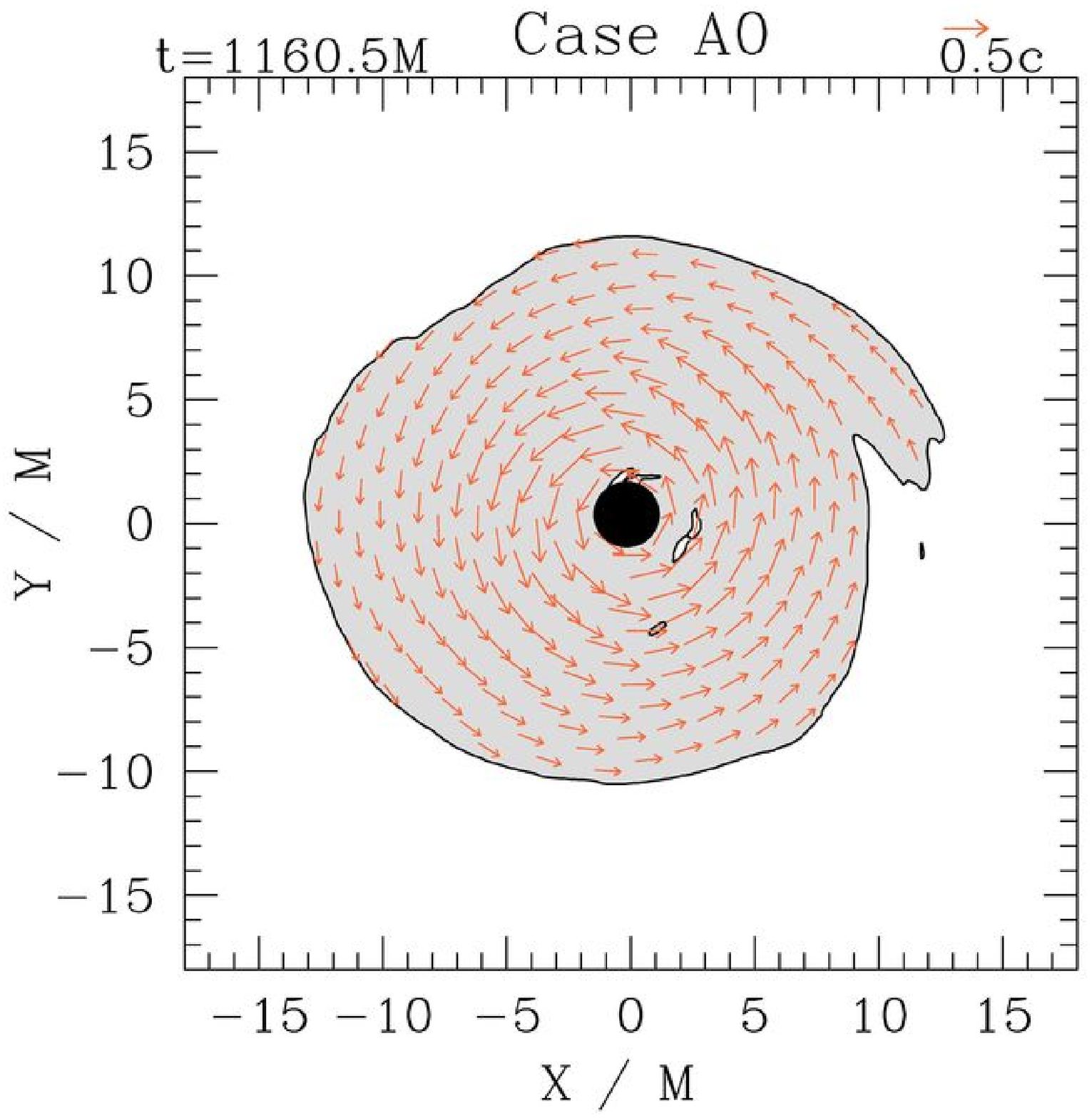}
\epsfxsize=3.5in
\leavevmode
\epsffile{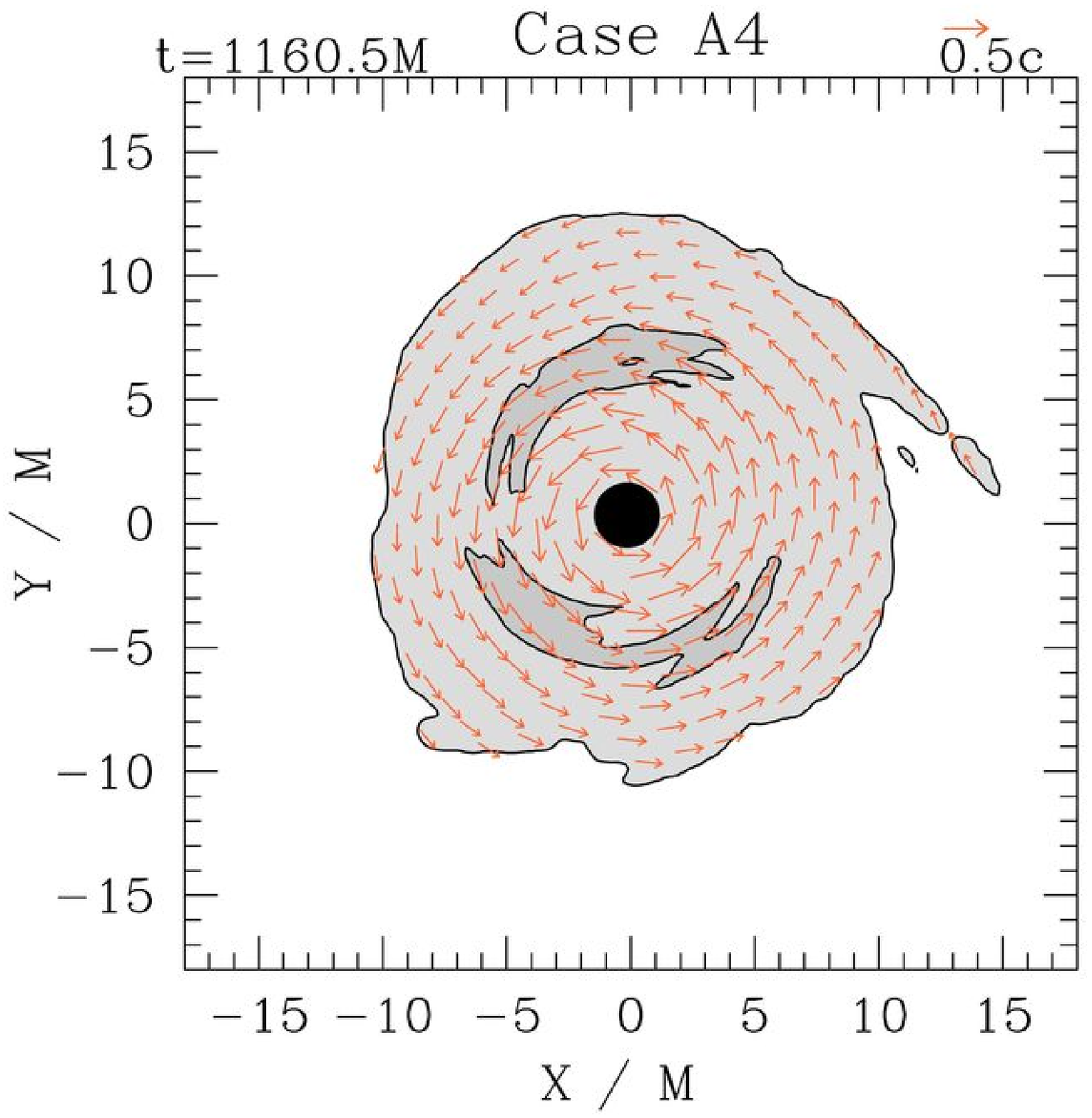}\\
\epsfxsize=3.5in
\leavevmode
\epsffile{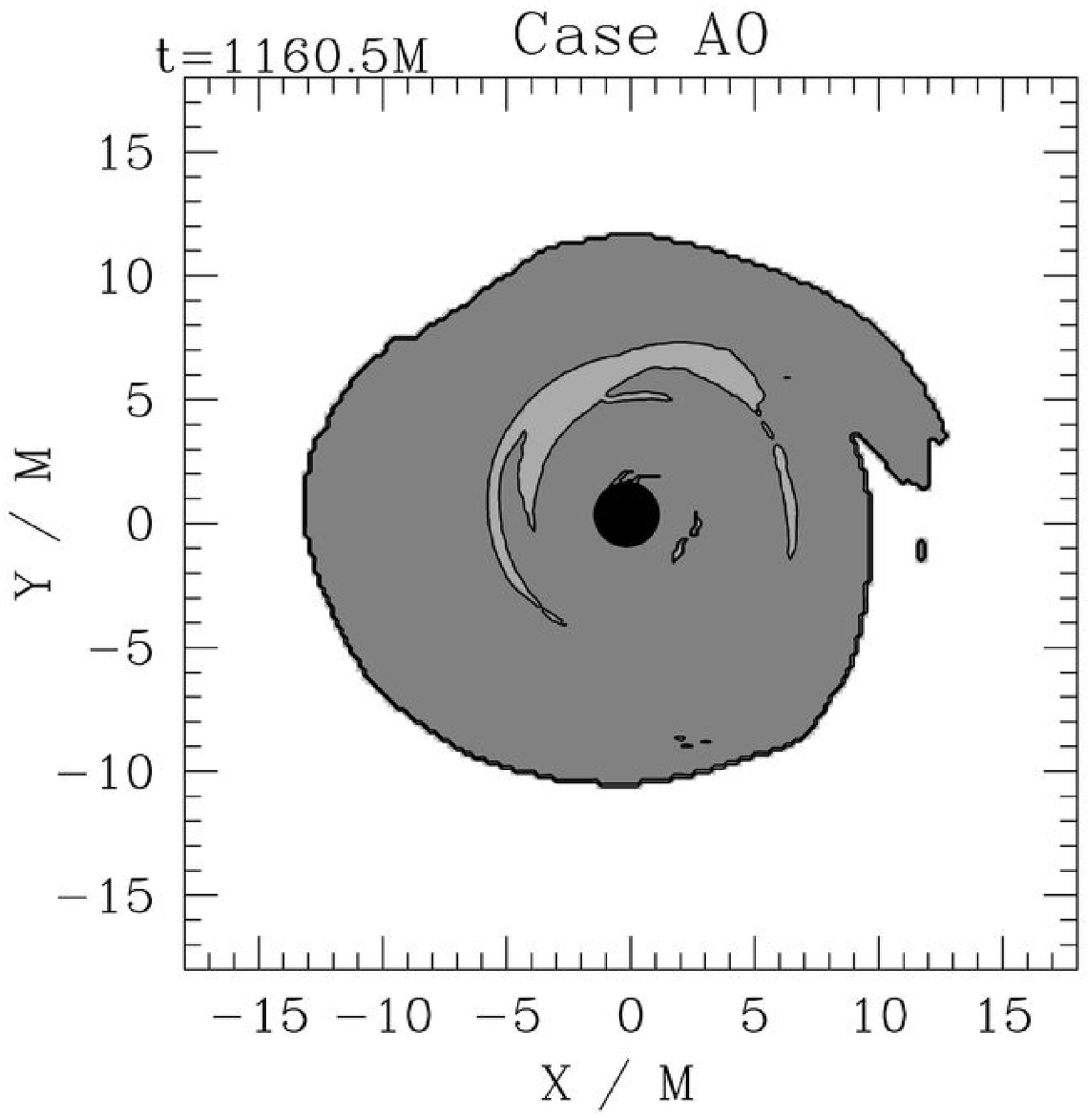}
\epsfxsize=3.5in
\leavevmode
\epsffile{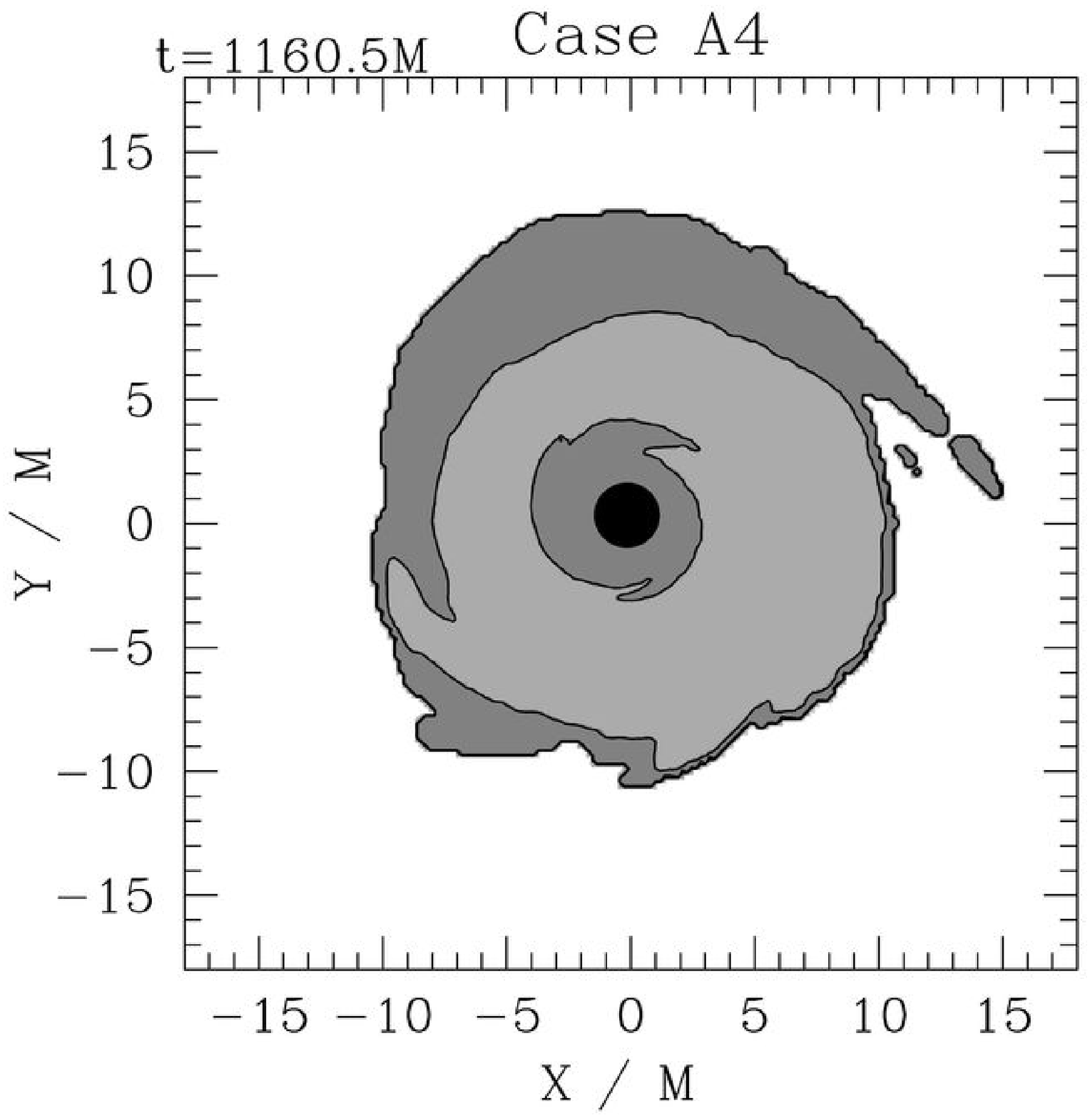}
\caption{Top two plots: Rest-mass density and velocity snapshots
  shortly after disk formation for cases~A0 (left) and A4 (right). Density contours are
  plotted in the orbital plane according to $\rho_0 = \rho_{0,{\rm
      max}} (10^{-0.92j})$,  ($j$=0, 1, ... 5), with darker
  greyscaling for higher density. 
  The maximum initial NS density is $\kappa \rho_{0,{\rm max}} =
  0.126$, 
  or $\rho_{0,{\rm max}}=9\times 10^{14}\mbox{g
    cm}^{-3}(1.4M_\odot/M_0)^2$.  
  Arrows represent the velocity field
  in the orbital plane.  The black hole AH interior is marked 
  by a filled black circle.  The ADM mass for this case is
  $M=2.5\times 10^{-5}(M_0/1.4M_\odot)$ s$=7.6(M_0/1.4M_\odot)$km. \\
  Bottom two plots: Snapshots of the entropy parameter $K$ contours for cases A0 (left) and
  A4 (right).  The light grey regions correspond to $1.4 < \log_{10} K
  < 2.6$, and the dark grey region corresponds to $2.6 < \log_{10} K
  < 3.8$.}
\label{A:rho_K_disk}
\end{center}
\end{figure*}

Compared to case A0, the final disk in case A4 is more massive, and the accretion rate is
lower (Fig.~\ref{A:accretion_history}).
Figure~\ref{A:rho_K_disk} compares the distribution of rest-mass
density and entropy ($\log K$) in the remnant disks of cases A0 and A4, at
the time shortly after disk formation.  The disk in
case A0 is of roughly uniform rest-mass density in the orbital plane.
Though the
disk volume in cases A4 and A0 are comparable, the case A4 disk is
about twice as massive as A0.  
Entropy in case A4 is lower, and entropy contours
fall off rapidly with density. On the other hand, the entropy in case
A0 is much more uniform in the disk, with only a slight drop near
the BH.  Thus in the nonspinning case, adding strong seed magnetic
fields to the NS results in colder, denser, more massive disks.

\begin{figure*}
\vspace{-4mm}
\begin{center}
\epsfxsize=3.5in
\leavevmode
\epsffile{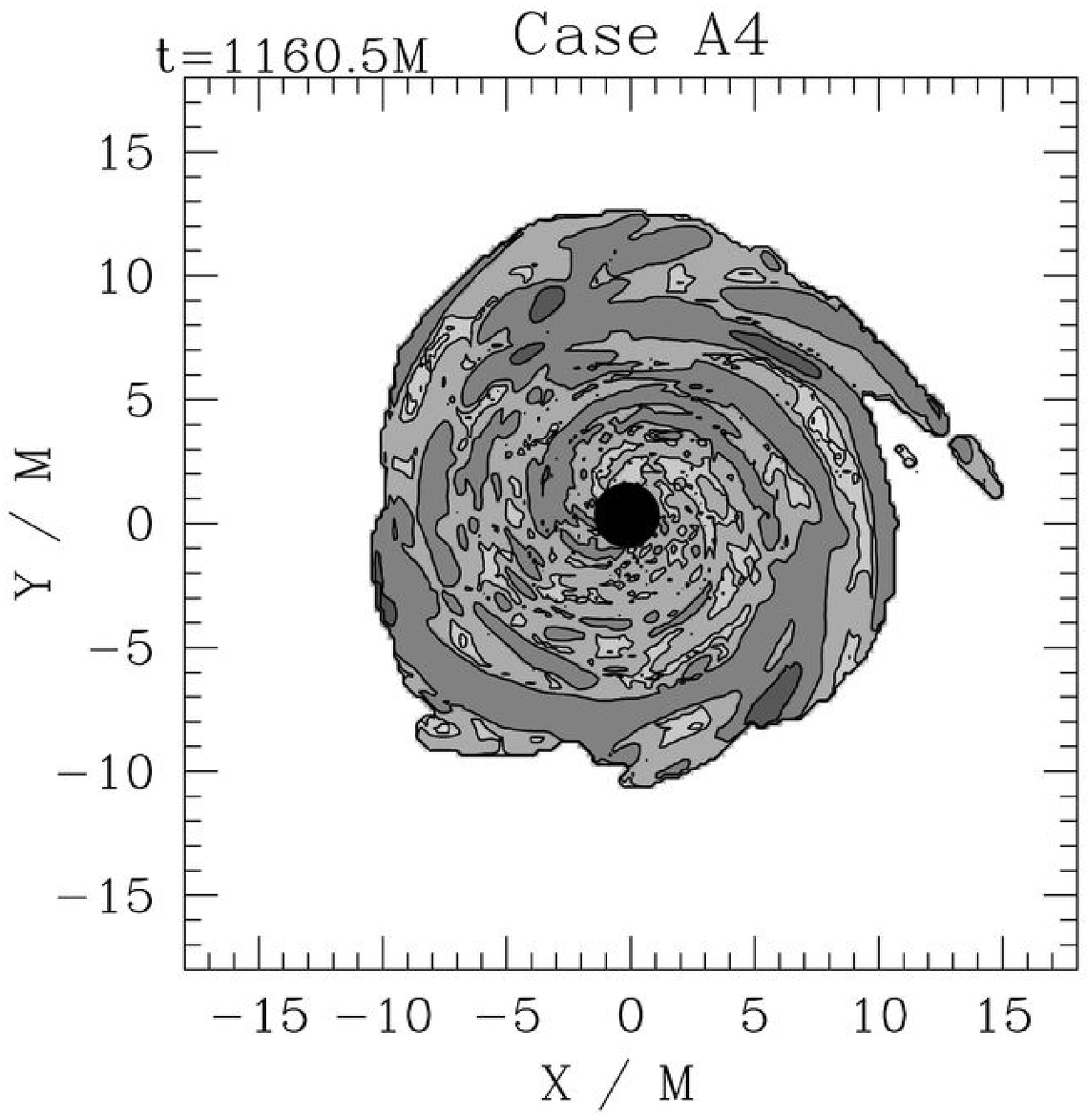}
\epsfxsize=3.5in
\leavevmode
\epsffile{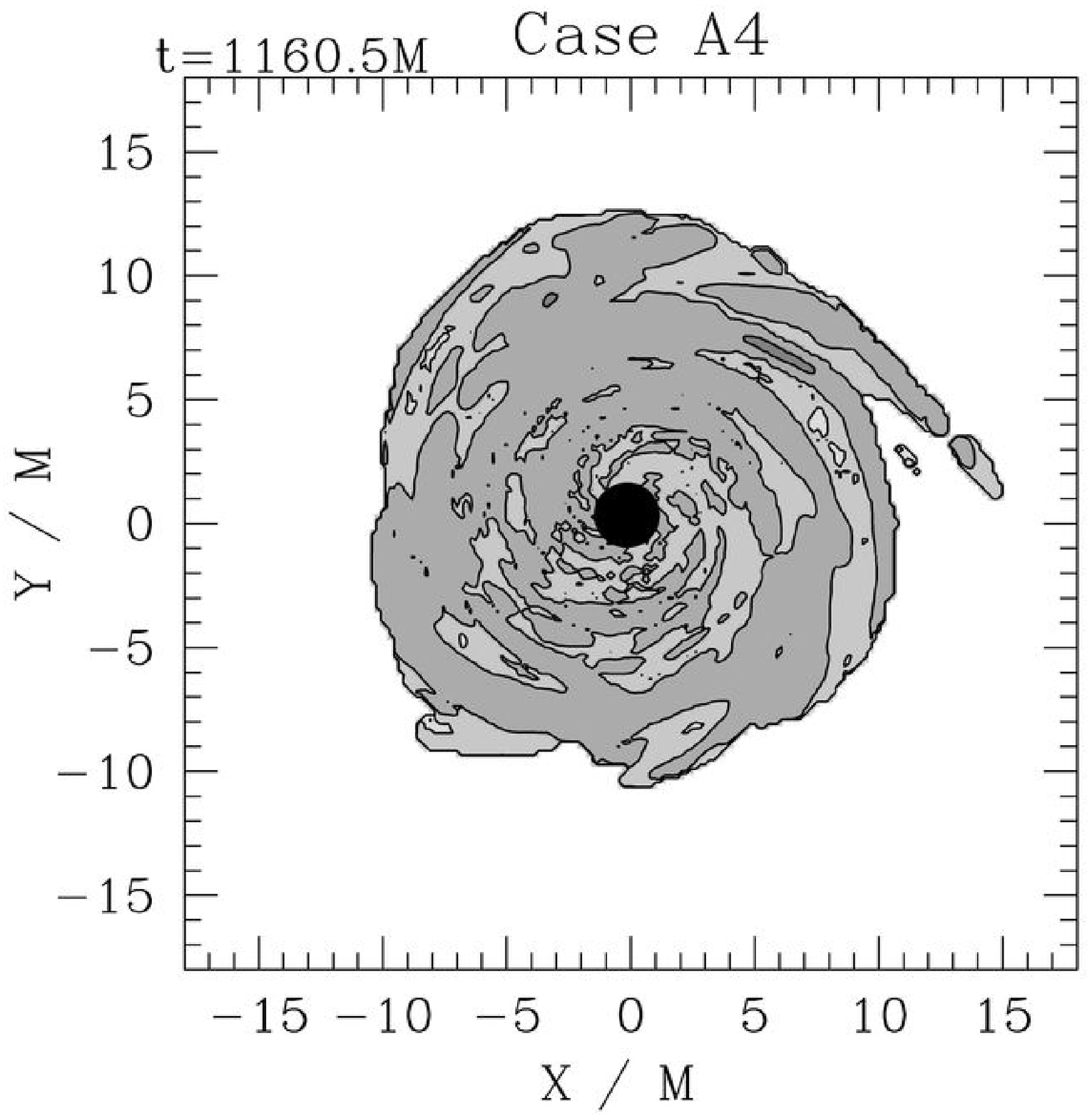}
\caption{$b^2/(2P)$ (left) and $b^2$ (right) contours, at the same
  time as Fig.~\ref{A:rho_K_disk}, plotted according to
  $b^2/(2P)=10^{-1.5}(10^{-1.3j})$, ($j$=0, 1, ... 5), and
  $\kappa b^2 =10^{-5}(10^{-2.2j})$, ($j$=0, 1, ... 5). 
    $b^2/(2P)$
  and $b^2$ contours are only plotted for regions with densities
  higher than the lowest-density $\rho_0$ contours in
  Fig.~\ref{A:rho_K_disk}. In cgs units, 
  $\kappa^{-1}=6\times 10^{36}{\rm erg\, cm}^{-3} (1.4 M_\odot/M_0)^2$.}
\label{A:b2andb2overP_disk}
\end{center}
\end{figure*}

Figure~\ref{A:b2andb2overP_disk} shows $b^2$ profiles for
case A4's disk, plotted at the same time as in
Fig.~\ref{A:rho_K_disk}.  The strong magnetic fields in A4
enhance the final disk mass by more than a factor of two, even though
magnetic pressure never exceeds about 3\% of the gas pressure, with
typical values around 0.01\% (left plot)--about an order of magnitude
lower than during tidal disruption
(Fig.~\ref{A:b2andb2overP_disruption}, left frame).  Short-wavelength
variations in $b^2$ contours appear near the BH only, with
longer-wavelength variations outside.  This is likely a numerical
artifact, since the disk spans about four AMR refinement levels, which
are centered on the BH.  Each refinement level drops the resolution by
a factor of two, meaning that matter in the outer reaches of this disk is
more poorly resolved by a factor of 16 than the region near the
BH. This filtering of wavelengths due to AMR likely suppresses
magnetic-induced turbulence in the disk.

\begin{figure*}
\vspace{-4mm}
\begin{center}
\epsfxsize=3.5in
\leavevmode
\epsffile{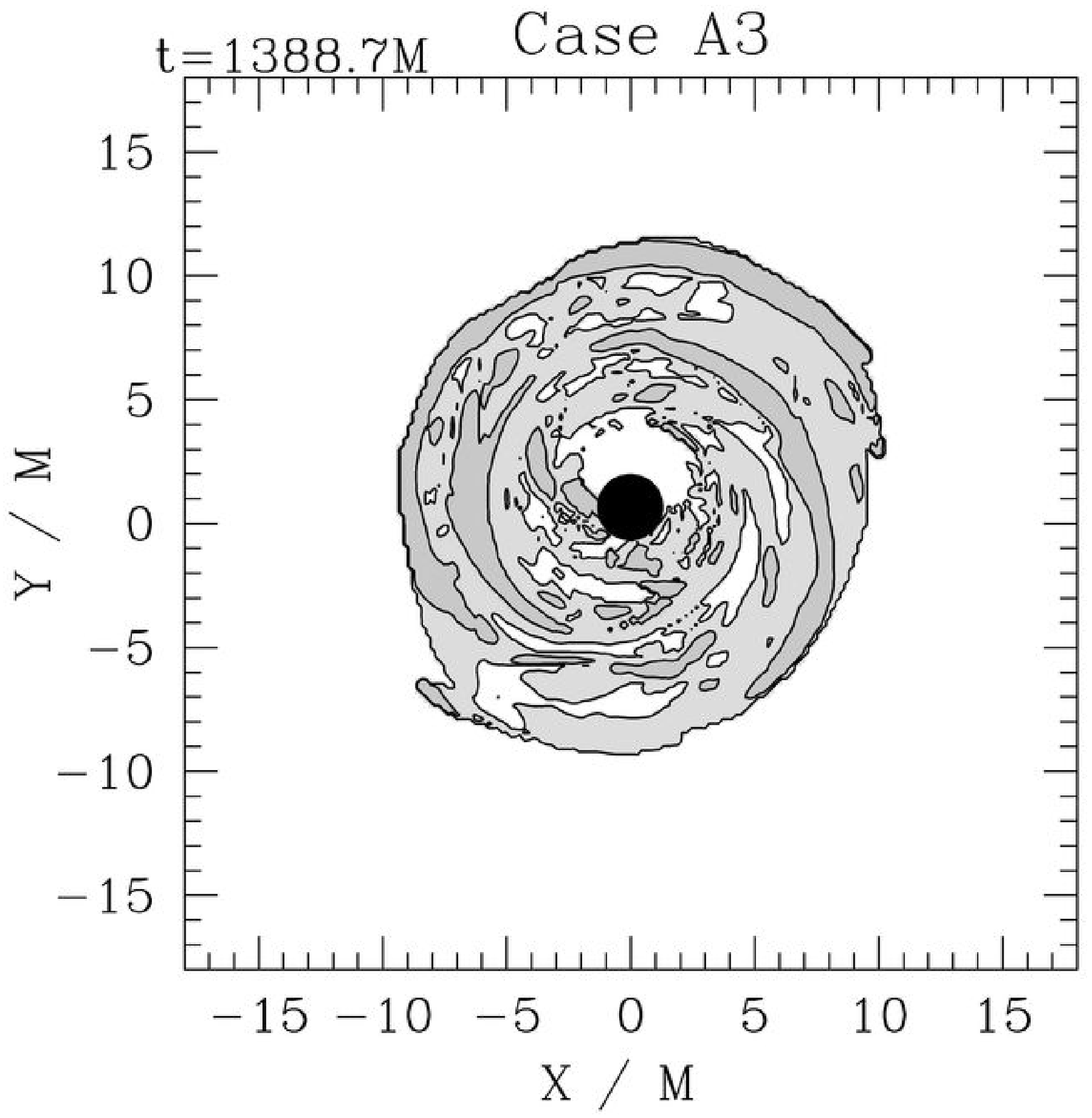}
\epsfxsize=3.5in
\leavevmode
\epsffile{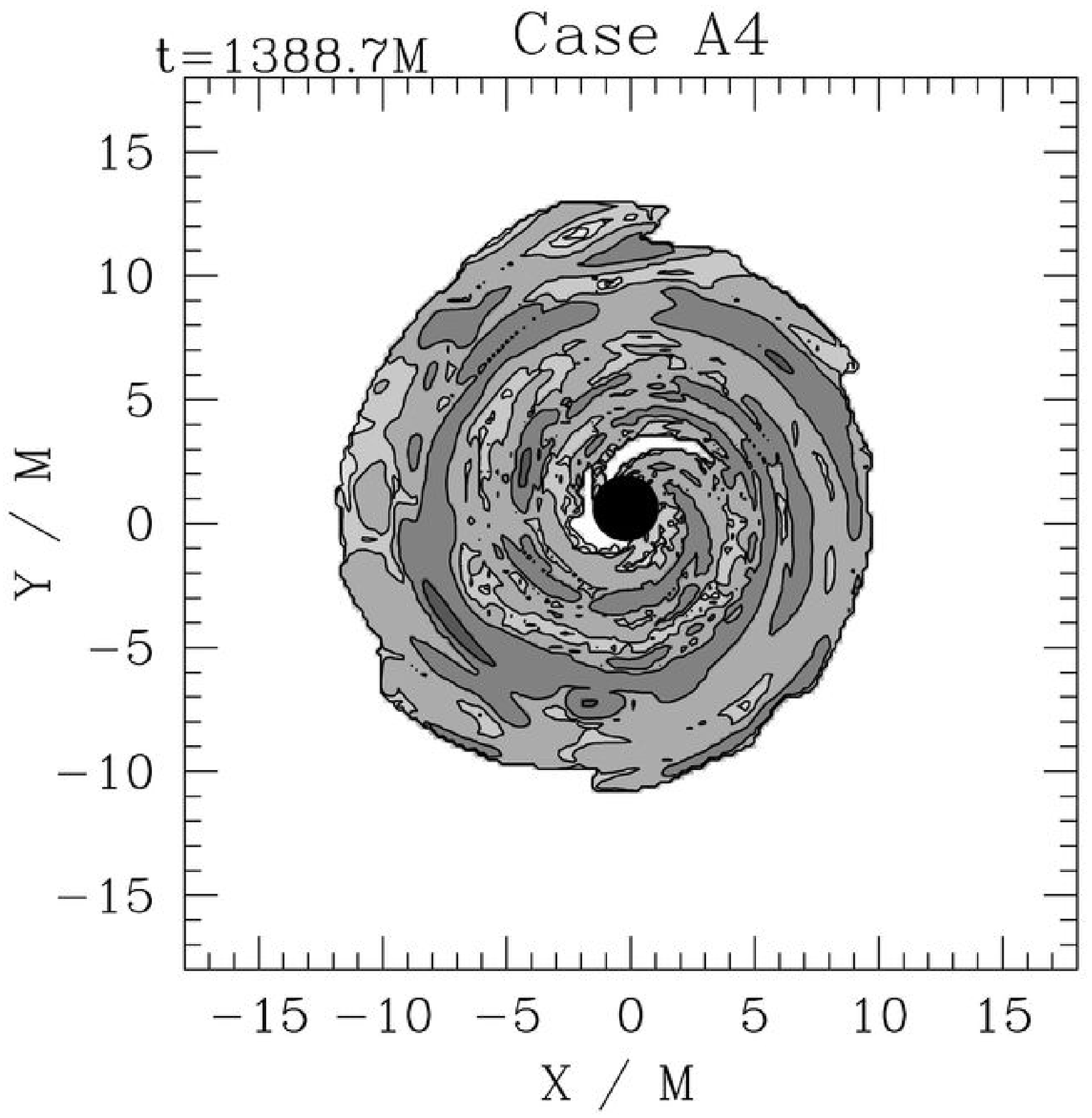}
\caption{$b^2/(2P)$ contours for cases A3 (left) and A4 (right),
  plotted according to $b^2/(2P)=10^{-1.5}(10^{-1.3j})$, ($j$=0, 1,
  ... 5).  Contours are only plotted for regions with densities
  above the low-density cutoff in Fig.~\ref{A:rho_K_disk}.}
\label{A:A3vsA4_b2over2P}
\end{center}
\end{figure*}

Figure~\ref{A:A3vsA4_b2over2P} contrasts $b^2/(2P)$ contours for
cases A3 and A4.  As $b^2/(2P)$ approaches unity, magnetic fields should
have more of a dynamical impact.  Case A3 is identical to A4, except
its seed magnetic fields are weaker by an order of magnitude.
Correspondingly, typical values of $b^2/(2P)$ in this late-time disk
are roughly 2--3 orders of magnitude lower in case A3 than A4.
This is consistent with the finding that the remnant disk mass and
accretion rate are unaffected by magnetic fields in case A3; pressure
from the magnetic fields is simply too small to be dynamically
relevant.  Thus, initial internal magnetic pressures of order 0.1\% of gas
pressure may be required for magnetic fields to have a significant impact
on the system's dynamics.

\subsection{Magnetic Field Study: Spinning Black Hole}

\begin{figure*}
\vspace{-4mm}
\begin{center}
\epsfxsize=2.3in
\leavevmode
\epsffile{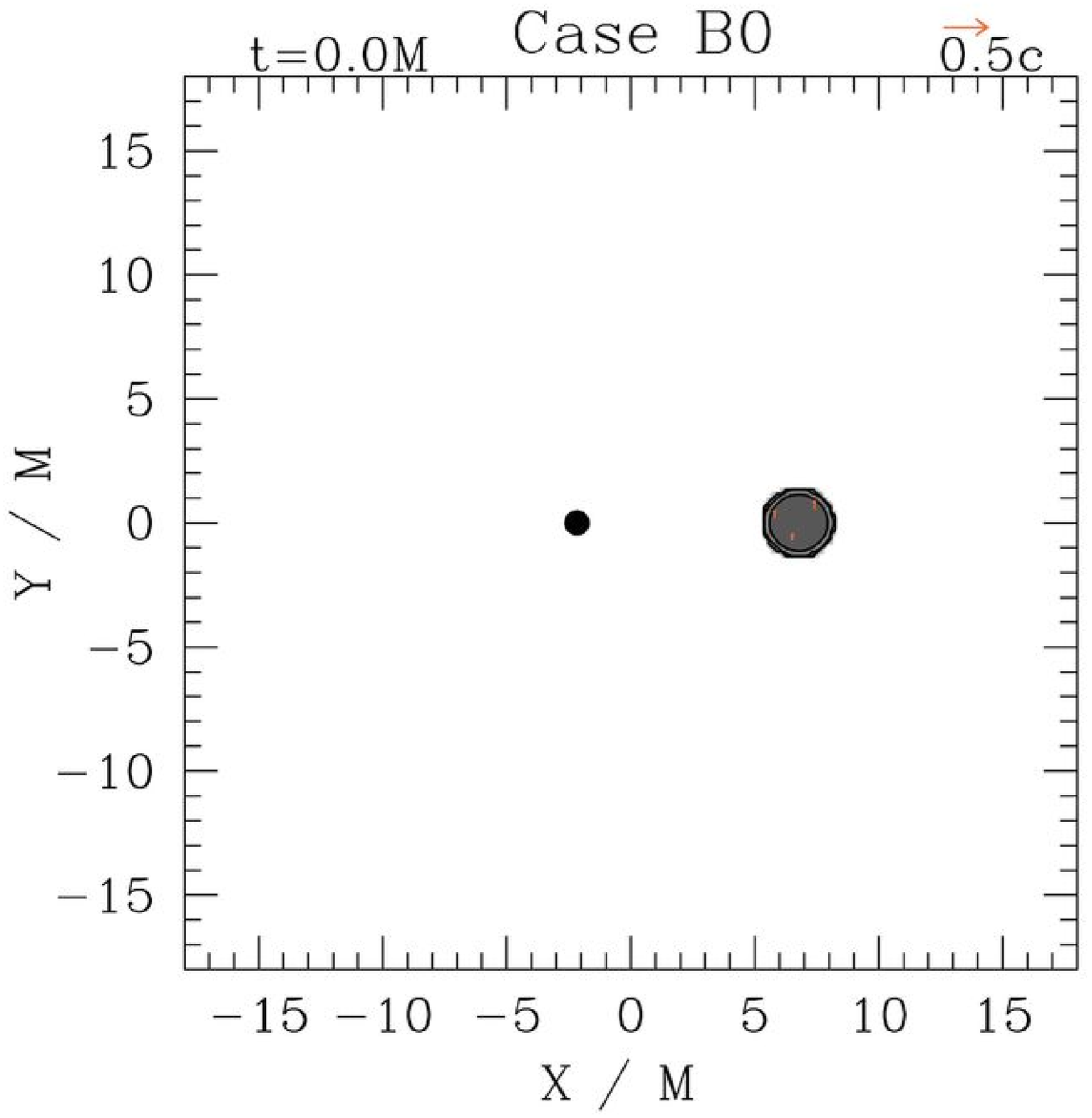}
\epsfxsize=2.3in
\leavevmode
\epsffile{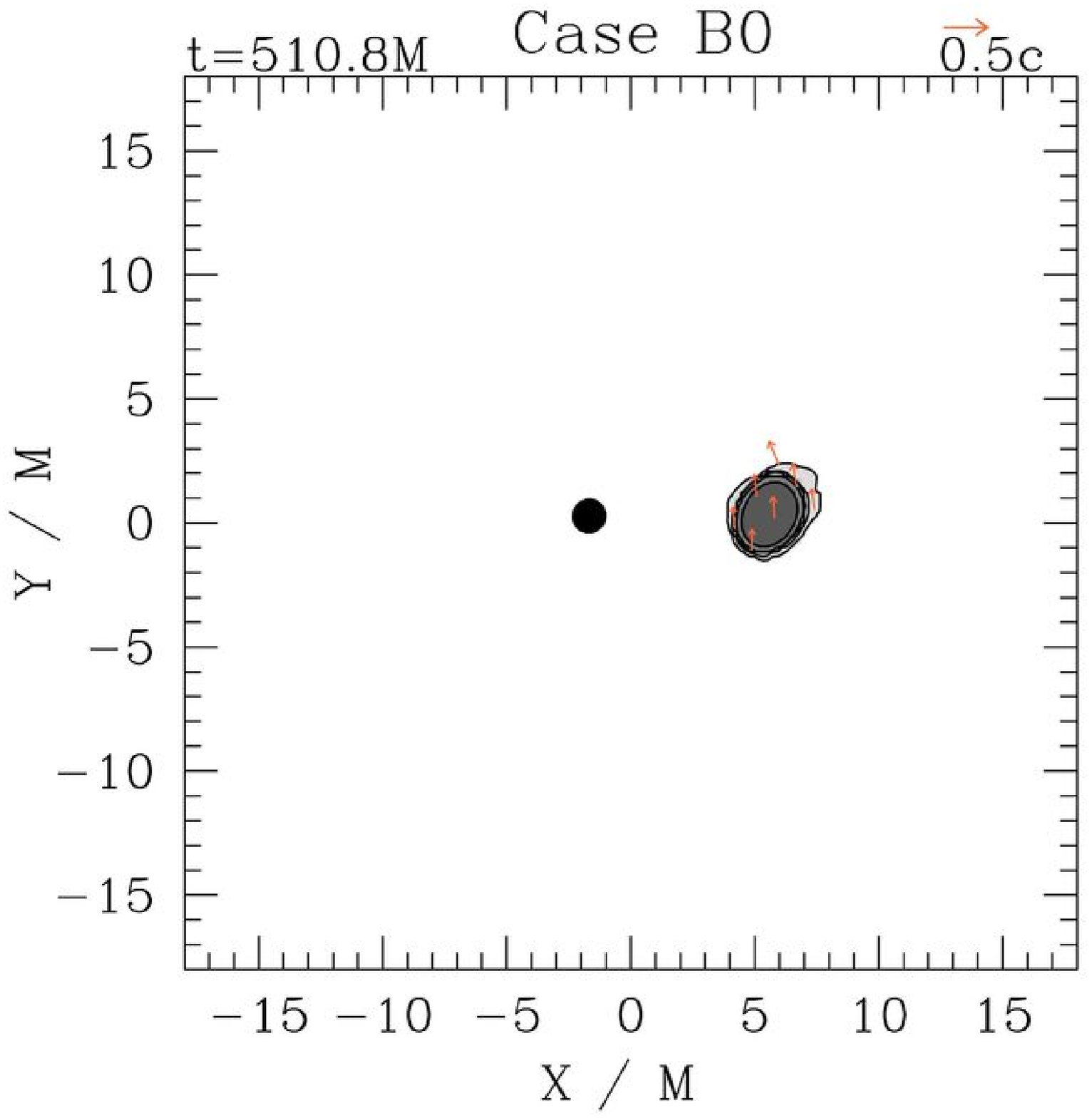}
\epsfxsize=2.3in
\leavevmode
\epsffile{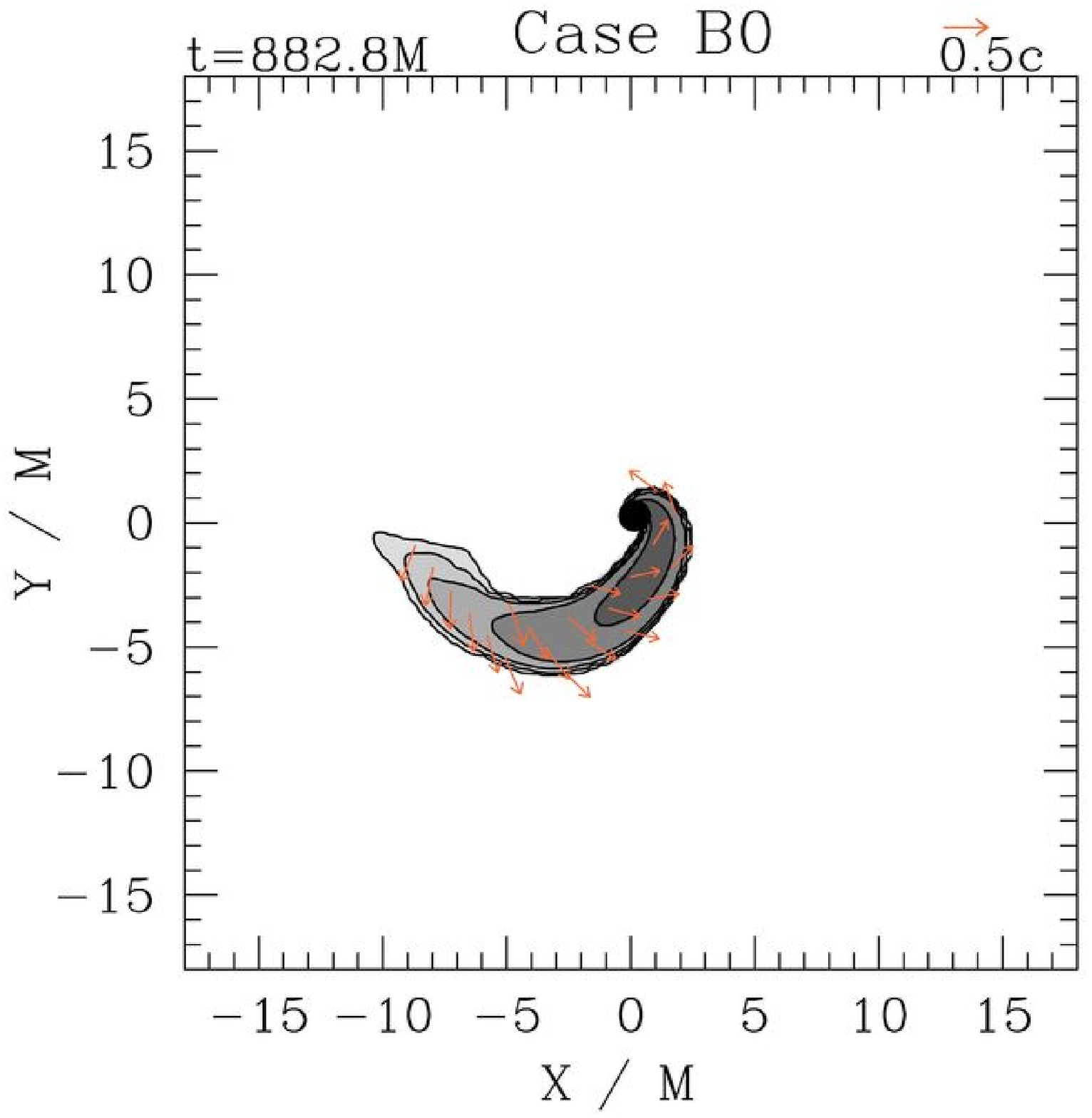}\\
\epsfxsize=2.3in
\leavevmode
\epsffile{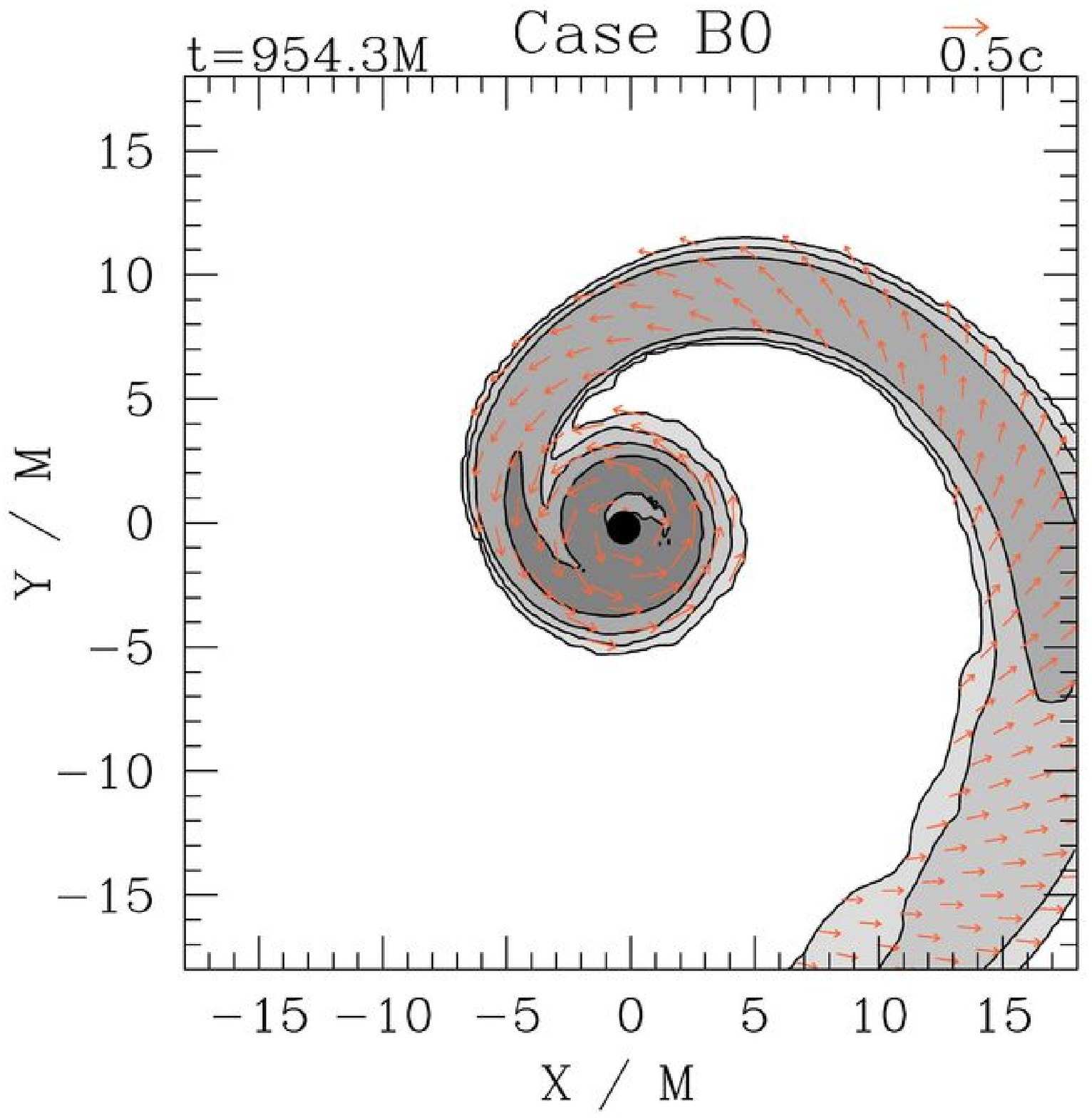}
\epsfxsize=2.3in
\leavevmode
\epsffile{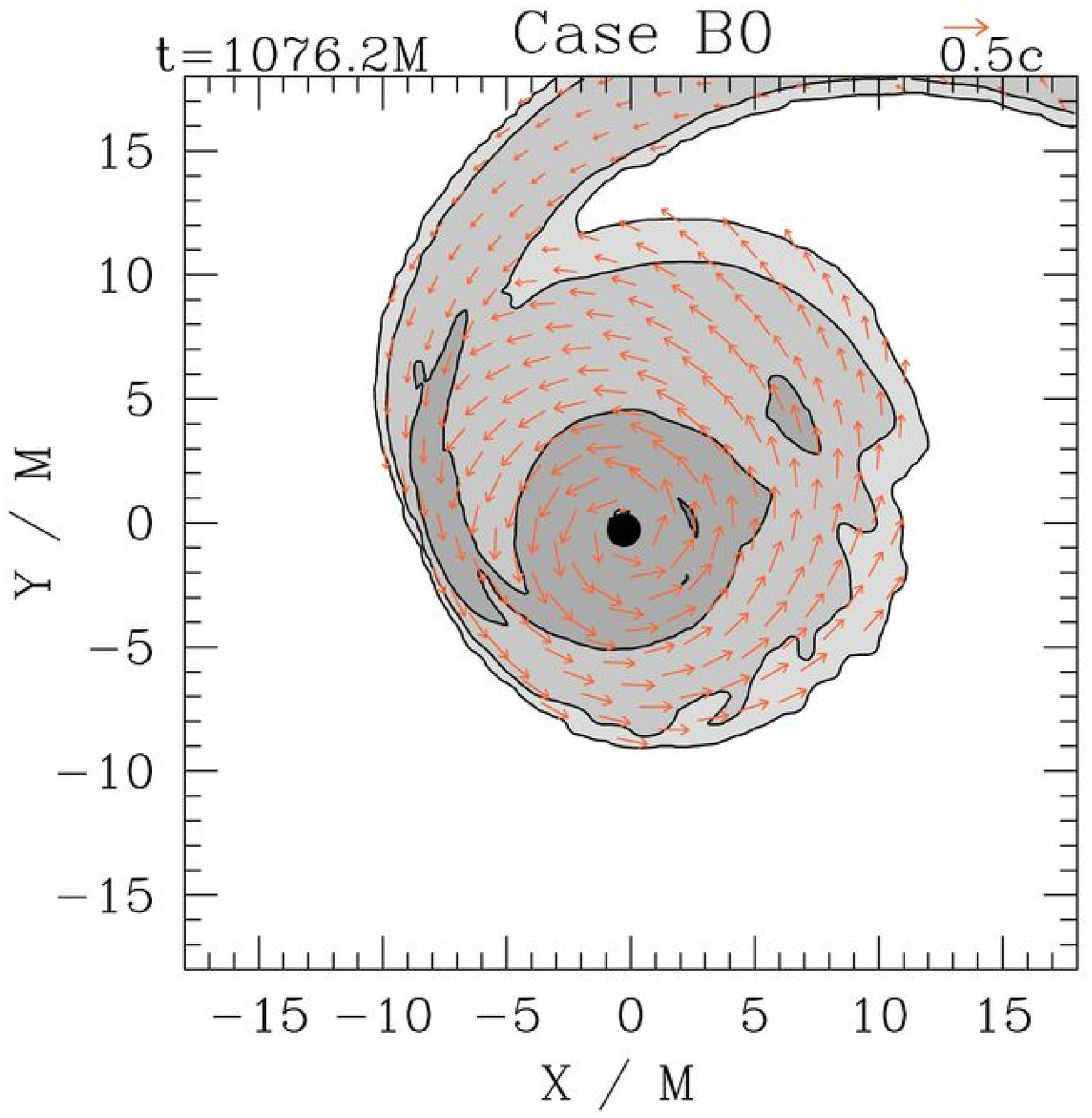}
\epsfxsize=2.3in
\leavevmode\
\epsffile{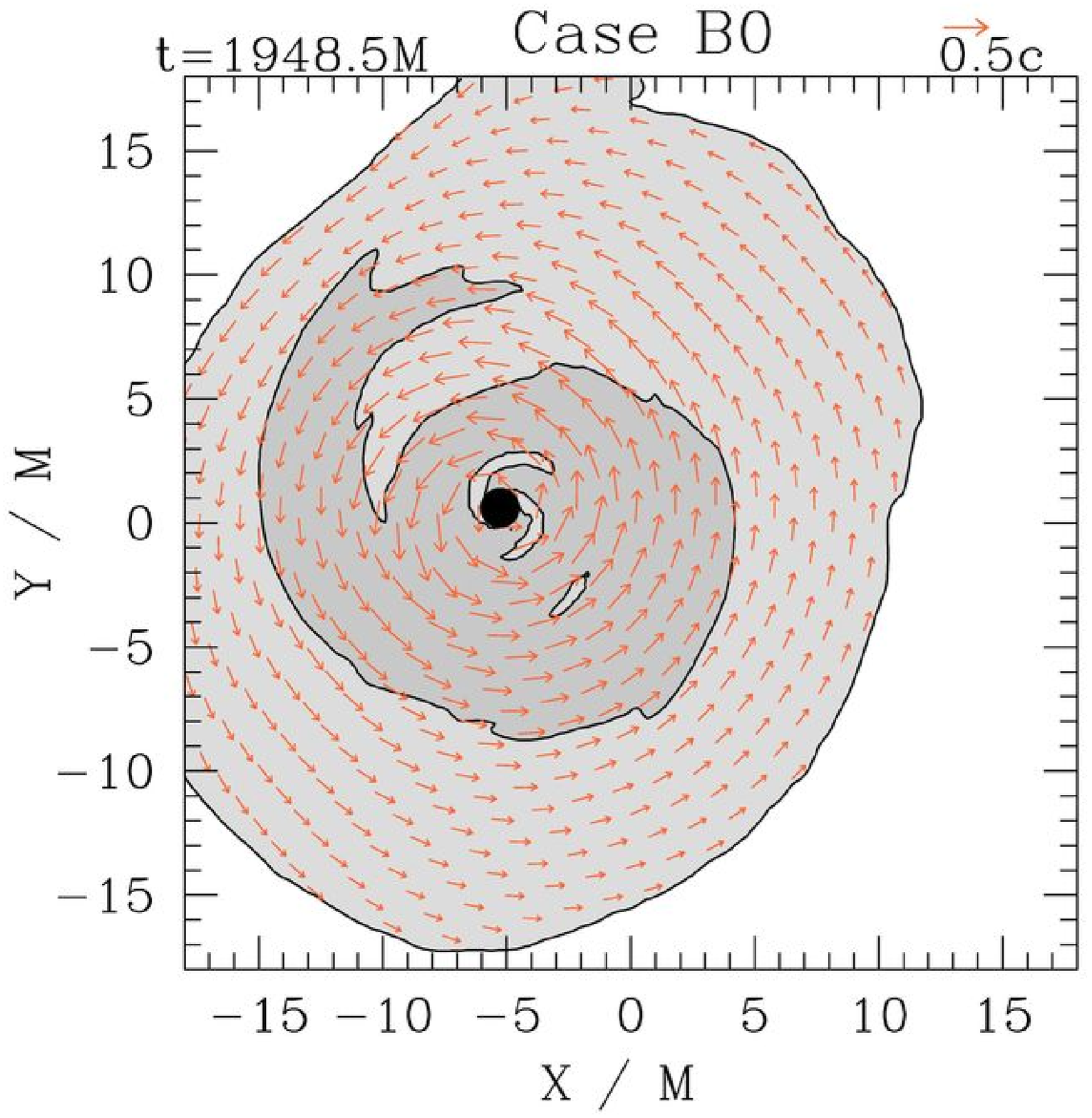}
\caption{Orbital-plane 
  density contours at selected times
  for case B0.  Density contours are plotted in
  the orbital plane, according to $\rho_0 = \rho_{0,{\rm max}} (10^{-0.92j})$,
  ($j$=0, 1, ... 5), with darker greyscaling for higher density.  The maximum initial NS
  density is $\kappa \rho_{0,{\rm max}} = 0.126$, or $\rho_{0,{\rm
      max}}=9\times 10^{14}\mbox{g cm}^{-3}(1.4M_\odot/M_0)^2$.
  Arrows in density contour plots represent the velocity field in the
  orbital plane, and the black hole AH interior is marked by a filled
  black circle.  The ADM mass for this case is
  $M=2.5\times 10^{-5}(M_0/1.4M_\odot)$s$=7.6(M_0/1.4M_\odot)$km.}
\label{B:evolution_story}
\end{center}
\end{figure*}

Figure~\ref{B:evolution_story} outlines the basic evolution scenario
for the fiducial $a_{\rm BH}/M_{\rm BH}$=0.75 (henceforth ``spinning'') case,
B0.  Unlike the nonspinning case, the NS is slightly distorted from
its equilibrium shape after about three orbits
(top-left plots, cf. top-left plots in Fig.~\ref{A:evolution_story}).
Although the spinning cases start with the same initial orbital
angular frequency as the nonspinning cases, all spinning cases require
about two more orbits before an accretion funnel forms (top-right
frame).  This is due to the well-known ``orbital hang-up'' effect.
Unlike the nonspinning case,
in which about 95\% of the NS matter immediately funnels into the BH,
strong frame-dragging in the spinning cases twists the accreting
funnel around the BH, promptly accreting only $\sim$70\% of the NS
matter. Accretion slows after the funnel twists
around the BH and intersects itself, generating shock heating and
producing a small, dense disk-like structure that expands and rarefies
as it orbits the BH. Notice that this ``disk'' is much smaller and
denser than in the nonspinning cases (bottom-left plot,
cf. Fig.~\ref{A:evolution_story}). Attached to this small ``disk'', the
outer layers of the NS have formed a long tidal tail, which is ejected to a
large radius before slowly falling back onto the expanding but accreting
``disk'' (bottom-middle frame). The disk continues its expansion as
much of the tail falls into it.
Near the time when the simulation is stopped, we find no indication
of a cavity near the BH (bottom-right frame). Longer simulations may be required 
for a cavity to appear, which might indicate the presence of an ISCO. 
Note that in contrast with the expected long-term evolution of this
unmagnetized case, stresses in magnetized disks may suppress the formation of hollow
regions around the ISCO \cite{kh02,bhk08}.

\begin{figure}
\epsfxsize=3.4in
\leavevmode
\epsffile{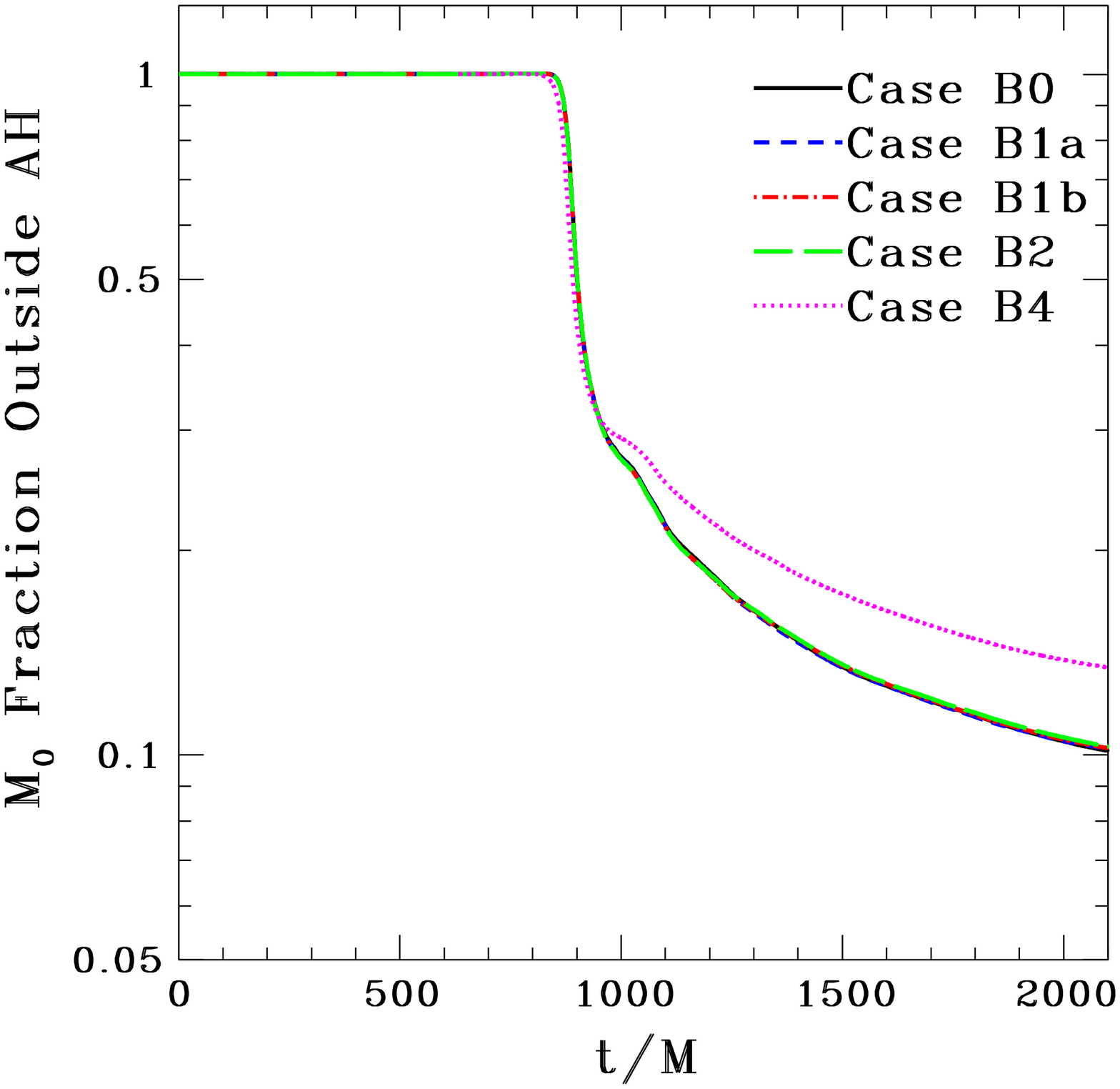}
\caption{Rest-mass fraction outside the AH for 
all cases in which the initial BH possesses spin parameter 
of $a_{\rm BH}/M_{\rm BH}$=0.75.}
\label{B:accretion_history}
\end{figure}

The corresponding accretion history for all cases
in which the BH initially has spin is shown in
Fig.~\ref{B:accretion_history}.  Similar to the nonspinning cases,
only the most strongly magnetized case, B4, has an
appreciably different accretion history.  The NS in case B4 possesses
the same magnetic field geometry and strength as the NS in the
(nonspinning) case A4: the seed magnetic fields are both
strong ($|B|_{\rm max}\sim10^{17}$G, initially) and pushed all the way
to the NS surface ($P_c=0.001$).
The magnetic fields in case B4 increase the disk mass from about 10\%
to 14\%, a net 40\% amplification in disk mass.  Notice also
that BH spin alone has a {\it very significant} influence on final
disk mass; increasing initial aligned BH spin from zero to $a_{\rm BH}/M_{\rm BH}$=0.75 increases
the final disk mass by about an order of magnitude ($\sim$0.9\% --
$\sim$10\%).  Based on the final accretion rate, the half-life
of the disk is roughly 5000$M$ , or $\sim 140 (M_0/1.4M_\odot)$ms.

\begin{figure}
\epsfxsize=3.4in
\leavevmode
\epsffile{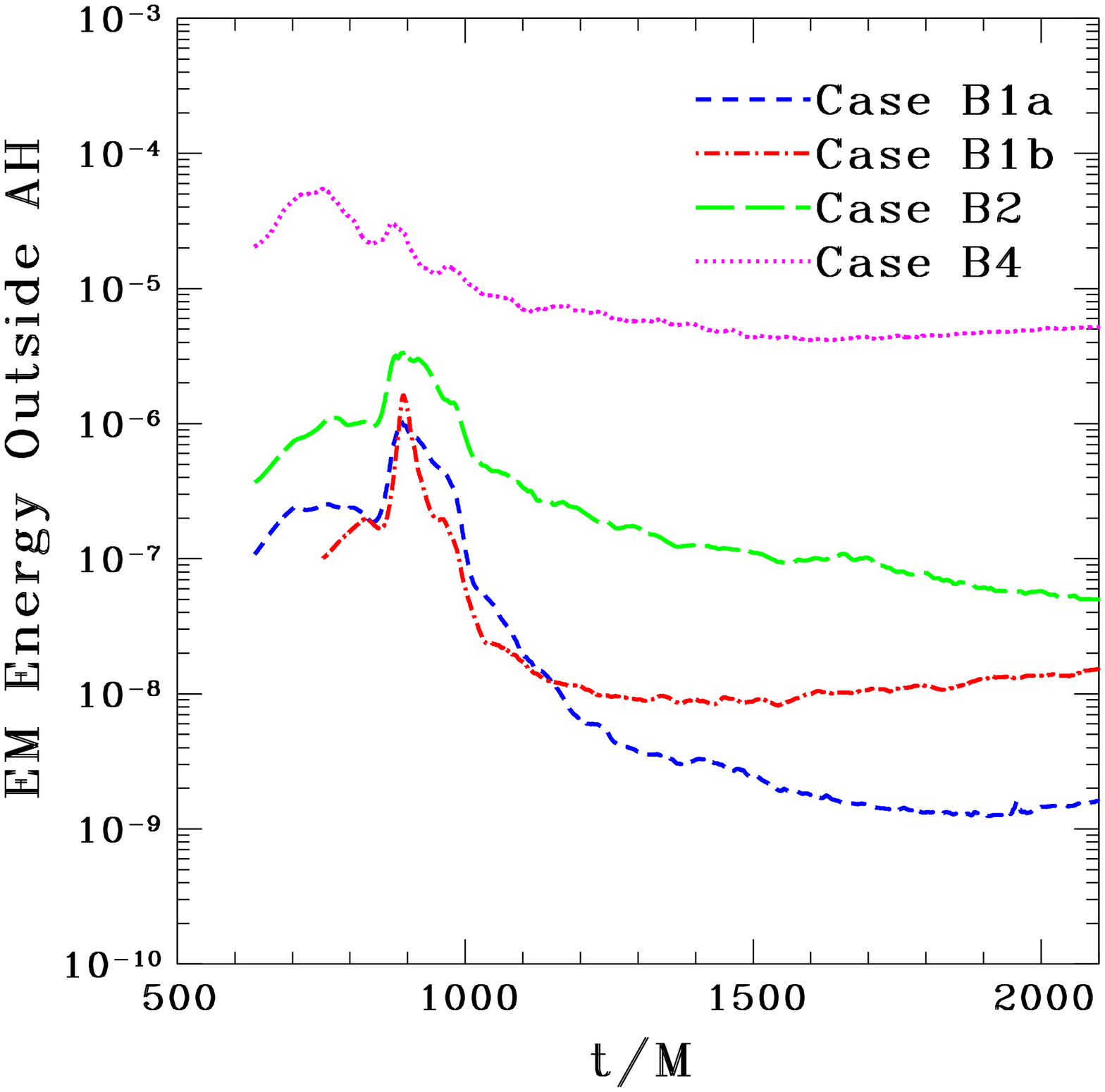}
\caption{Total magnetic energy outside the AH for all magnetized cases
  in which the initial BH possesses spin parameter of $a_{\rm BH}/M_{\rm
    BH}$=0.75, normalized by the ADM mass 
$M = 9.3 \times 10^{54} (M_0/1.4M_\odot)$erg..} 
\label{B:magnetic_energy_history}
\end{figure}

Figure~\ref{B:magnetic_energy_history} plots the magnetic energy outside
the AH versus time for all spinning, magnetized cases.
During merger there are two competing effects: magnetic energy will
increase as the NS tidally disrupts and the field lines are stretched,
while magnetic energy outside the AH will decrease as magnetized NS matter is accreted
into the BH.  This major accretion event occurs at
$t\approx900M$ for all spinning cases, corresponding to a spike
in magnetic energy at that time.  At $t=1000M$, only about 20\% of the
NS matter remains outside the BH, but the magnetic energy in all cases
is {\it higher} than when the seed fields were added to the NS, indicating
that a large amplification in magnetic field strength has occurred.
Only case B4 has sufficiently strong magnetic fields to severely
impact the disk dynamics; the disk mass is amplified by about 40\%,
and the magnetic energy in this case is 1--2 orders of magnitude
stronger than any other case at all times.  

To determine the sensitivity of our results on the time at which the 
magnetic fields were added to the NS, two simulations were
performed: cases B1a and B1b.  These simulations are identical except
the seed magnetic fields were added to the NS about half an orbit
later in case B1b.  The accretion histories of cases B1a and B1b
overlap completely, implying that the bulk dynamics are unaffected by
when the magnetic fields were inserted into the NS.  During and
directly after the merger  ($t\approx900M$), magnetic energy in cases
B1a and B1b overlap.  After merger, only a tiny fraction of the fields
in the outer layers of the NS remains outside the AH.  Correspondingly, the
magnetic energy plummets in both cases to negligibly small
values, yet agree to within an order of magnitude even at late
times. Thus we conclude that the final result is largely insensitive
to when the seed magnetic fields were added.

\begin{figure*}
\vspace{-4mm}
\begin{center}
\epsfxsize=2.2in
\leavevmode
\epsffile{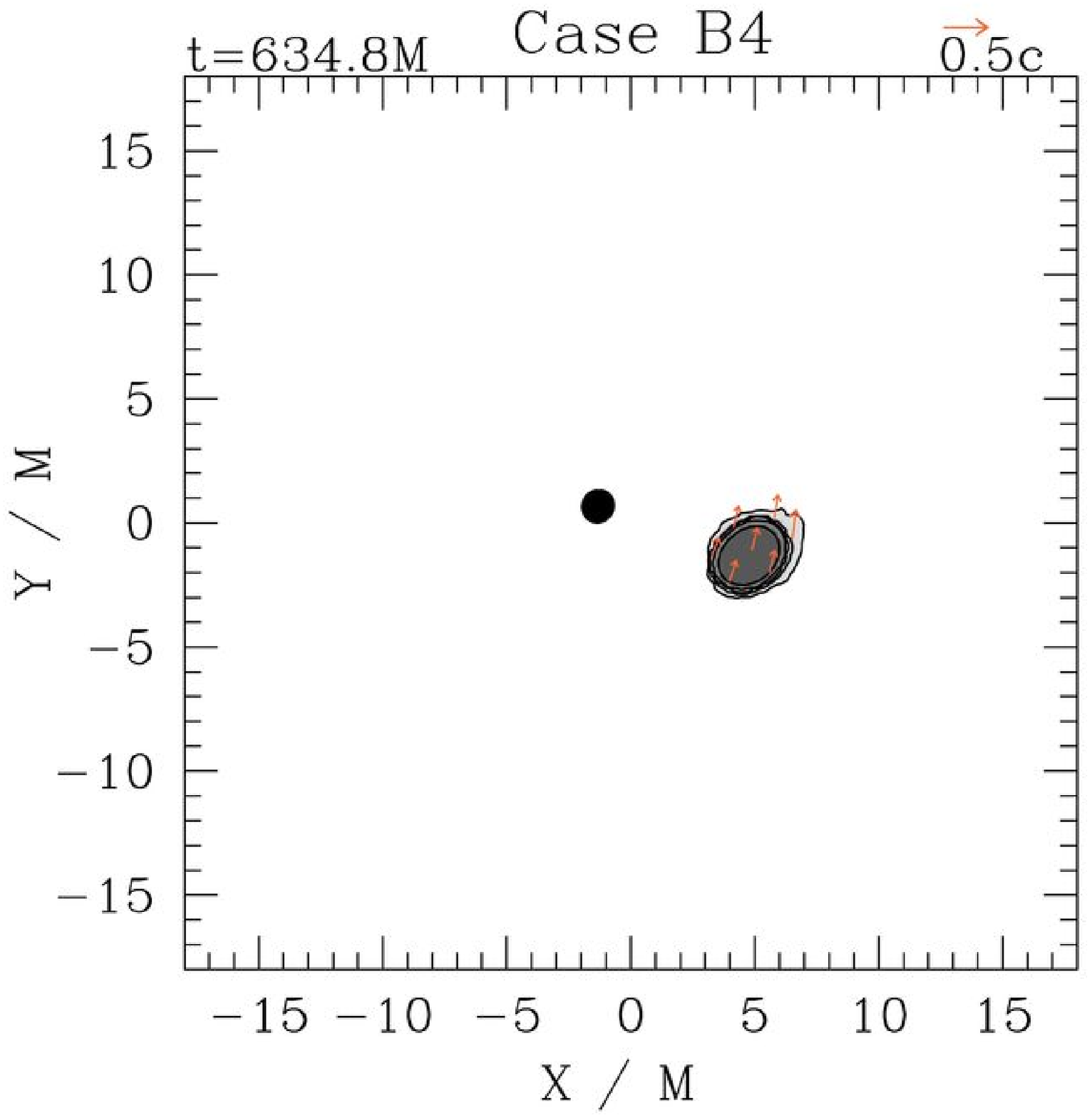}
\epsfxsize=2.3in
\leavevmode
\epsffile[20 -60 575 437]{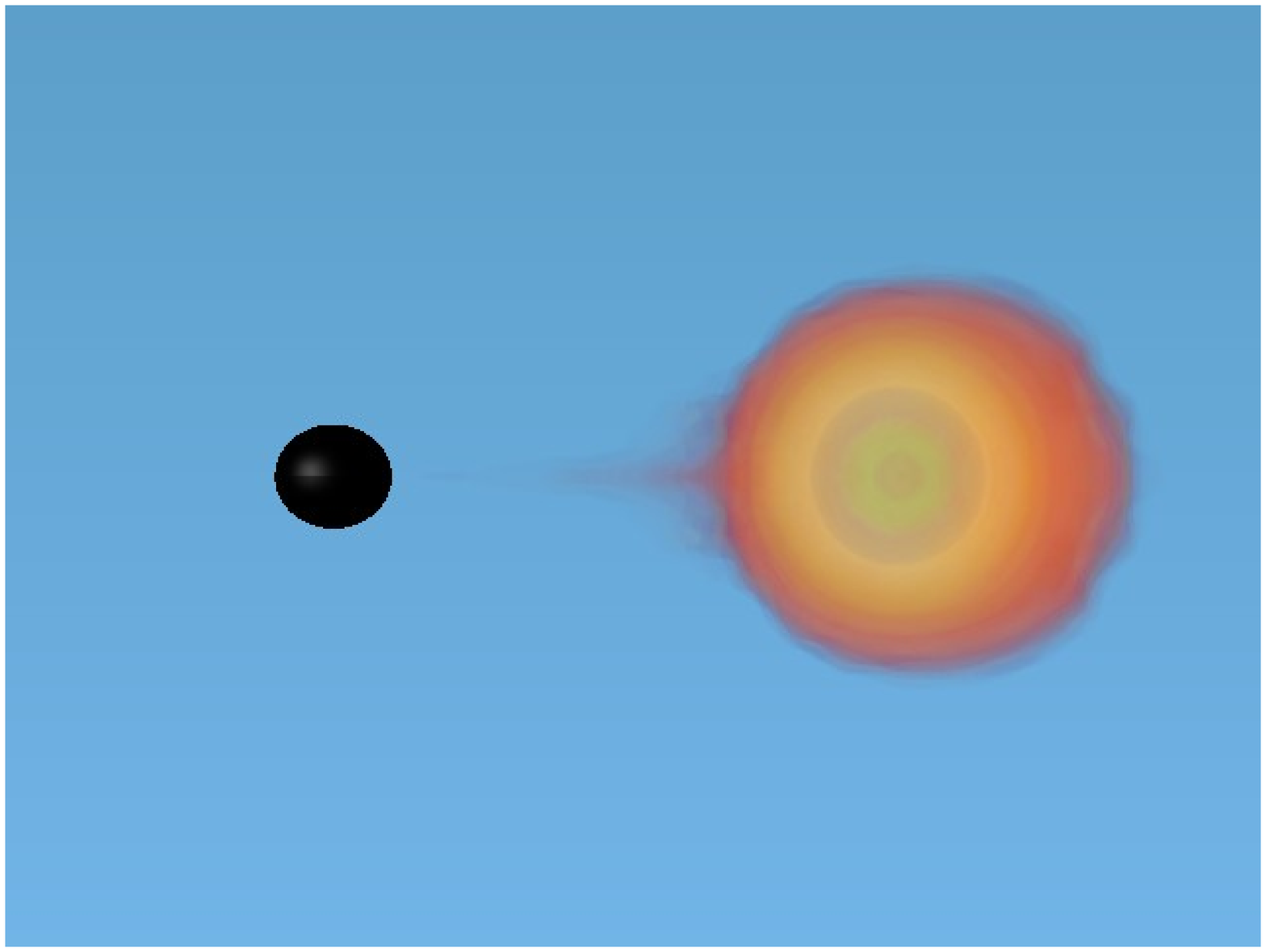}
\epsfxsize=2.3in
\leavevmode
\epsffile[20 -60 575 437]{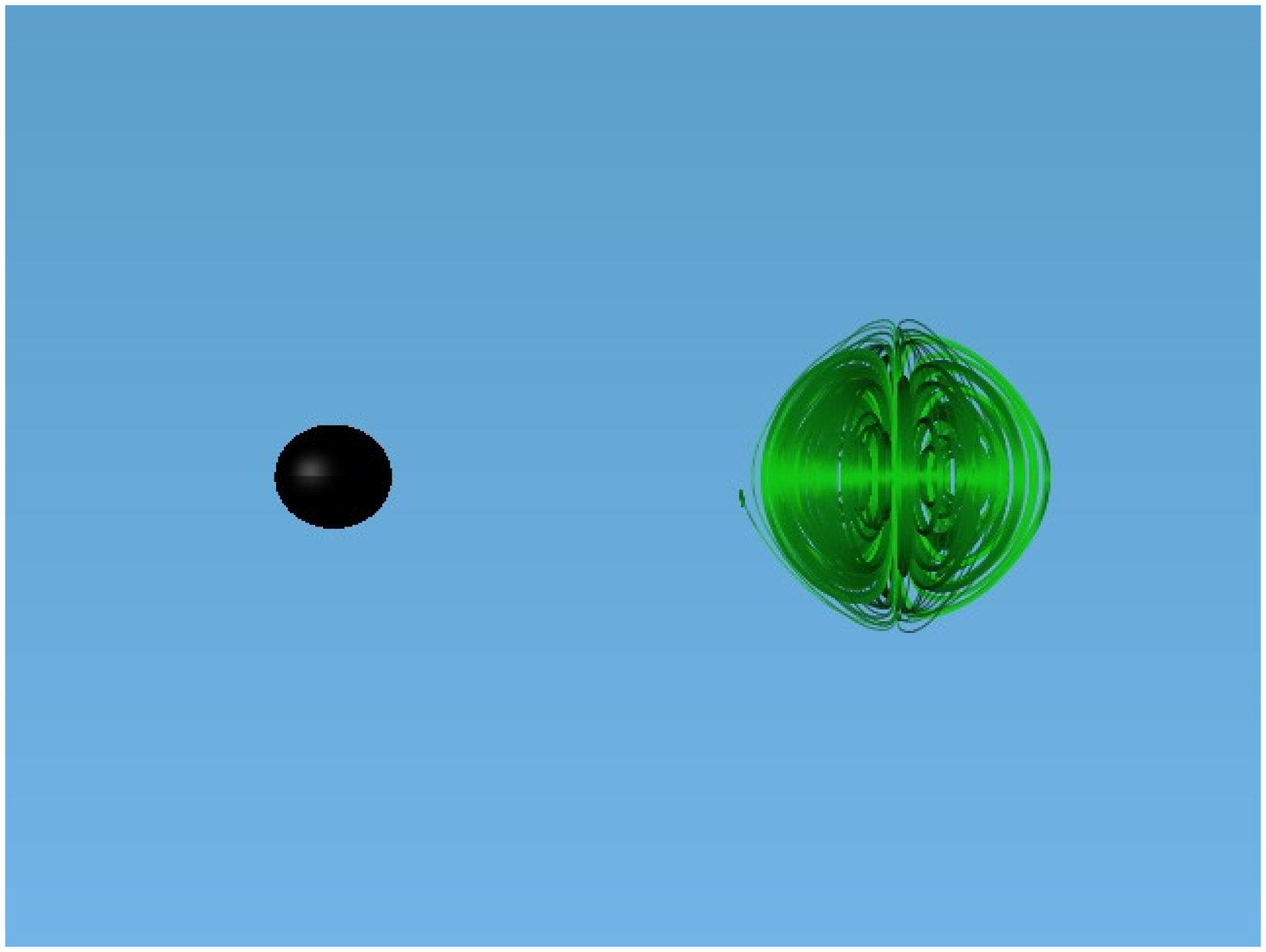}\\
\epsfxsize=2.2in
\leavevmode
\epsffile{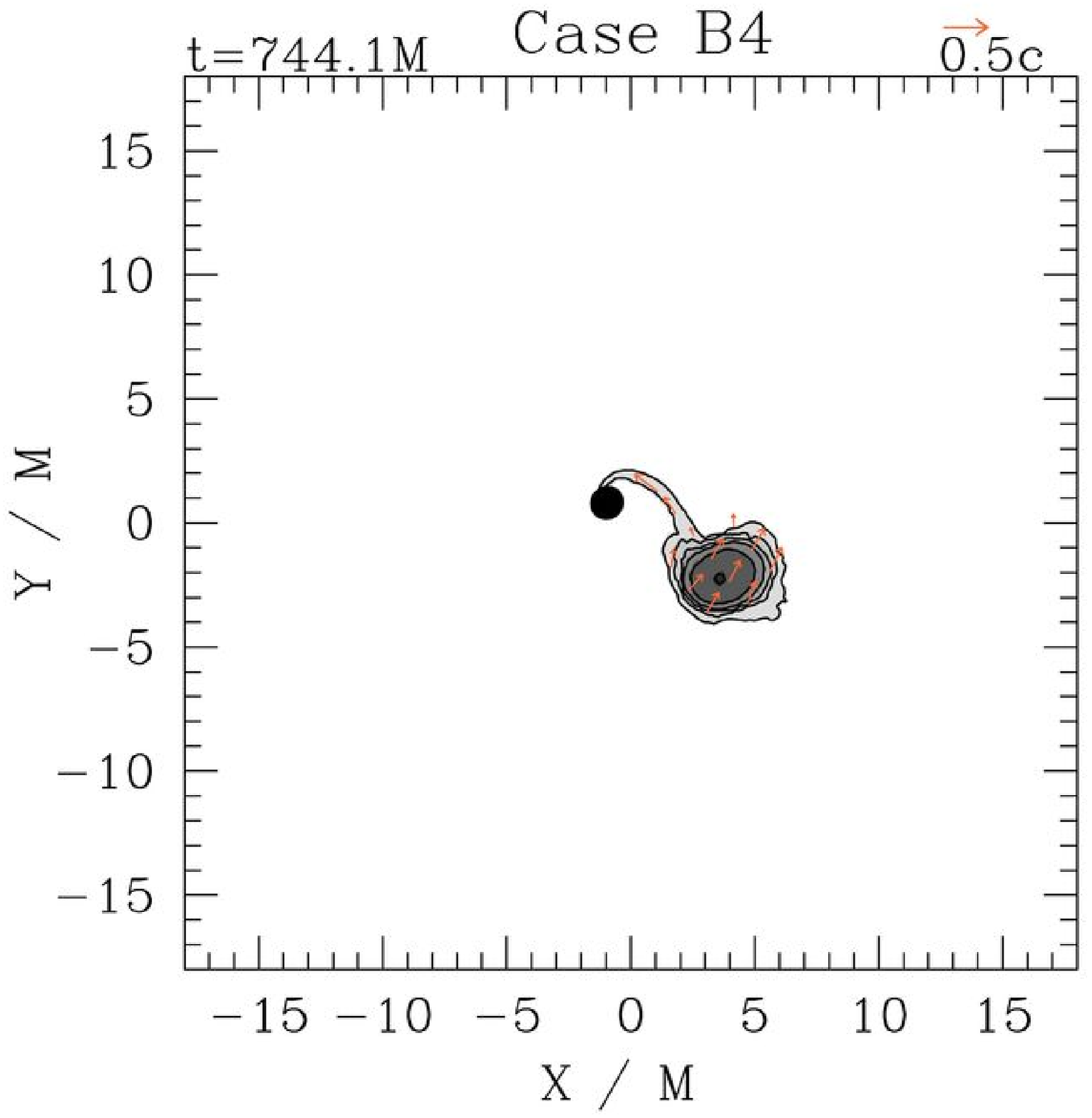}
\epsfxsize=2.3in
\leavevmode
\epsffile[20 -60 575 437]{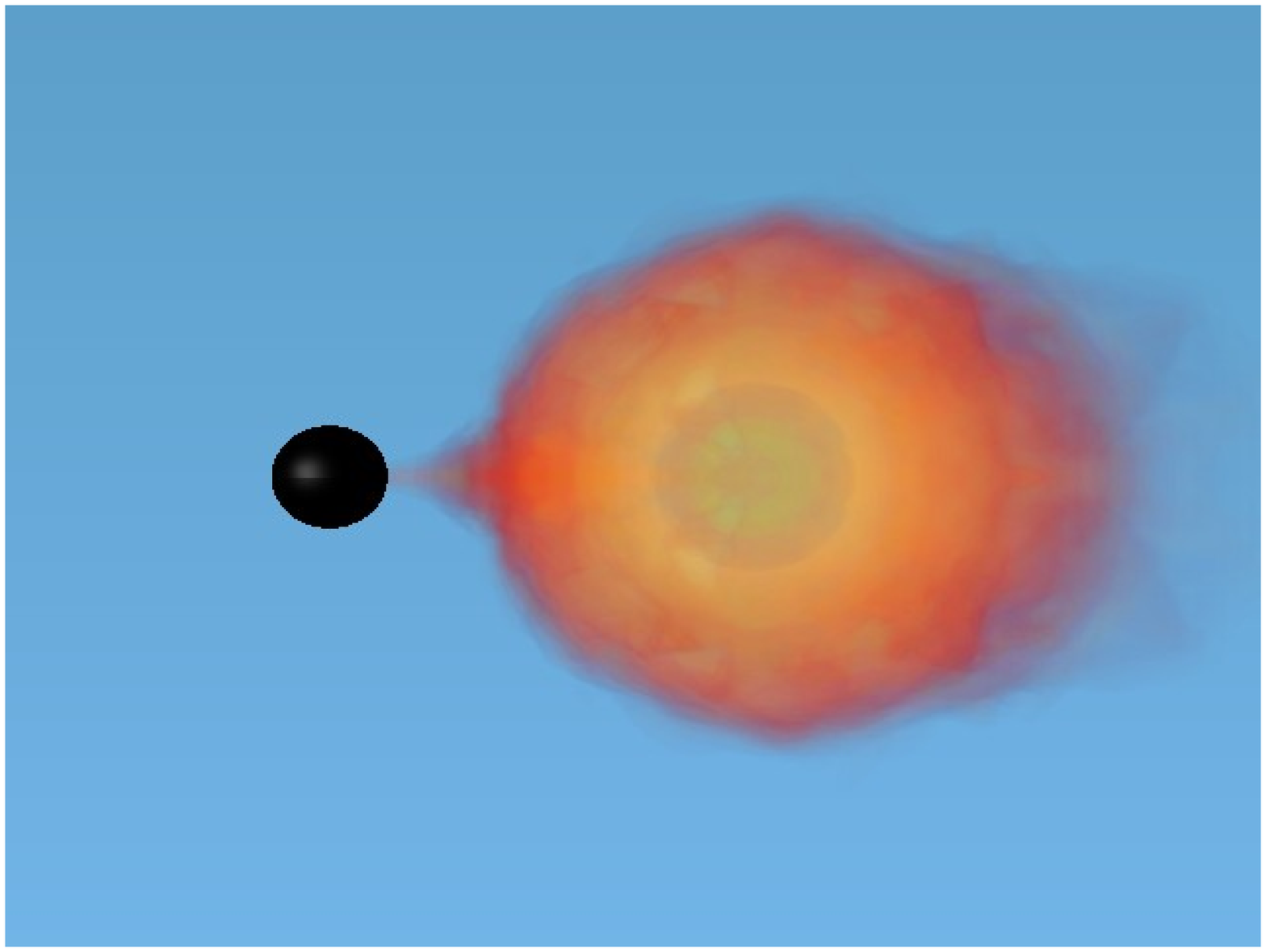}
\epsfxsize=2.3in
\leavevmode
\epsffile[20 -60 575 437]{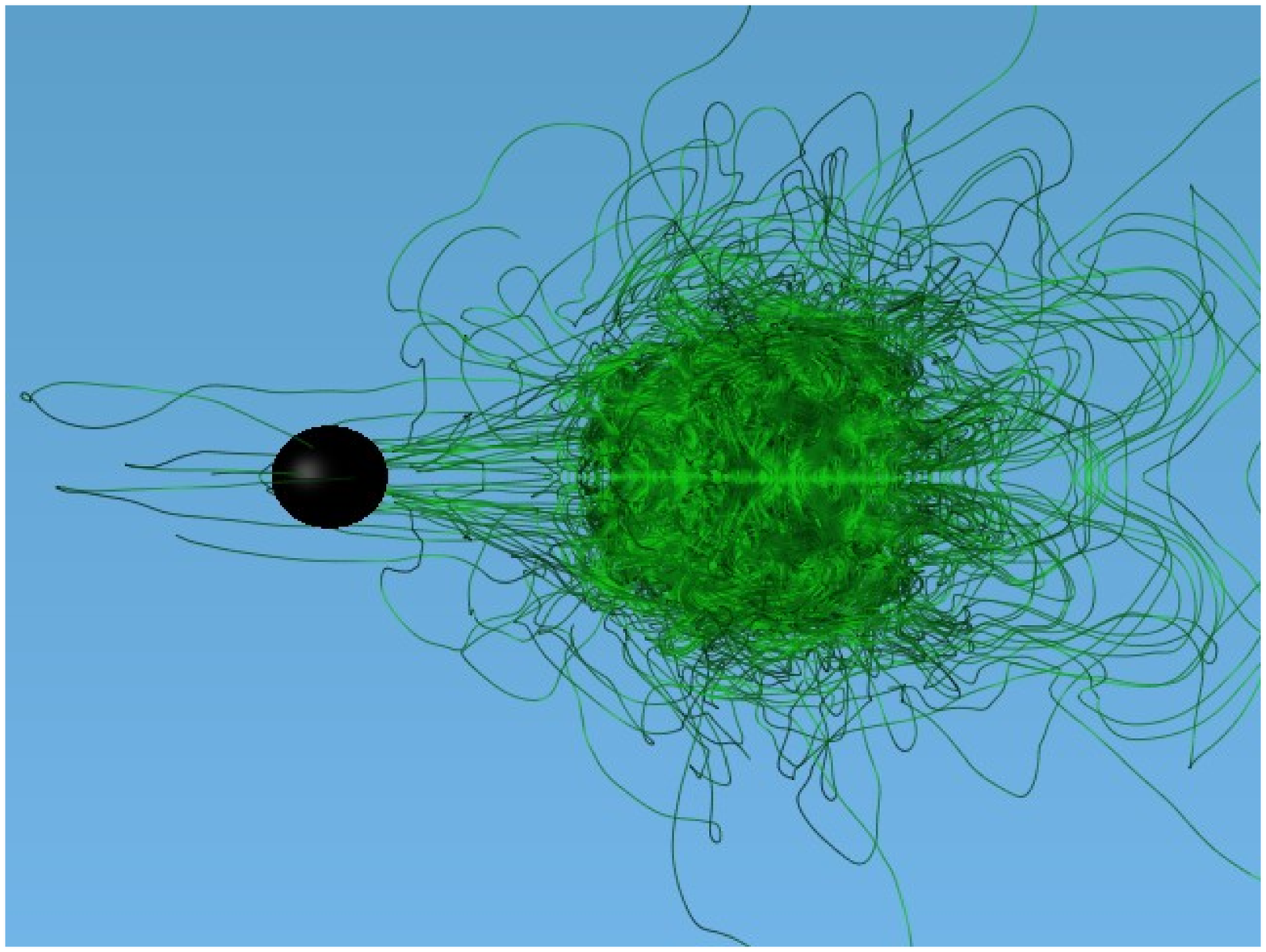}\\
\epsfxsize=2.2in
\leavevmode
\epsffile{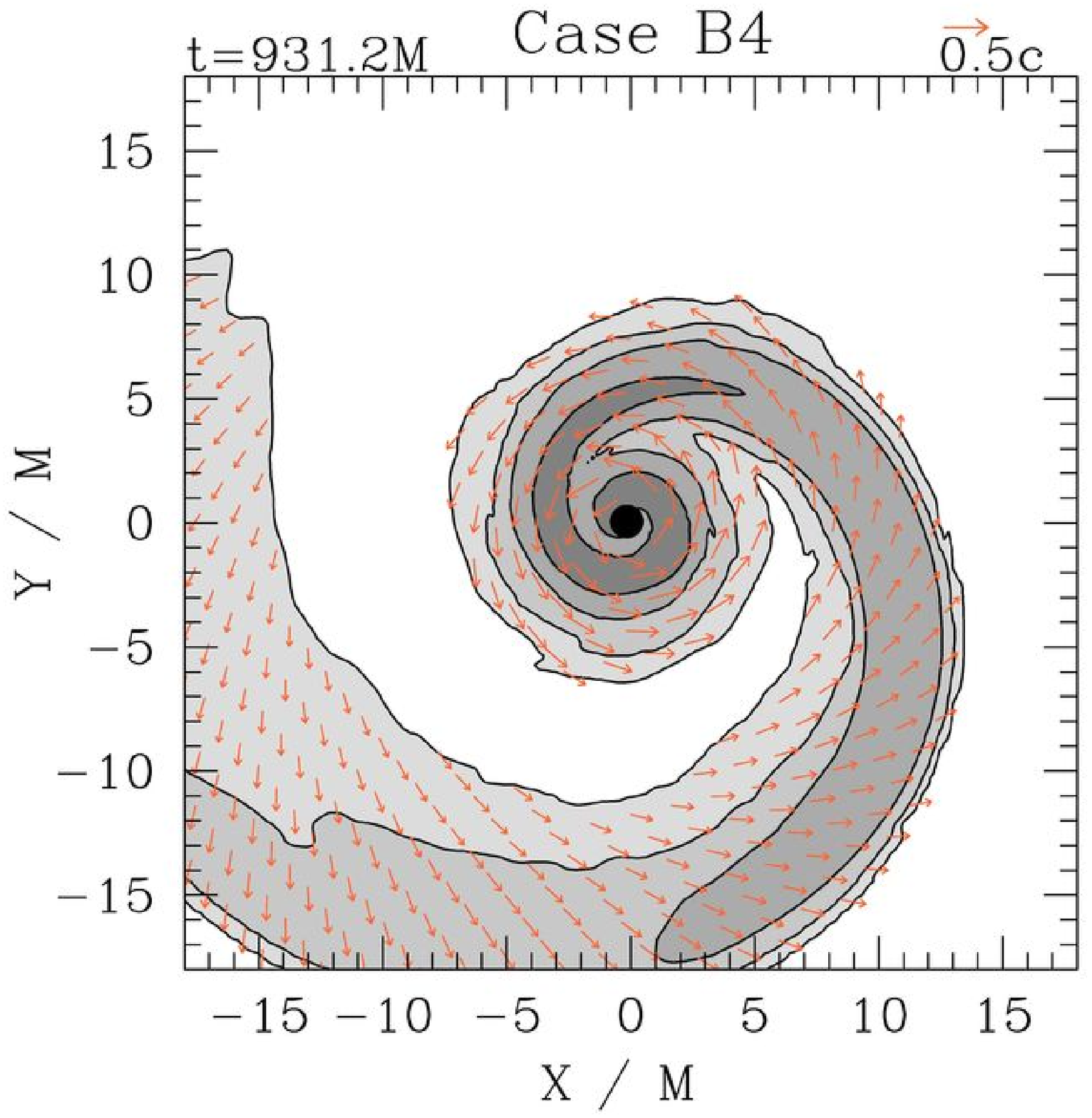}
\epsfxsize=2.3in
\leavevmode
\epsffile[20 -60 575 437]{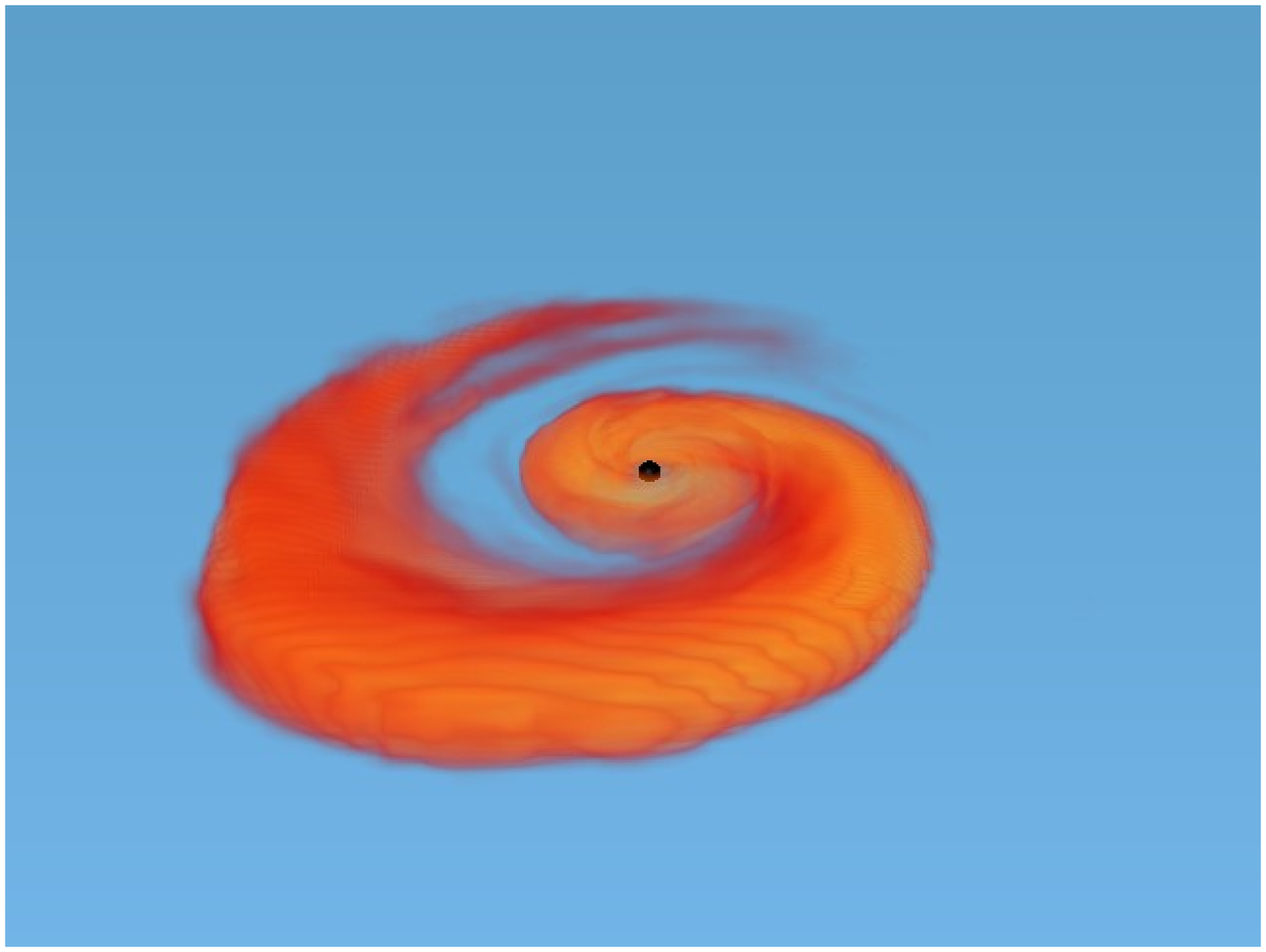}
\epsfxsize=2.3in
\leavevmode
\epsffile[20 -60 575 437]{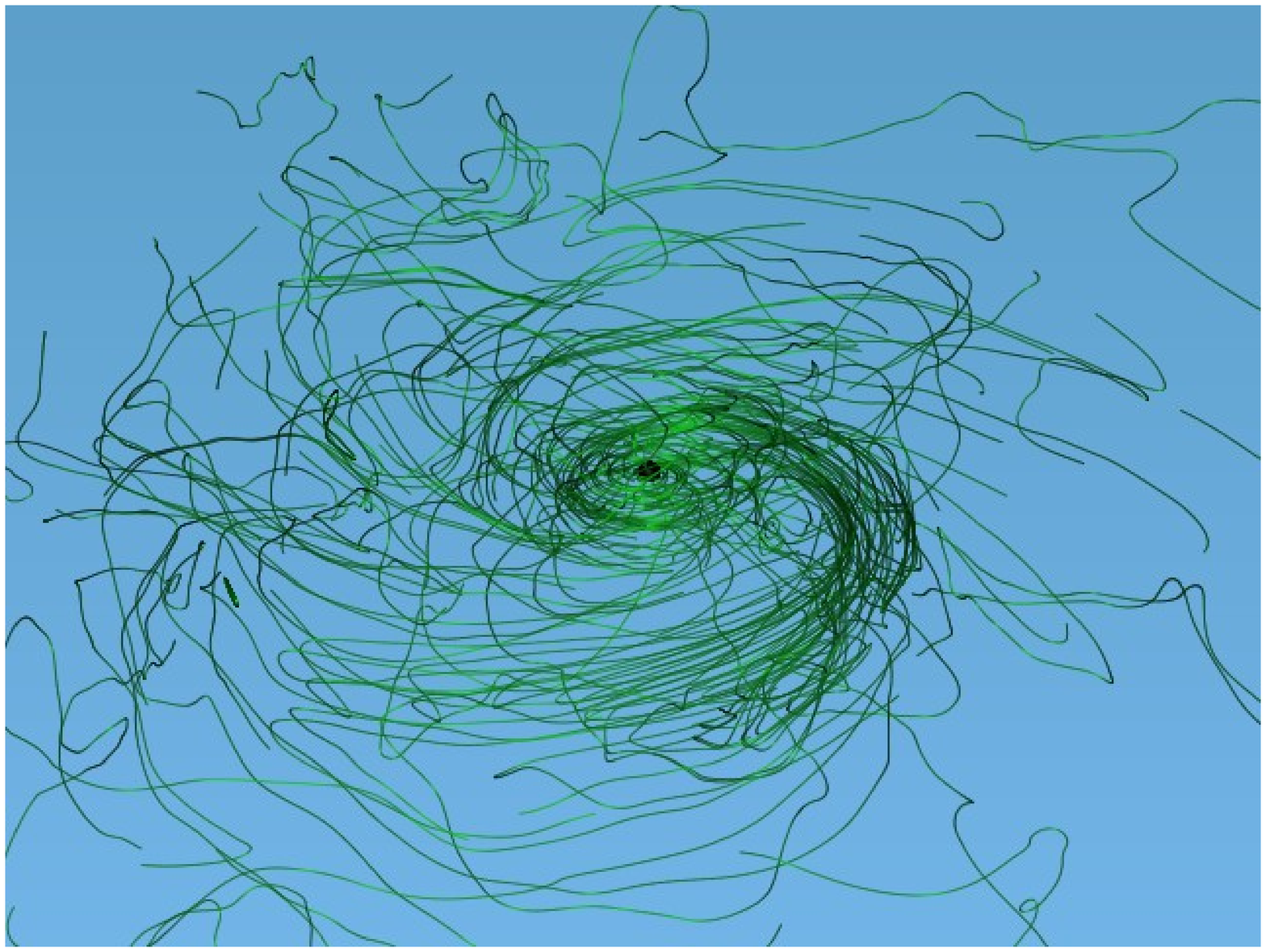}\\
\epsfxsize=2.2in
\leavevmode
\epsffile{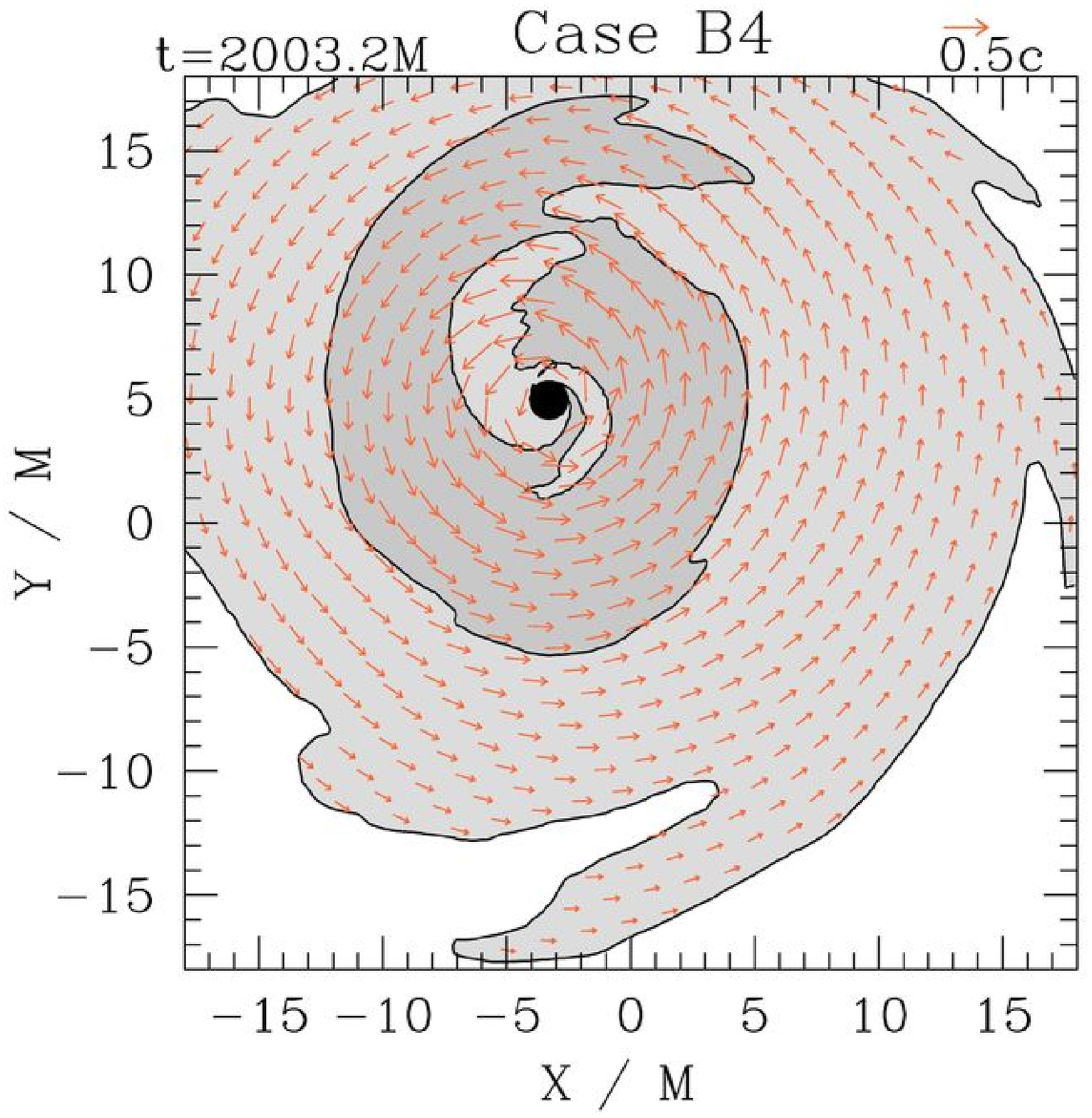}
\epsfxsize=2.3in
\leavevmode
\epsffile[20 -60 575 437]{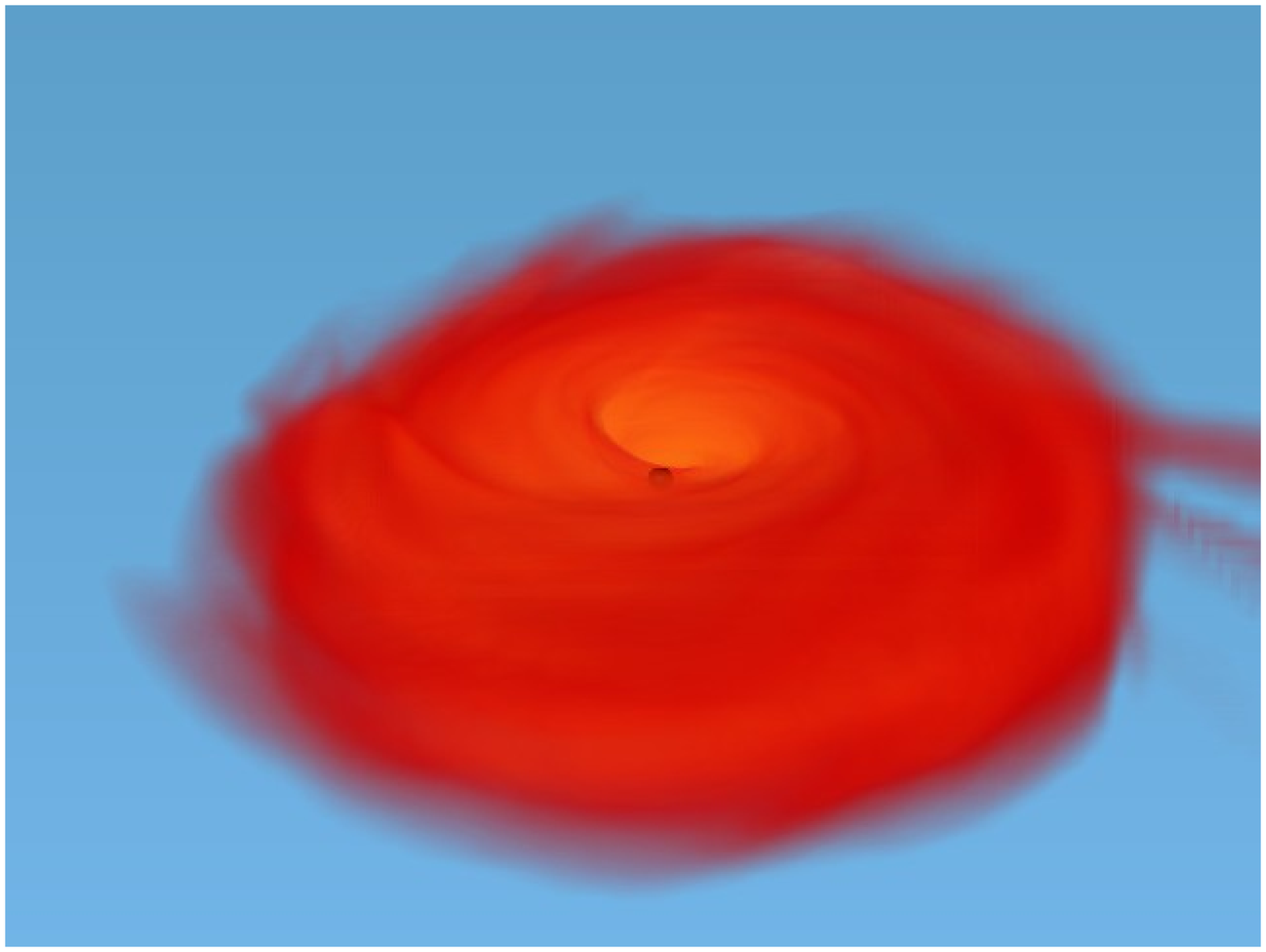}
\epsfxsize=2.3in
\leavevmode
\epsffile[20 -60 575 437]{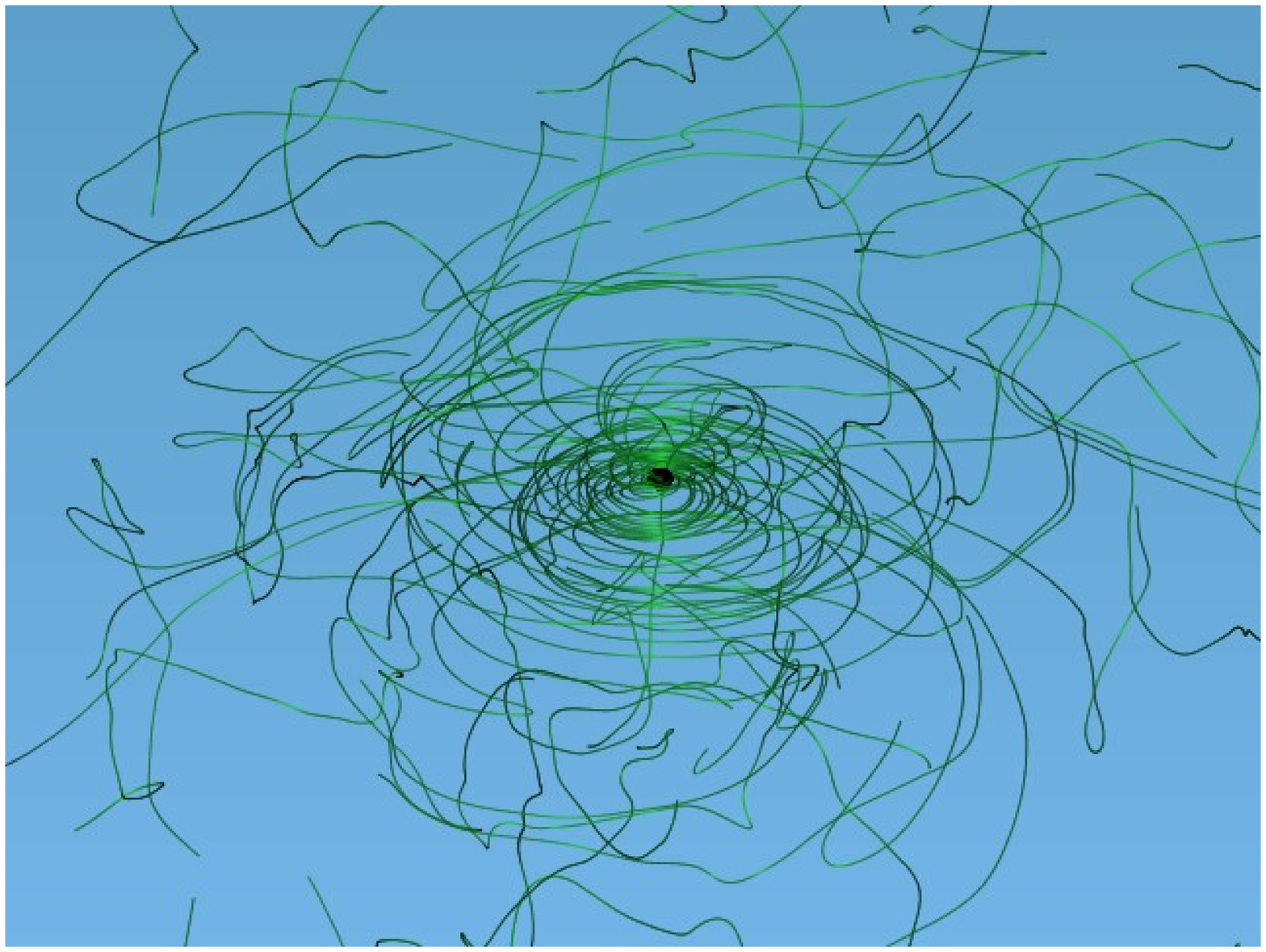}
\caption{Orbital-plane density contours (left column), 3D density
  profiles (middle column) and
  3D magnetic field lines (right column) at four selected times
  for case B4.  The times in the rows (top to bottom) are
  $t/M=$634.8, 744.1, 931.2, and 2003.   Density
  contours in the orbital plane (left column) are plotted according to
  $\rho_0 = \rho_{0,{\rm max}} (10^{-0.92j})$,  ($j$=0, 1, ... 5),
  with darker greyscaling for higher density.  The maximum initial NS
  density is $\kappa \rho_{0,{\rm max}} = 0.126$, or $\rho_{0,{\rm
      max}}=9\times 10^{14}\mbox{g cm}^{-3}(1.4M_\odot/M_0)^2$.
  Arrows in density contour plots represent the velocity field in the
  orbital plane, and the black hole AH interior is marked by a filled
  black circle.
  Magnetic fields are plotted as streamlines of the magnetic field
  vector $B^i$,
  distributed in proportion to $|B^i|$.
  The ADM mass for this case is
  $M=2.5\times 10^{-5}(M_0/1.4M_\odot)$s$=7.6(M_0/1.4M_\odot)$km.
The 3D visualizations were produced
using the ZIBamira software system \cite{Stalling:AmiraVDA-2005}.
}
\label{B:B4_magnetic_geometry}
\end{center}
\end{figure*}

Next we analyze how magnetic fields
evolve over time in case B4, starting with the time at which they were seeded
into the NS (top-right frame of Fig.~\ref{B:B4_magnetic_geometry}). At
the onset of tidal disruption (second row),
the magnetic field structure has changed significantly,
even within the non-disrupted regions of the NS.
Apparently the magnetic fields have undergone some slight
rearrangement since they were added to the NS.  

After the accretion funnel has wrapped around the BH and intersected
itself, it forms a small disk-like structure around the BH
(bottom frames).  In this ``disk'' region, the magnetic fields
wind around the BH.  The B4 simulation is continued for about 30$(M_0/1.4M_\odot)$ms after tidal disruption,
and then it is
stopped.  At this time, the BH+disk system has
drifted significantly (bottom plots of~\ref{B:B4_magnetic_geometry}), 
due to gauge effects and the
gravitational-wave kick at merger.  There is a great deal of winding
of magnetic fields threading the disk at this time (green lines), but
no strong evidence of collimation around the BH poles or magnetic field
turbulence in the disk.  
The lack of magnetic field turbulence may be due to insufficient resolution
in the disk, which artificially suppresses instabilities like MRI.
Insufficient resolution, coupled with the termination of the
simulation after only $30(M_0/1.4M_\odot)$ms may explain why no magnetic field
collimation was observed.
The bottom left and center frames of the figure show that a cavity has not formed around the BH 
by the end of the simulation. This result appears to be consistent with studies which suggest 
that stresses in magnetized disks may suppress the presence of an ISCO
\cite{kh02,bhk08}. However, longer, more accurate disk evolutions will be necessary to 
fully assess the agreement between these studies and simulations in full GR.

\begin{figure*}
\vspace{-4mm}
\begin{center}
\epsfxsize=3.5in
\leavevmode
\epsffile{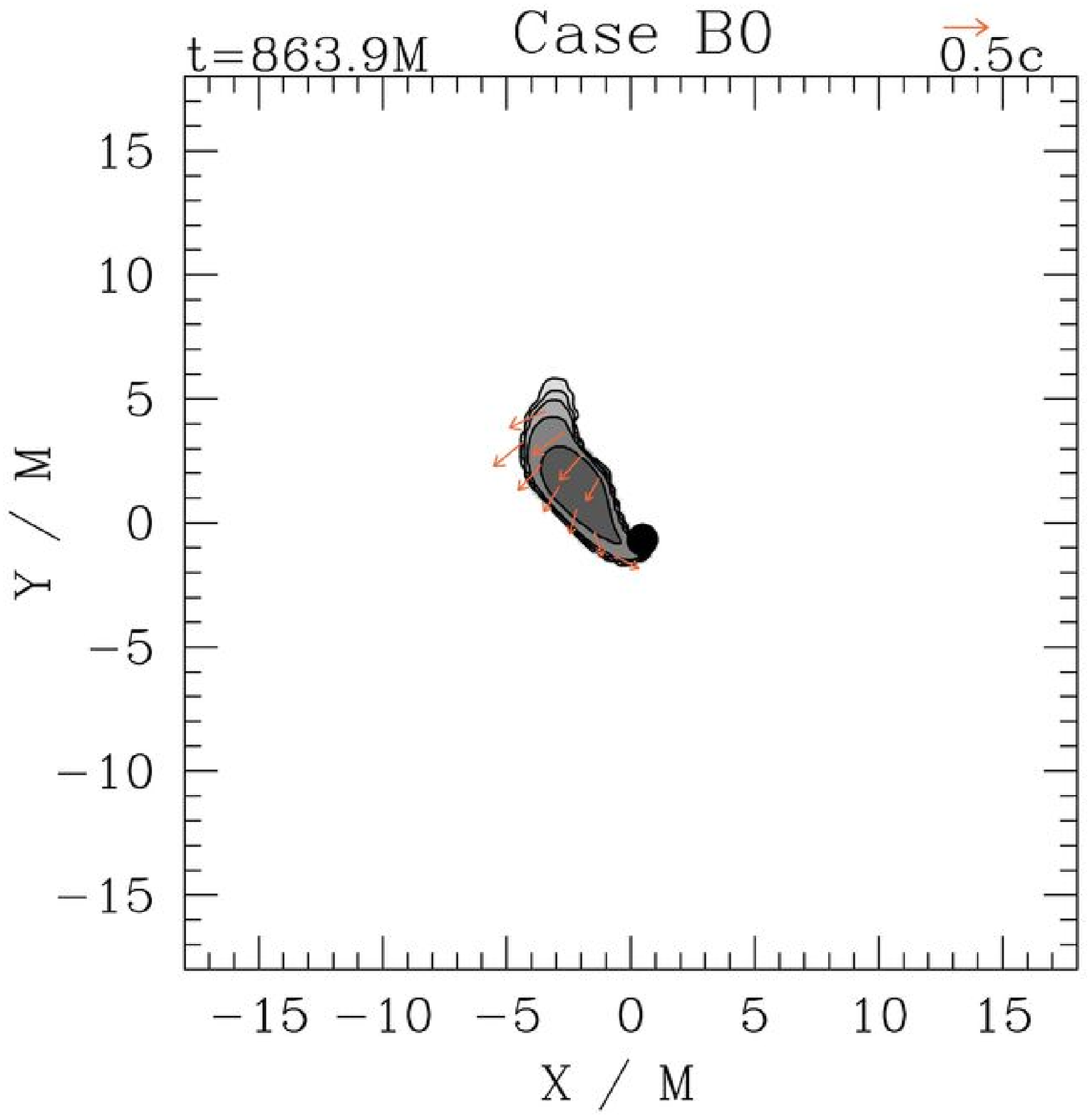}
\epsfxsize=3.5in
\leavevmode
\epsffile{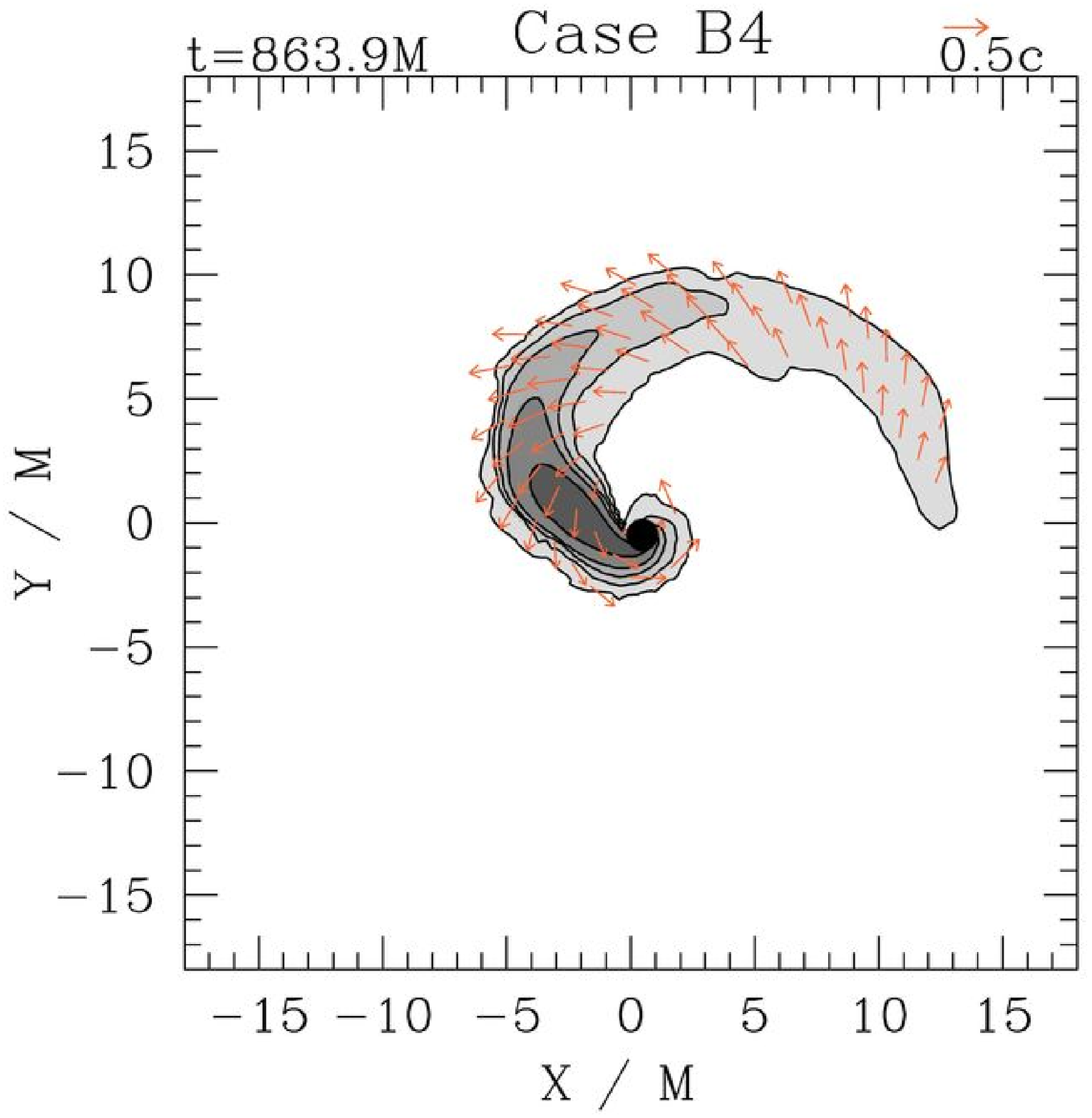}
\caption{Density and velocity profile snapshots during NS tidal disruption
  for cases~B0 (left) and B4 (right). Density contours are
  plotted in the orbital plane according to $\rho_0 = \rho_{0,{\rm
      max}} (10^{-0.92j})$,  ($j$=0, 1, ... 5), with darker
  greyscaling for higher density. 
  The maximum initial NS density is $\kappa \rho_{0,{\rm max}} =
  0.126$, 
  or $\rho_{0,{\rm max}}=9\times 10^{14}\mbox{g
    cm}^{-3}(1.4M_\odot/M_0)^2$.  
  Arrows represent the velocity field
  in the orbital plane, and the black hole AH interior is marked 
  by a filled black circle.  The ADM mass for this case is
  $M=2.5\times 10^{-5}(M_0/1.4M_\odot)$ s$=7.6(M_0/1.4M_\odot)$km.}
\label{B:rho_disruption}
\end{center}
\end{figure*}

\begin{figure*}
\vspace{-4mm}
\begin{center}
\epsfxsize=3.5in
\leavevmode
\epsffile{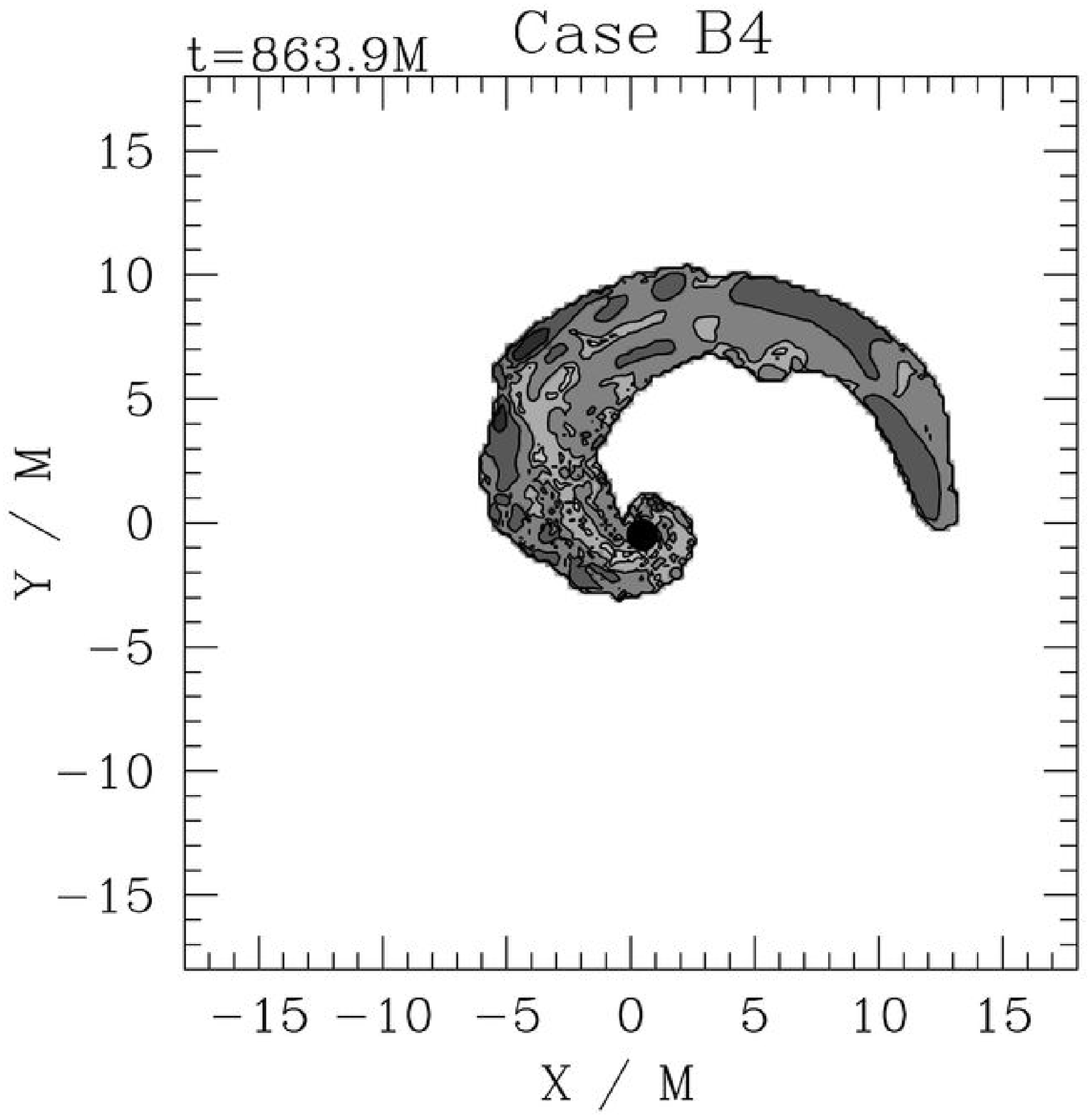}
\epsfxsize=3.5in
\leavevmode
\epsffile{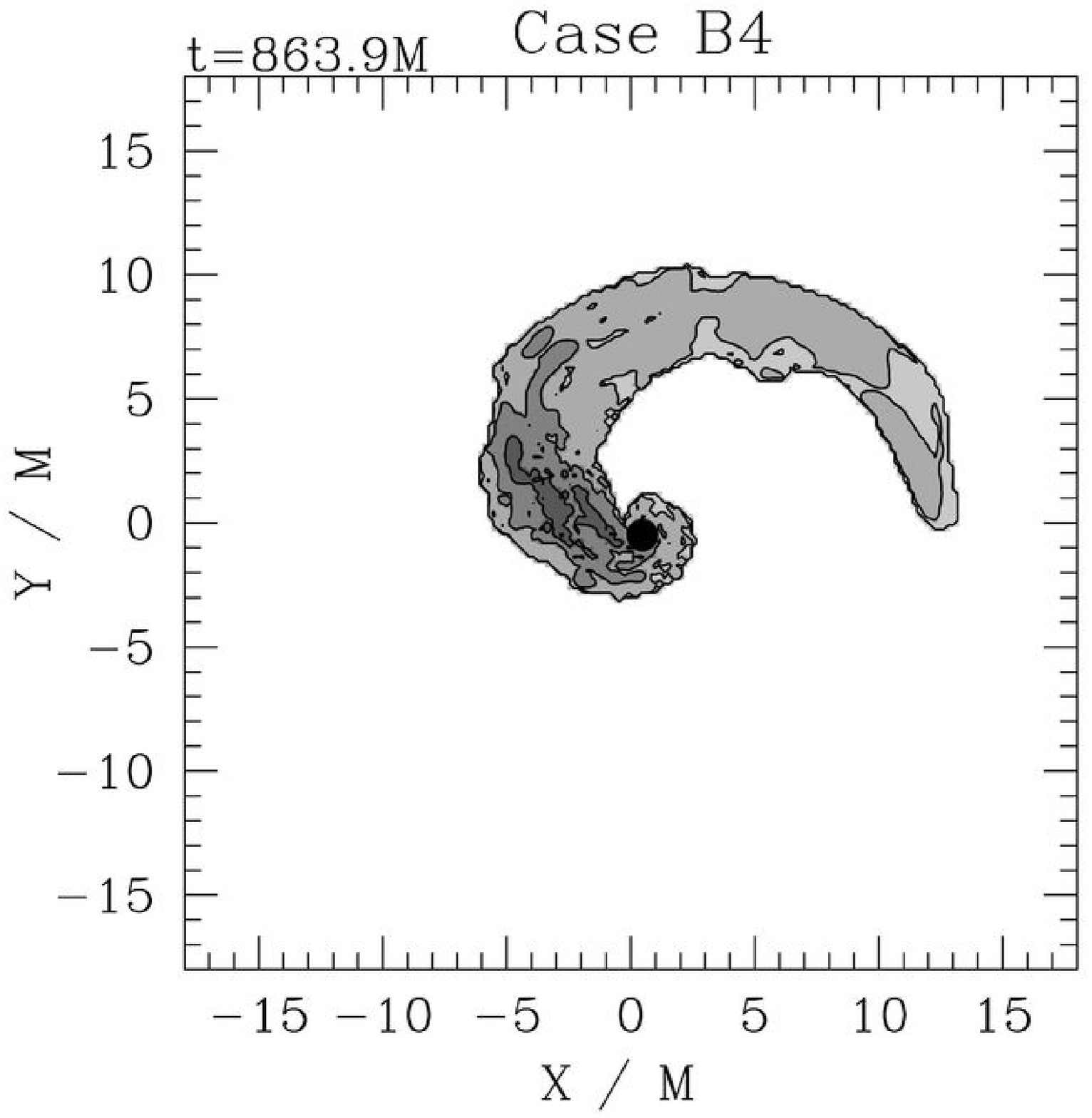}
\caption{Pressure ratio $b^2/(2P)$ (left) and magnetic pressure
   $b^2/2$ (right) contours during NS tidal disruption, at the same
  time as Fig.~\ref{B:rho_disruption}, plotted according to
  $b^2/(2P)=10^{-1.5}(10^{-1.3j})$, ($j$=0, 1, ... 5), and
  $\kappa b^2 =10^{-5}(10^{-2.2j})$, ($j$=0, 1, ... 5). 
Darker greyscaling denotes higher values. Contours are only plotted 
for regions with densities higher than the lowest-density $\rho_0$ contours in
  Fig.~\ref{B:rho_disruption}. In cgs units,
  $\kappa^{-1}=6\times 10^{36}{\rm dyn\, cm}^{-2} (1.4 M_\odot/M_0)^2$.}
\label{B:b2andb2overP_disruption}
\end{center}
\end{figure*}

The strong seed magnetic fields in case B4 amplify the disk mass
significantly, similar to case A4. Figure~\ref{B:rho_disruption}
demonstrates how the NS density contours are affected by such strong magnetic
fields, comparing case B4 (right plot) with the unmagnetized case
(left plot) shortly after tidal disruption.  Although the seed
magnetic fields are the same strength and geometry as in case A4
(cf. Fig.~\ref{A:rho_disruption}), the NS outer layers in case B4 
are pushed out much more than in case A4.
The magnetic pressure in case B4 does not rise above about 3\% of the
gas pressure in the orbital plane, as shown in
Fig.~\ref{B:b2andb2overP_disruption} (left plot).  This is consistent
with case A4.  Although $b^2/(2P)$ varies strongly throughout the NS,
$b^2$ tends to be stronger in the densest regions of the NS (right
plot), again similar to case A4 (cf. Fig.~\ref{A:b2andb2overP_disk}).

\begin{figure*}
\vspace{-4mm}
\begin{center}
\epsfxsize=3.5in
\leavevmode
\epsffile{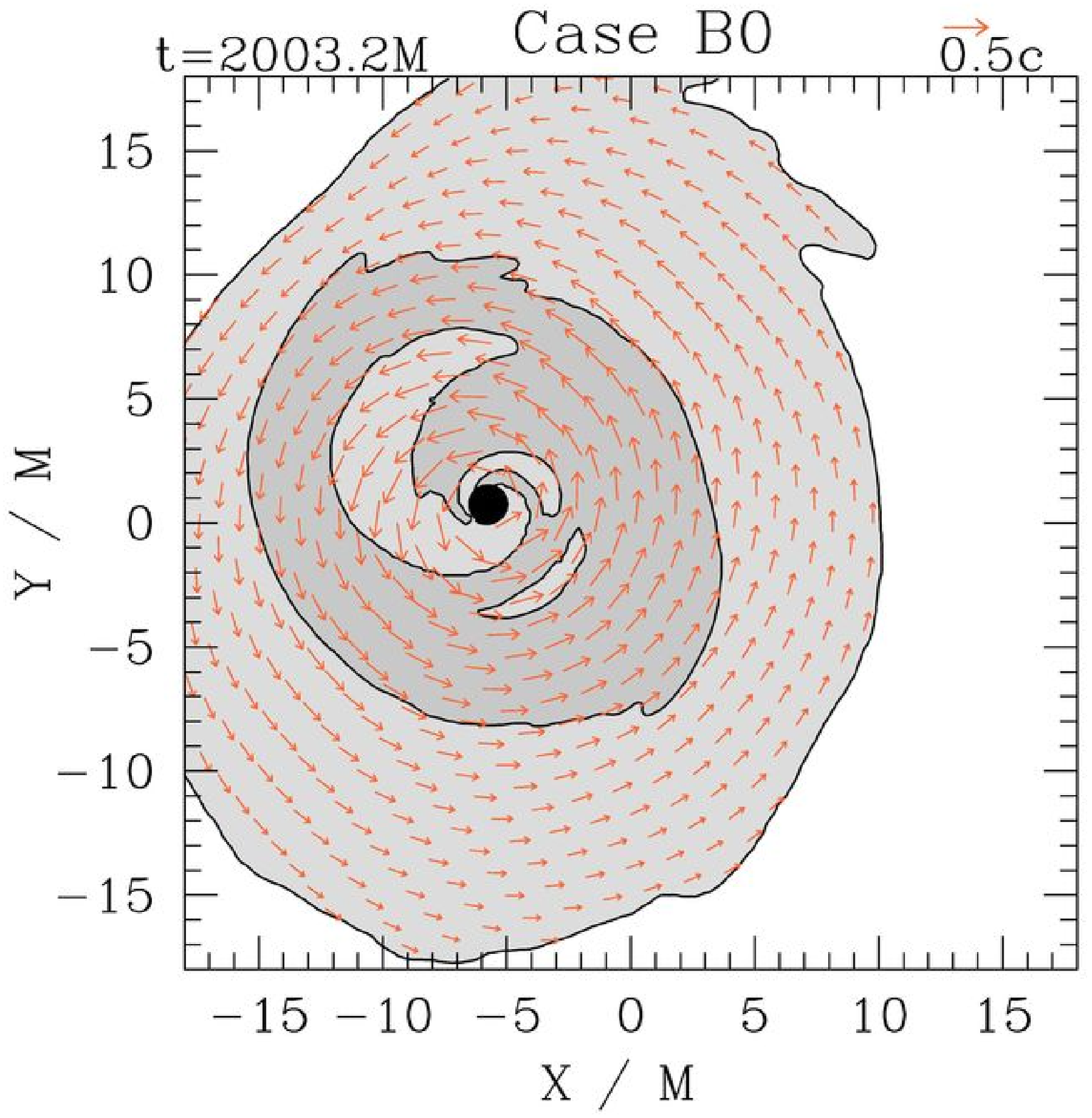}
\epsfxsize=3.5in
\leavevmode
\epsffile{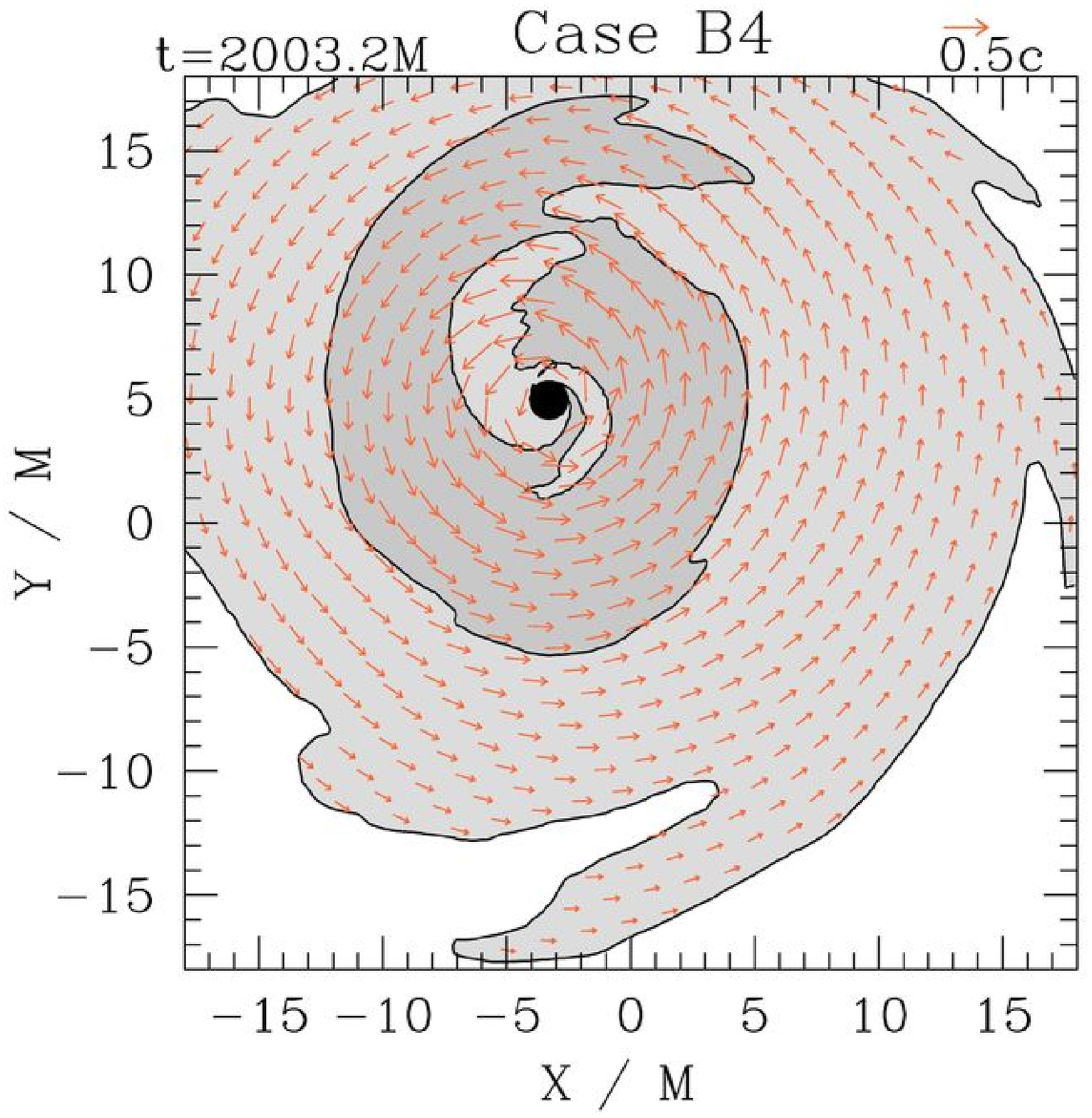}\\
\epsfxsize=3.5in
\leavevmode
\epsffile{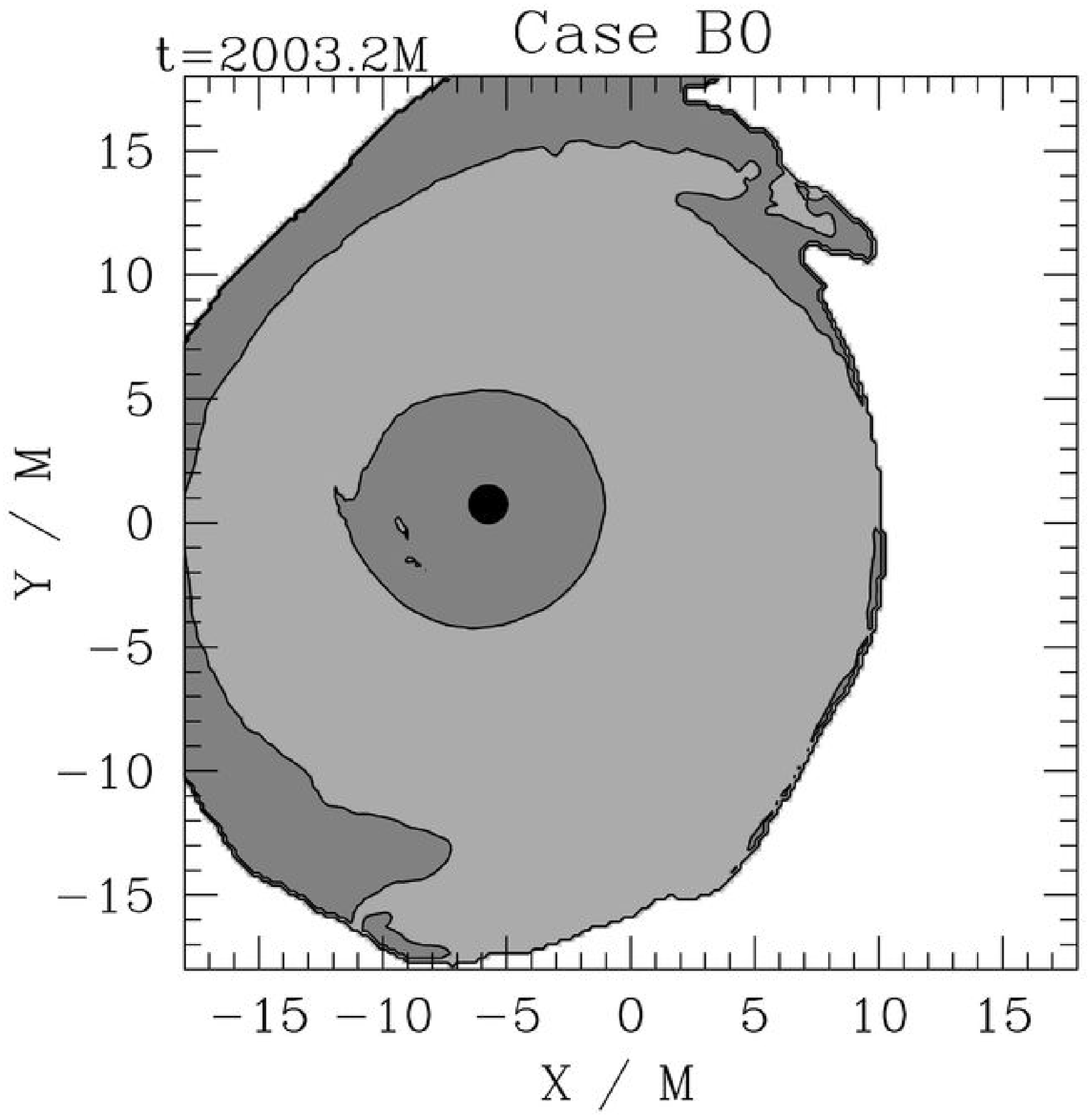}
\epsfxsize=3.5in
\leavevmode
\epsffile{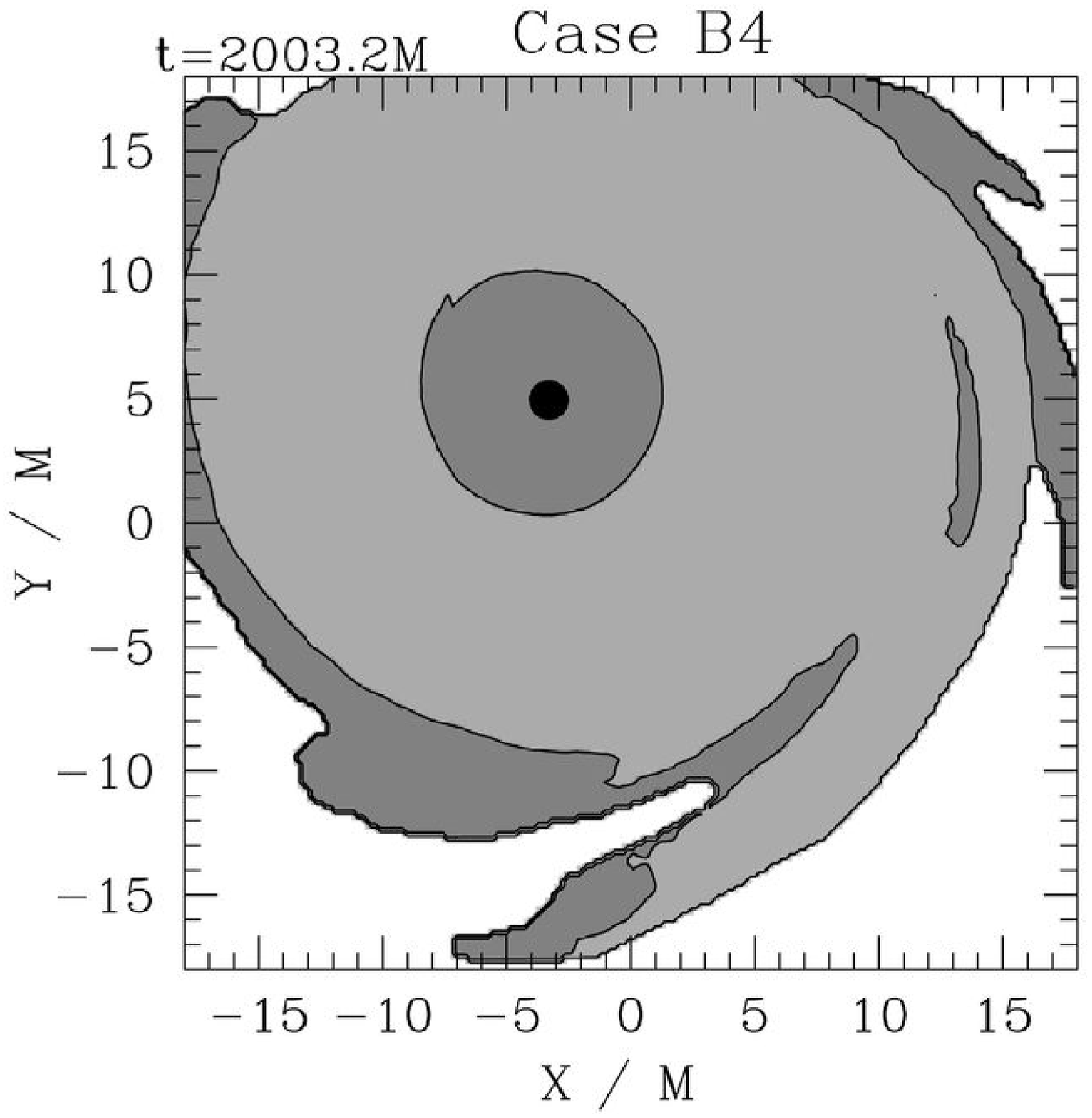}
\caption{Top two plots: Density and velocity profile snapshots
  at the time in which the B0 simulation is stopped, for cases~B0
  (left) and B4 (right). The contours represent the density in the
  orbital plane, plotted according to $\rho_0 = \rho_{0,{\rm
      max}} (10^{-0.92j})$,  ($j$=0, 1, ... 5), with darker greyscaling for higher density.  
  The maximum initial NS density is $\kappa \rho_{0,{\rm max}} =
  0.126$, 
  or $\rho_{0,{\rm max}}=9\times 10^{14}\mbox{g
    cm}^{-3}(1.4M_\odot/M_0)^2$.  
  Arrows represent the velocity field
  in the orbital plane.  The black hole AH interior is marked 
  by a filled black circle.  The ADM mass for this case is
  $M=2.5\times 10^{-5}(M_0/1.4M_\odot)$s$=7.6(M_0/1.4M_\odot)$km. \\
  Bottom two plots: Snapshots of entropy parameter $K$ contours for cases B0 (left) and
  B4 (right).  The light grey regions correspond to $1.4 < \log_{10} K
  < 2.6$, and the dark grey region corresponds to $2.6 < \log_{10} K
  < 3.8$.}
\label{B:rho_K_finaldisk}
\end{center}
\end{figure*}

\begin{figure*}
\vspace{-4mm}
\begin{center}
\epsfxsize=3.5in
\leavevmode
\epsffile{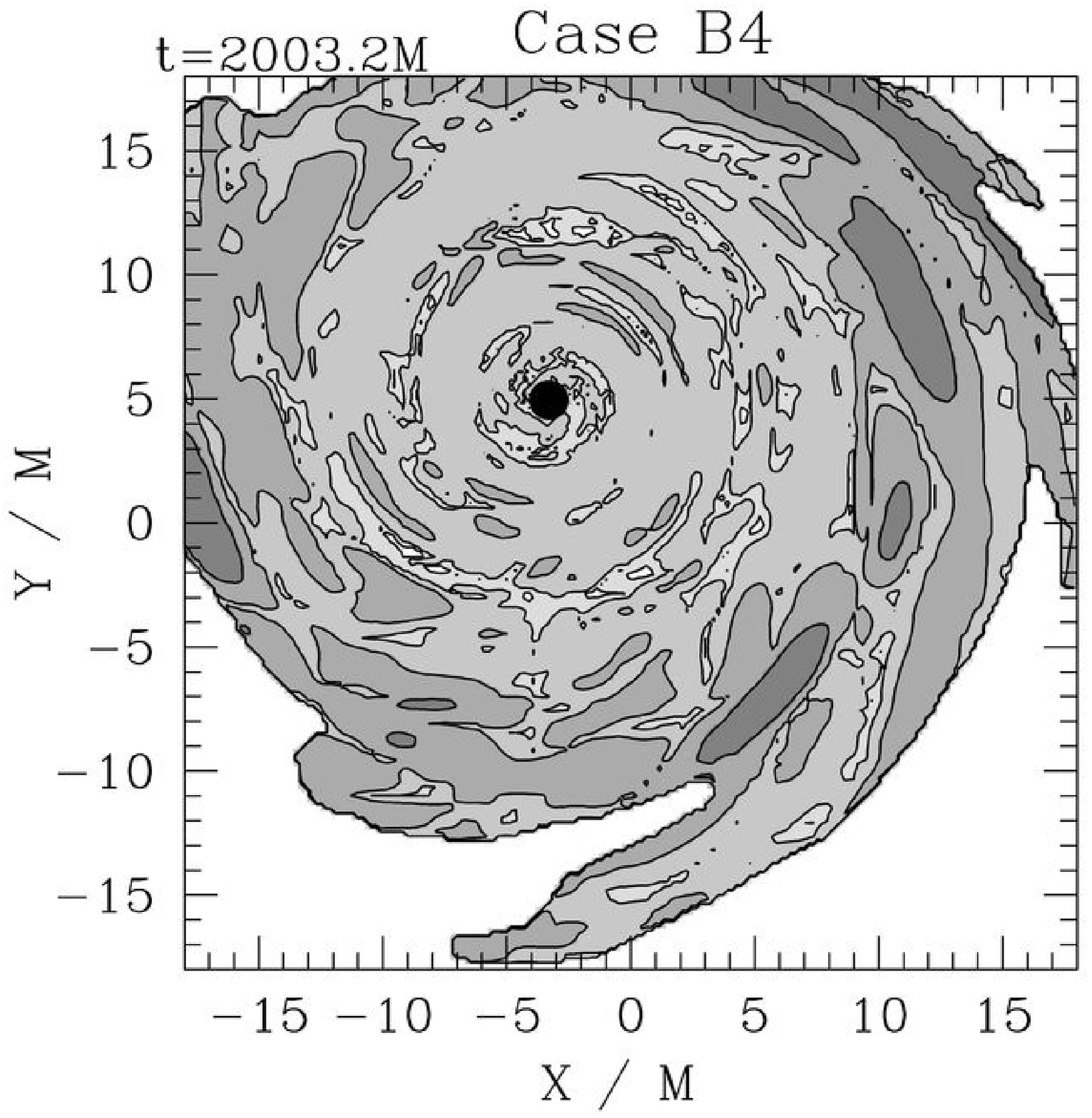}
\epsfxsize=3.5in
\leavevmode
\epsffile{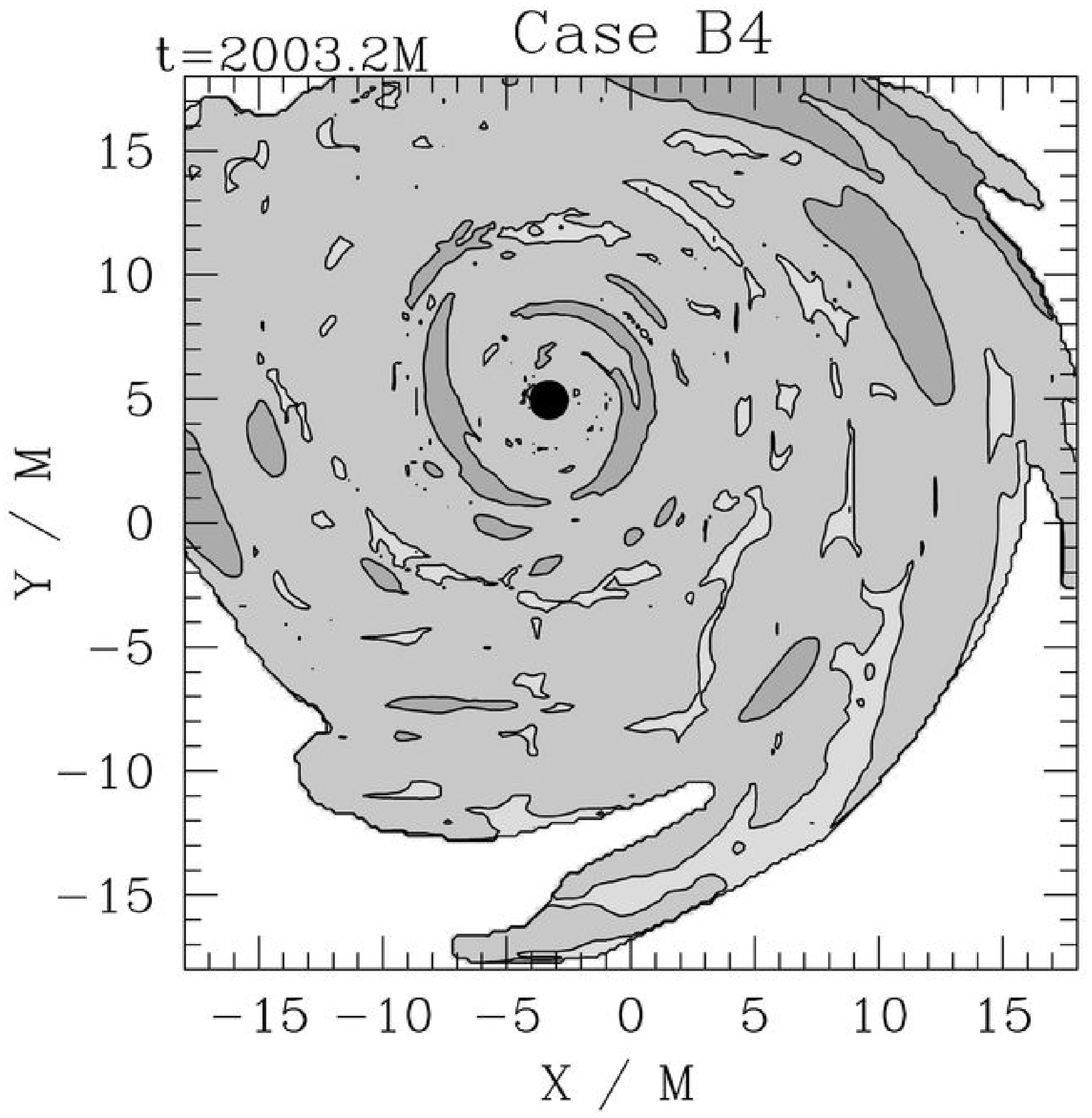}
\caption{Pressure ratio $b^2/(2P)$ (left) and magnetic pressure $b^2/2$ (right) contours, at the same
  time as Fig.~\ref{B:rho_K_finaldisk}, plotted according to
  $b^2/(2P)=10^{-1.5}(10^{-1.3j})$, ($j$=0, 1, ... 5), and
  $\kappa b^2 =10^{-5}(10^{-2.2j})$, ($j$=0, 1, ... 5). 
    $b^2/(2P)$
  and $b^2$ contours are only plotted for regions with densities
  higher than the lowest-density $\rho_0$ contours in
  Fig.~\ref{B:rho_K_finaldisk}. In cgs units,
$\kappa^{-1}=6\times 10^{36}{\rm dyn\ cm}^{-2} (1.4 M_\odot/M_0)^2$.}
\label{B:b2andb2overP_finaldisk}
\end{center}
\end{figure*}

\begin{figure}
\epsfxsize=3.4in
\leavevmode
\epsffile{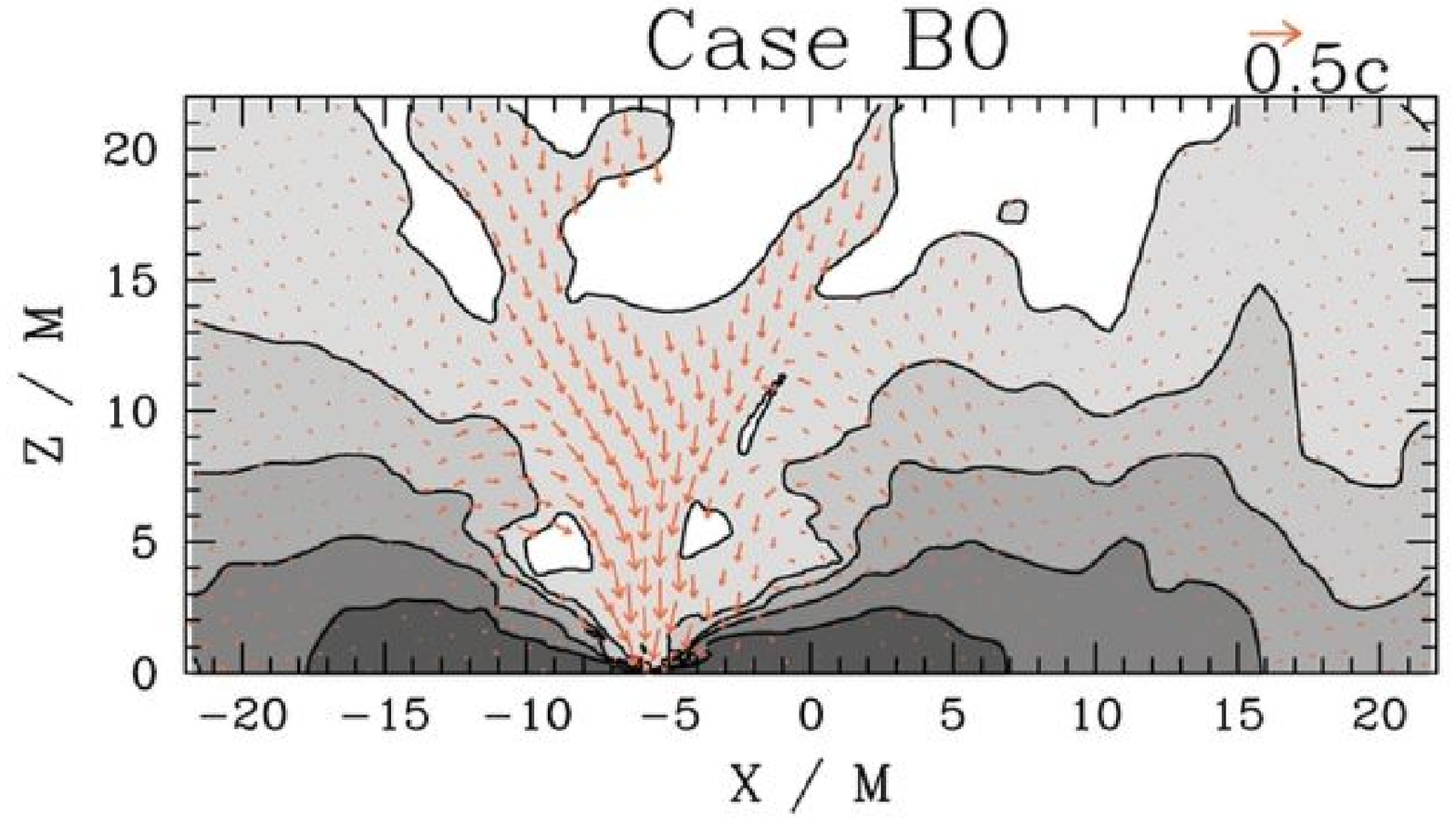}\\
\epsfxsize=3.4in
\leavevmode
\epsffile{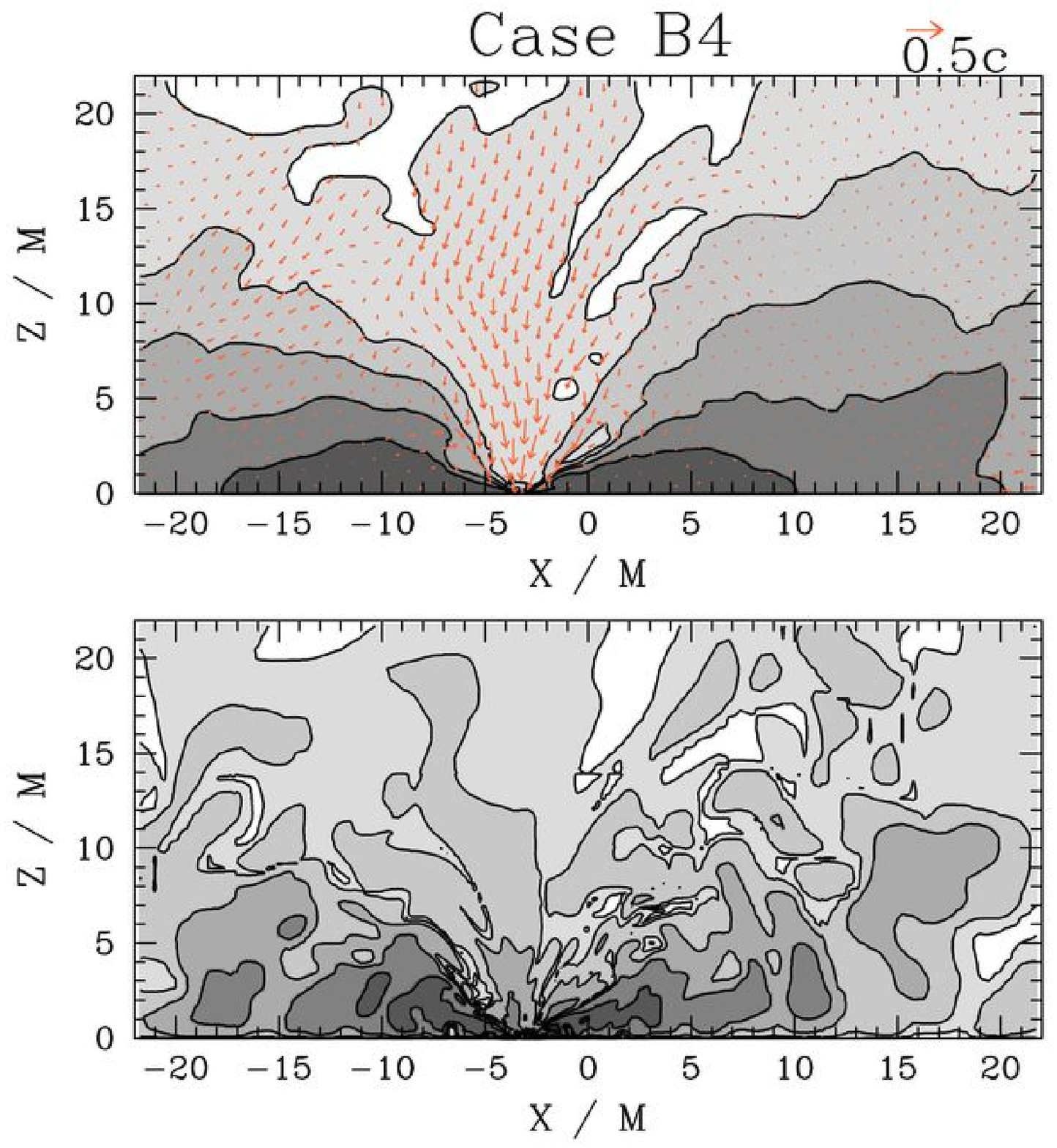}
\caption{Rest-mass density profile for case B0 (top), 
B4 (middle) and magnetic energy density $b^2$ (bottom)
profiles for B4 in the meridional plane at the end of simulation 
($t=2003.2M$). Density contours are plotted according to 
$\rho_0 = \rho_{0,{\rm max}} 10^{-7.6+0.717j}$, ($j=0,...,6$), 
with darker greyscaling for higher density. $b^2$ contours are plotted 
according to $\kappa b^2 =10^{-12+0.833j}$, ($j=0,...,6$).
  The maximum initial NS density is $\kappa \rho_{0,{\rm max}} =
  0.126$, 
  or $\rho_{0,{\rm max}}=9\times 10^{14}\mbox{g
    cm}^{-3}(1.4M_\odot/M_0)^2$.  
  Arrows represent the velocity field
  in the meridional plane.  
In cgs units, the 
  ADM mass for this case is $M=2.5\times 10^{-5}(M_0/1.4M_\odot)$
  s$=7.6(M_0/1.4M_\odot)$km, and 
$\kappa^{-1}=6\times 10^{36}{\rm erg\ cm}^{-3} (1.4 M_\odot/M_0)^2 
= 7\times 10^{15}\mbox{g cm}^{-3}(1.4M_\odot/M_0)^2$.}
\label{B:meridional}
\end{figure}

Figure~\ref{B:rho_K_finaldisk} contrasts the distribution of rest-mass
density and entropy ($\log K$) in the remnant disks of cases B0 and B4, at
the time in which the B4 simulation was stopped.  In the nonspinning
cases A0 and A4 (Fig.~\ref{A:rho_K_disk}), the sizes of the
remnant disks in the orbital plane are remarkably similar, though
the disk mass in case A4 is about twice A0's disk mass.  The extra disk
mass in case A4 may be explained by the existence of a high-density
ring of matter in the disk.  
Unlike the nonspinning cases however, the
high-density regions of the remnant disks in cases B0 and B4 are 
remarkably similar (upper plots).  Despite this, B4's final disk
possesses about 40\% more mass than B0. Apparently the excess
mass in case B4's disk comes from its larger volume.  The size
difference between B0 and B4's final disks is seen more clearly in the
entropy contour plots (bottom two plots), which demonstrate the disks
are hotter close to the BH and in the lowest-density outer regions.

Magnetic field amplitude contour plots for case B4's disk at the same
time as Fig.~\ref{B:rho_K_finaldisk} are displayed in
Fig.~\ref{B:b2andb2overP_finaldisk}.  As in case A4, the final disk
mass is greatly increased by strong magnetic fields, even though
magnetic pressure never exceeds about 3\%
of the gas pressure, with typical values around 0.1\% (left plot of Fig.~\ref{B:b2andb2overP_finaldisk}).
Bubbles of enhanced $b^2$, corresponding to order-of-magnitude
increases in $|b|$, appear in a small ring around the BH and in the
disk's outer regions (right plot of Fig.~\ref{B:b2andb2overP_finaldisk}). 

Finally, Fig.~\ref{B:meridional} compares B0 and B4 profiles 
in the meridional plane. As in the nonspinning cases, the 
geometry of the disk appears to be mostly unchanged by the addition of
magnetic fields. There is 
small amount of material flowing into the BH from the poles, and 
the magnetic fields are mainly confined inside the disk.

\subsection{Gravitational Waves Study}

\begin{figure}
\epsfxsize=3.4in
\leavevmode
\epsffile{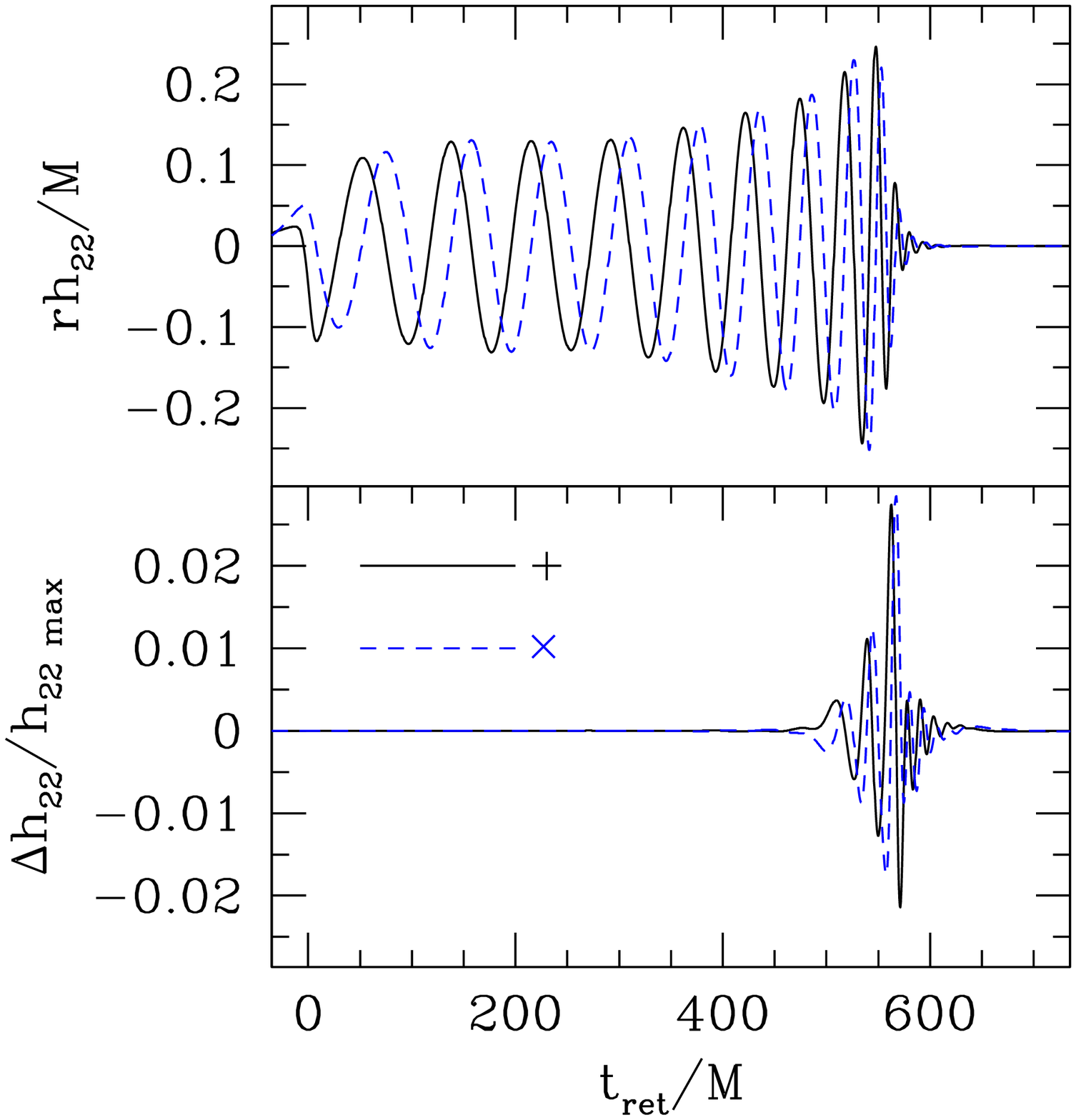}
\caption{Upper panel: (2,2) mode of the gravitational wave strain
  $h^+_{22}$ (solid line) and $h^\times_{22}$ (dashed line) versus
  retarded time $t_{\rm ret}$ for case A0.  \\
  Lower panel: difference in $h^+_{22}$ (solid line) and
  $h^\times_{22}$ (dashed line) between cases A0 and A4 versus $t_{\rm
    ret}$, normalized by the maximum value of $|h^+_{22}-ih^\times_{22}|$
  for case A0 over time, $h_{\rm 22\, max}$. 
}
\label{fig:rhphc_A0-A4}
\end{figure}

\begin{figure}
\epsfxsize=3.4in
\leavevmode
\epsffile{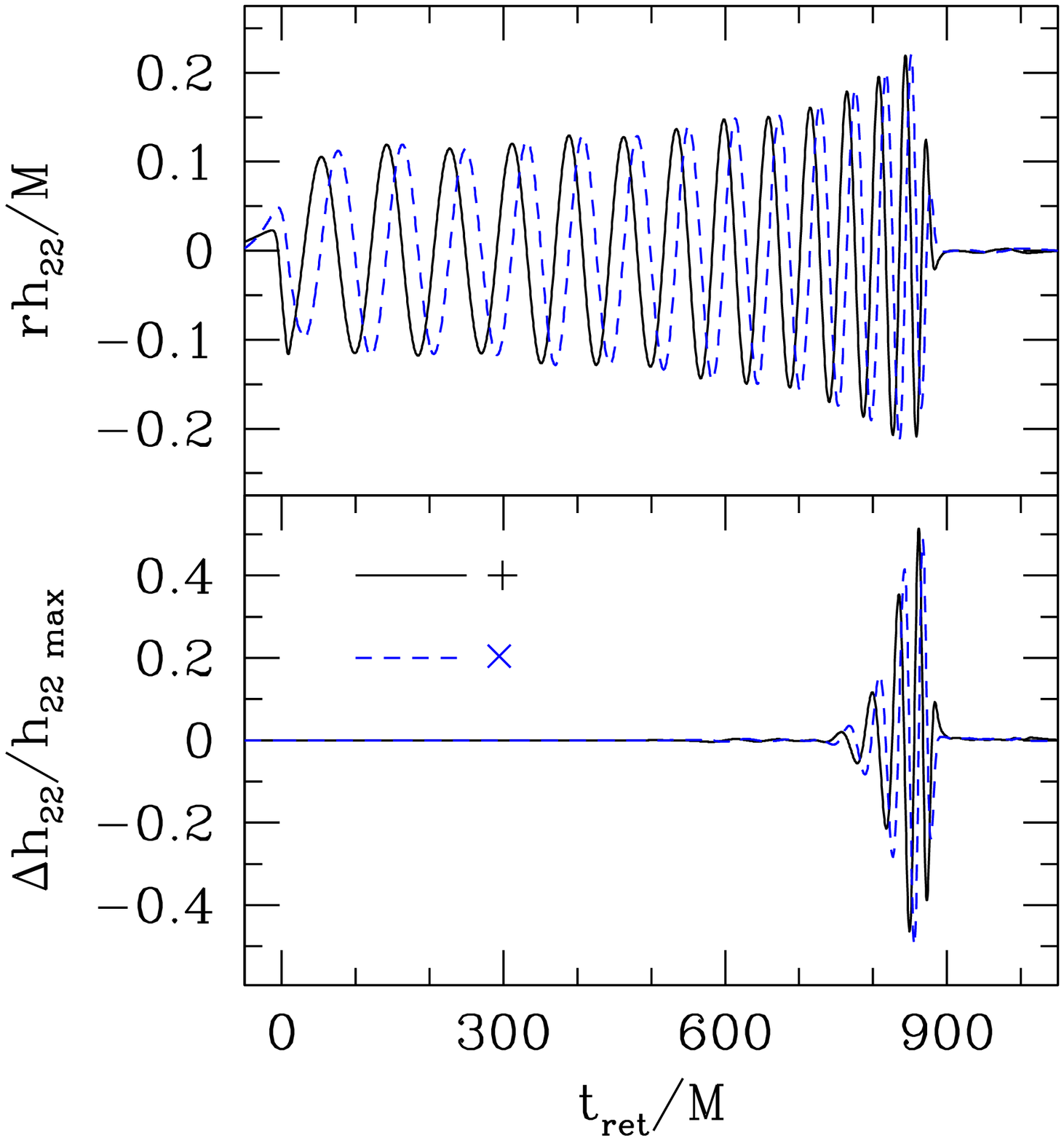}
\caption{Same as Fig.~\ref{fig:rhphc_A0-A4} but for cases B0 and
  B4. 
}
\label{fig:rhphc_B0-B4}
\end{figure}
\begin{figure}
\epsfxsize=3.4in
\leavevmode
\epsffile{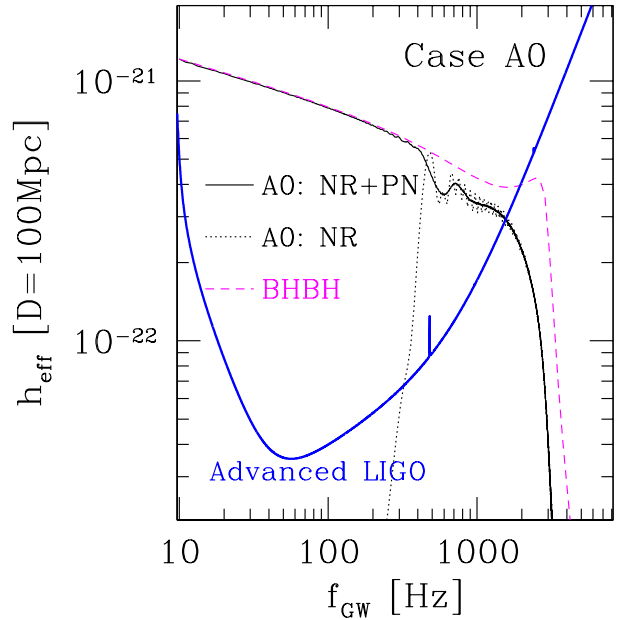}
\caption{Gravitational-wave power spectrum for case A0
computed as in Sec.~IIIF of~\cite{eflstb08}.
The solid curve shows the power spectrum of the numerical signal
(dotted curve) stitched to the post-Newtonian waveform, including only
the dominant $(2,2)$ and $(2,-2)$ modes.  The dashed curve shows the
power spectrum expected from BHBH binaries with the same total mass and mass ratio
as the BHNS, derived in~\cite{Ajith2} from analysis of multi-orbit,
non-precessing BHBH binaries.
The heavy solid curve displays the effective strain of the Advanced
LIGO detector, defined such that $h_{\rm LIGO}(f)\equiv
\sqrt{fS_h(f)}$. The noise curve used here corresponds to the
Advanced LIGO configuration ZERO\_DET\_HIGH\_P~\cite{AdvLIGnoisecurve}. To set
physical units, a NS rest mass of $M_0=1.4M_\odot$ and a
source distance of $D$=100Mpc is assumed.}
\label{fig:heff_A0}
\end{figure}

\begin{figure}
\epsfxsize=3.4in
\leavevmode
\epsffile{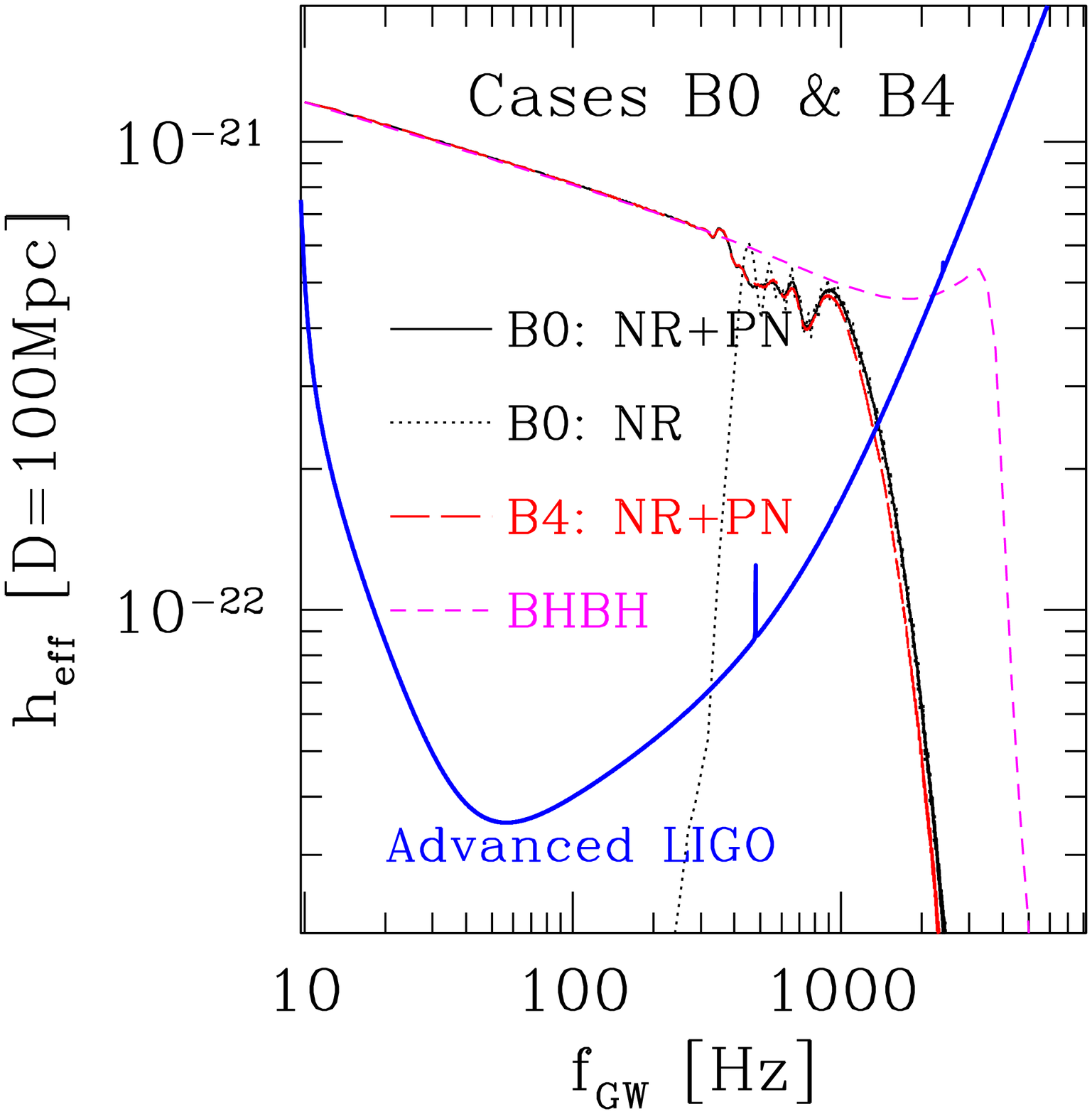}
\caption{Same as Fig.~\ref{fig:heff_A0} but for Cases B0 and B4.}
\label{fig:heff_B}
\end{figure}

Gravitational waves are extracted at several radii between
$50M$ and $130M$. These radii are sufficiently far from the binary 
for the waves to overlap very well when plotted against retarded time, 
after accounting for the amplitude fall-off with radius.  The upper
panel of Fig.~\ref{fig:rhphc_A0-A4} shows the dominant (2,2) mode of
the gravitational waveform as a function of retarded 
time $t_{\rm ret}$ for case A0, and the difference between
the two data sets is displayed in the lower panel. The
difference in amplitude between A0 and A4 waveforms is 
less than 3\%, indicating that magnetic fields 
explored here do not significantly impact
the global dynamics during BHNS inspiral and merger.
By contrast, the difference between the B0 and B4 waveforms
during merger is substantial (Fig.~\ref{fig:rhphc_B0-B4}).  The more
highly spinning BHs of cases B0 and B4 possess smaller ISCOs, enabling
NS to orbit the BH more closely before accreting, resulting
in more gravitational wave cycles.  Further, at smaller separations
from the BH, the effects of frame dragging are much more pronounced,
twisting NS matter around the BH and reducing the BH accretion rate,
ultimately giving the magnetic fields more time to amplify and affect
the global dynamics of NS matter.

However, the {\it observability} of magnetic effects on gravitational
waveforms depends on the response of the GW detectors to the time series data.
One way to assess the detectability is to compute the mismatch
\beq
  MM \equiv 1 - \frac{ (h_{B0} | h_{B4})}
{\sqrt{ (h_{B0} | h_{B0}) (h_{B4} | h_{B4})}} \ ,
\eeq
between the waveforms of B0 and B4, assuming the NS mass is 1.4 $M_\odot$,
where
\beq
  (h_1 | h_2 )
= 4\, {\rm Re} \int_0^{\infty} \frac{\tilde{h}_1(f) \tilde{h}^*_2(f)}{S_h(f)} df .
\eeq
Here $h=h_+ - i h_\times$, $\tilde{h}$ is the Fourier transform of $h(t)$,
and $S_h(f)$ is the power spectral density of the Advanced LIGO
noise. 
Using the Advanced LIGO broadband configuration HIGH\_DET\_HIGH\_P,
the minimum mismatch (by varying the time and phase shifts between the 
two waveforms) between B0 and B4 waveforms is only 0.004,
indicating that it may be challenging for Advanced LIGO broadband to
detect the strong internal magnetic fields of case B4.  

Complete GW spectra in the frequency domain require the creation of
hybrid waveforms, which stitch together numerical and post-Newtonian
(PN) waveforms.  We generate hybrids by first computing the minimum of
\begin{equation}
\int_{t_{\rm i}}^{t_{\rm f}} dt \sqrt{\sum_{i \in \{+,\times\}} (h_i^{\rm NR} -
  h_i^{\rm PN})^2}
\end{equation}
via the Nelder-Mead algorithm, using as free parameters initial PN phase,
amplitude, and orbital angular frequency.  Here, $h_{+,\times}^{\rm
  NR}$ and $h_{+,\times}^{\rm PN}$ specify our numerical BHNS waveforms
and the TaylorT1 PN waveforms of~\cite{nijidataformat}, respectively.
The integration bounds
were chosen to be $t_{\rm i}\approx 200M$ and $t_{\rm f}\approx 400M$.
The hybrid waveform consists of a linear combination of the PN and NR
waveforms, as in~\cite{Ajith3}.

Effective GW strains of the hybrid waveforms in frequency space are
plotted in Fig.~\ref{fig:heff_A0} for A0, and Fig.~\ref{fig:heff_B}
for cases B0 and B4.
Assuming the NS has a rest mass of $1.4M_\odot$ and binary distance of
100Mpc, we plot these strains against the Advanced LIGO sensitivity curve
$h_{\rm LIGO}(f)\equiv \sqrt{fS_h(f)}$.  Within this distance, the
BHNS event rate is estimated to be
$5\times 10^{-3}$--$10$ per year, assuming an overall rate of
0.05--100 mergers per Myr per Milky Way-equivalent galaxy (and a density of
$0.1~{\rm gal/Mpc}^3$)~\cite{aetal10}. Cases A1--A4 and
B1--B3 are not shown in the figures because their effective GW strains
are almost indistiguishable from cases A0 and B0, respectively.  Advanced LIGO
in the chosen configuration may be able to marginally distinguish
between BHNS waveforms and those produced by BHBH mergers at high frequencies
(500--1000Hz). However,
the effects of even the strongest magnetic fields chosen here (cases A4 and B4) on
the waveforms are quite small and may be challenging for Advanced LIGO
to detect in a broadband configuration. This is
consistent with the result from the minimum mismatch calculation
between B0 and B4, as mentioned above. Nevertheless, 
recent innovations exploring `squeezed light' effects may reduce 
quantum noise and increase the sensitivity in this very domain~\cite{squeezedlight}.

It has been suggested that the differences between BHNS and BHBH
waveforms during late inspiral and merger may be used to extract the
tidal deformability of the NS~\cite{lksbf11}, which can in turn be
used to constrain the NS EOS.  Our results suggest that seeding the NS
with magnetic fields of the configurations and strengths explored 
here do not alter gravitational waveforms significantly. Further 
studies at higher resolution may be required to confirm this finding.

\subsection{Energy and angular momentum conservation}

\begin{figure}
\epsfxsize=3.4in
\leavevmode
\epsffile{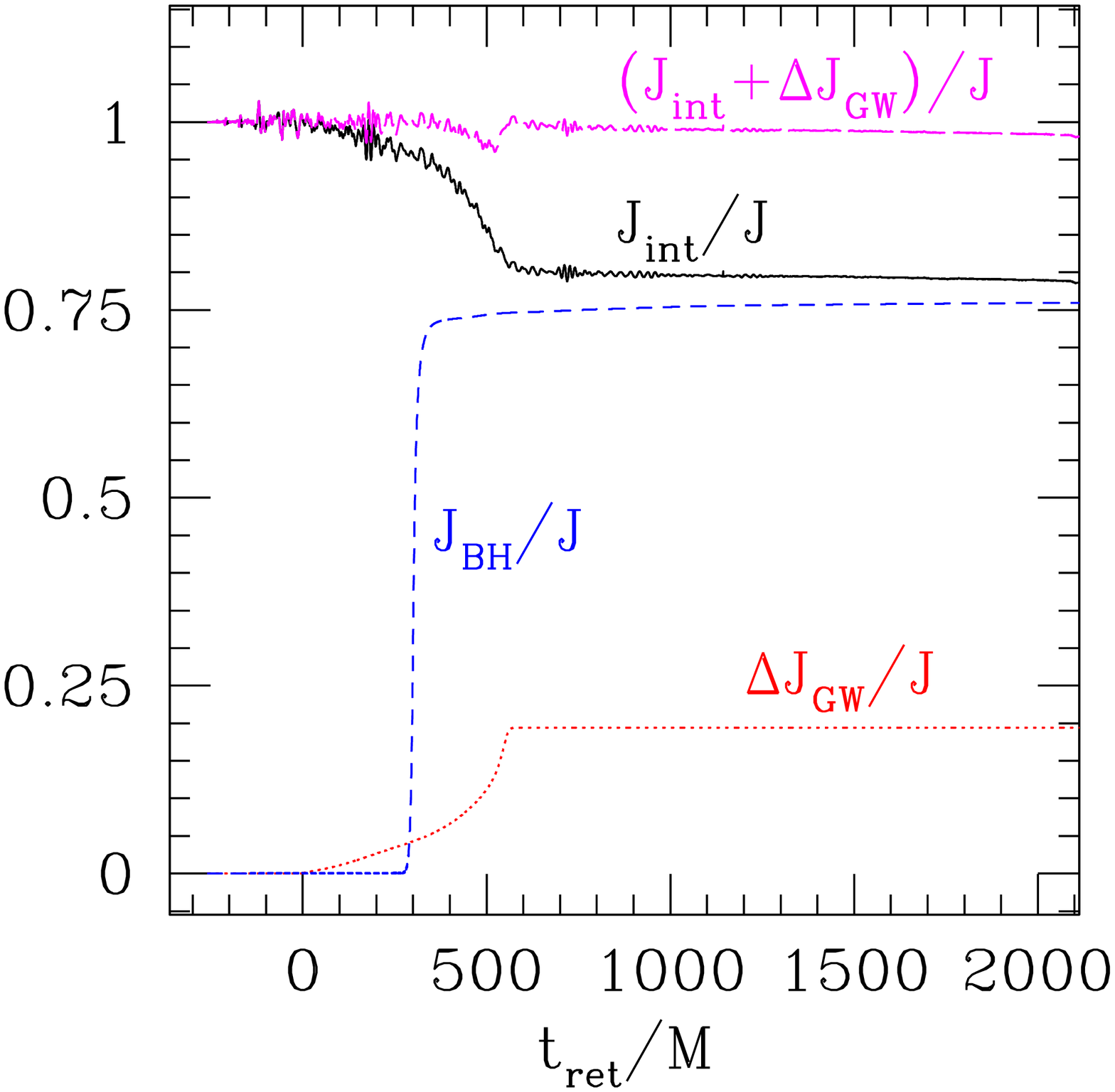}
\caption{Evolution of interior angular momentum $J_{\rm int}$,
 angular momentum carried off by GW $\Delta J_{\rm GW}$, and BH's angular momentum
$J_{\rm BH}$ for case~A4. All quantities are normalized by the ADM angular
momentum of the binary $J$.}
\label{fig:J_A4}
\end{figure}

\begin{figure}
\epsfxsize=3.4in
\leavevmode
\epsffile{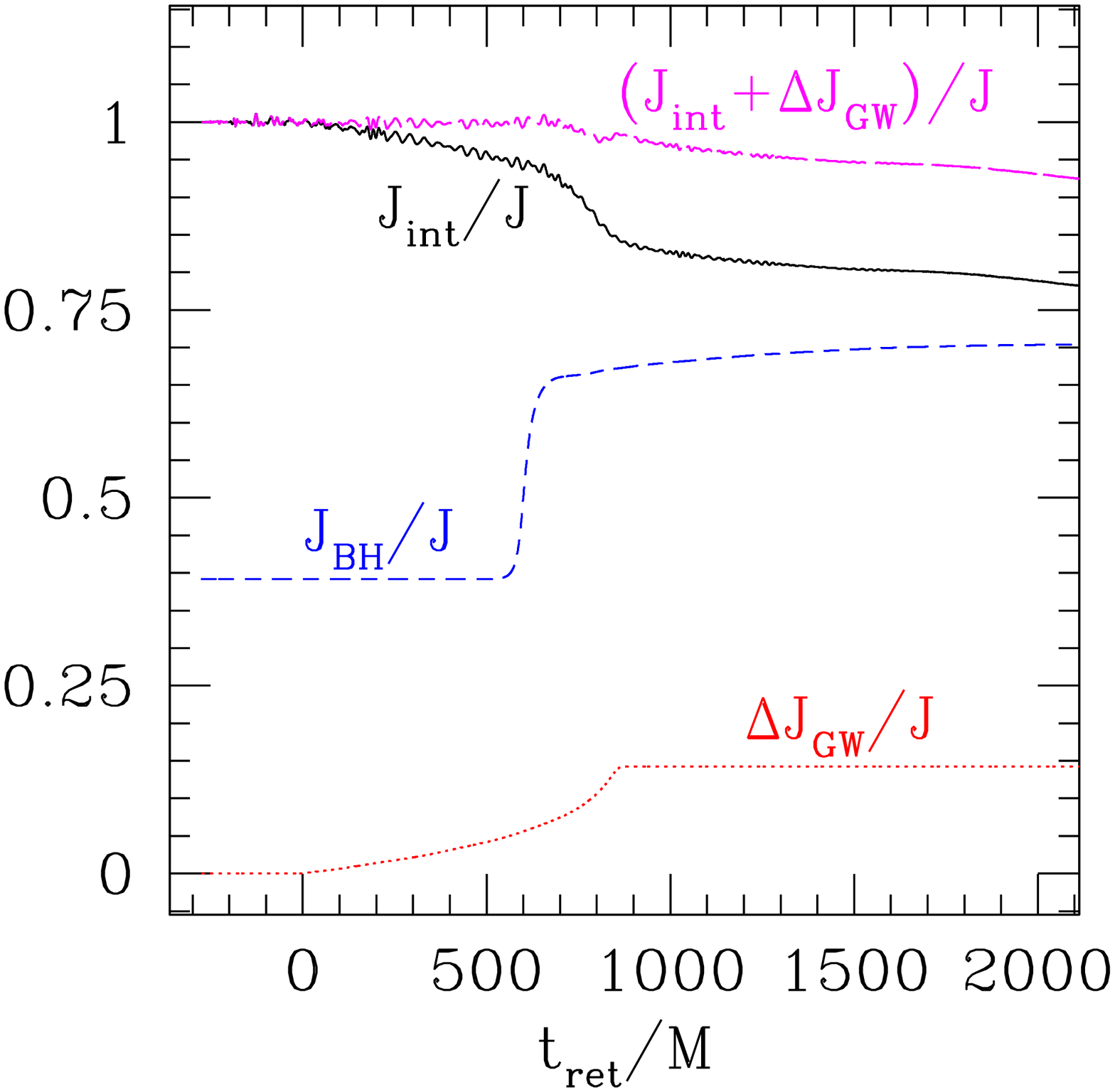}
\caption{Same as Fig.~\ref{fig:J_A4} but for case~B4.}
\label{fig:J_B4}
\end{figure}

We compute the energy $\Delta E_{\rm GW}$ and angular 
momentum $\Delta J_{\rm GW}$ carried away by 
the GWs, as well as the GW recoil velocity $v_{\rm kick}$.  Violation
of energy $\delta E$ and angular momentum $\delta J$,
as defined in Eqs.~(\ref{eq:deltae}) and (\ref{eq:deltaj}),
respectively, is also monitored. Results are given in
Table~\ref{table:results}. For cases A0--A4, $\delta E \approx$ 0.2\%
and $\delta J \approx$ 2\%. Whereas for cases B0--B4, $\delta E
\approx$ 0.6\% and $\delta J \approx$ 8\%. In all cases, a fraction 
of $E$ and $J$
are lost spuriously, and the situation is slightly worse in spinning
BH cases. 

To further study the issue of $E$ and $J$ loss, we calculate the
angular momentum inside the computational domain, $J_{\rm int}$,
using Eq.~(\ref{eq:Jint_sur_vol}), as well as the accumulated angular
momentum carried away by the GWs, $\Delta J_{GW}$. Figures~\ref{fig:J_A4}
and \ref{fig:J_B4} show the evolution of $J_{\rm int}$ and $\Delta J_{GW}$ 
for representative cases (A4 and B4). The corresponding plots for the
various components of energy are similar.  Conservation 
of angular momentum implies that $J_{\rm sum}=J_{\rm int}+\Delta J_{GW}$ 
is constant in time and is equivalent to the ADM angular momentum of
the binary.  Notice that the total angular momentum $J_{\rm sum}$ is
conserved well during inspiral for both cases ($t_{\rm ret} \leq 500M$
for case A4 and $t_{\rm ret} \leq 700M$ for B4). Post-merger, $\Delta J_{GW}$ 
flattens as GW emission subsides in all cases. $J_{\rm int}$ also
flattens after merger in case A4, as expected. Hence the spurious loss
of $J$ ($\approx$ 2\%) occurs primarily during merger for case A4. The
same behavior is observed in cases A0--A3 as well. However, in cases
B0--B4, $J_{\rm int}$ decreases on a secular timescale after
merger. The value of $J_{\rm sum}$ deviates from its initial value by 
about 2\% after the merger at $t_{\rm ret} \approx 900M$, but the deviation 
increases slowly and reaches $\approx$ 7\% at the end of simulation 
($t_{\rm ret} =2000M$), indicating that the spurious loss of $J$
continues after merger in case B4. 

The BH spin parameter for case B4 
increases from $a_{\rm BH}/M_{\rm BH}=$0.75 to $\approx 0.85$ 
during the simulation (see Table~\ref{table:results}). A secular
decrease of total angular momentum as found in case B4 is commonly observed in rapidly spinning
vacuum BH simulations (see e.g., \cite{mtbgs08}), which find that spurious
$J$ loss is reduced with increasing resolution.  

Figures~\ref{fig:J_A4} and~\ref{fig:J_B4} also show the evolution of
the BH's angular momentum $J_{\rm BH}$, computed using the isolated
and dynamical horizon formalism. $J_{\rm BH}$ monotonically increases in both
A4 and B4, and is always less than $J_{\rm int}$.  This seems to
suggest that the spurious loss of $J$ in case B4 after merger may
occur mainly in the remnant disk, where a substantial amount of matter
is continuously crossing the refinement levels. However, the
possibility of spurious loss of $J$ in the strong-field region near
the BH cannot be ruled out completely. One way to resolve the issue
would be to compute the amount of $J$ flowing into the 
BH and check for the conservation of $J$ at the BH horizon (see
e.g., Sec.~4.2 of~\cite{IH_DH}).

\section{Summary and Future work}
\label{sec:summaryandfuturework}

We present preliminary results from magnetized simulations of BHNS
binaries with BH:NS mass ratio 3:1. We treat both initially
nonspinning (cases A) and moderately-spinning (cases B) BHs.  
For those magnetic field
configurations we consider, only initial NS 
magnetic fields with maximum (internal) strength of $\approx 10^{17}$G --
corresponding to average magnetic to gas pressure ratio of 0.5\% --
significantly impact the inspiral and merger dynamics.
During merger, most of the magnetized NS matter is
captured by the BH.  After disruption, the dynamics are followed for
about 30--50 $(M_0/1.4M_\odot)$ms.  Only in the cases in which
magnetic fields are strongest are magnetic effects dynamically significant, increasing the final disk mass by
up to a factor of two.  The strong magnetic fields help push out the outer
layers of the NS during tidal disruption, resulting in a gravitational
wave mismatch of 0.004 for the Advanced LIGO broadband configuration.
It may be challenging for the upcoming generation of
gravitational wave detectors to observe effects from such 
magnetic fields. Further studies with different field geometries, 
black hole spins and higher resolution may be required to confirm 
this finding.
 
Some SGRB models require a massive, hot, magnetized disk
around a BH with collimated magnetic fields to launch jets that generate 
$\gamma$-rays (see, e.g.~\cite{vk03a,vk03b}). In our BHNS
simulations, the remnant disk is hot ($T\sim 1$MeV) and massive
($M_{\rm disk} \sim 0.02M_\odot$ and $\sim 0.1M_\odot$ for cases A and
B, respectively), and possesses magnetic fields that are 
tightly wound.  However, evidence for magnetic collimation 
around the BH or magnetic field-induced turbulence in the disk is not
observed. Future analyses will focus on mode growth studies and 
B-field decomposition in poloidal and toroidal components with respect to the centre of mass of the NS, 
similar to previous studies of single rotating or magnetized stars \cite{Zink,Lasky,Ciolfi}.
The lack of collimation may be due to the short disk
evolution time before our simulation is terminated. The absence of
turbulence in the disk may be due to insufficient resolution in the
disk, thereby suppressing instabilities like MRI.  
Therefore, future work will focus on evolving the disk at
higher resolution, coupled with longer disk evolutions and different initial
magnetic field configurations to more thoroughly assess
the possibility of BHNS binaries as short-hard GRB progenitors.

\acknowledgments
The authors wish to thank Charles F.~Gammie for useful discussions. 
We also thank the Illinois Relativity Group's Research Experience for
Undergraduates (REU) team, including 
Gregory Colten,
Stephen Drake,
Miaotianzi Jin,
David Kolschowsky,
David Kotan, and Francis Walsh, for assistance in producing the 3D 
visualizations of our simulations.
This paper was supported in part by NSF Grants AST-1002667, and
PHY-0963136 as well as NASA Grant NNX10AI73G at the
University of Illinois at Urbana-Champaign. The 3D visualizations were produced
using the ZIBamira software system \cite{Stalling:AmiraVDA-2005}.


\appendix 

\section{Hydrodynamic and MHD inequalities} 
\label{sec:inequalities}

For a given set of ``primitive'' variables $(\rho_0,P,v^i,B^i)$ in the 
physical range (i.e.\ $\rho_0\geq 0$, $P\geq 0$ and $\epsilon \geq 0$), 
the corresponding ``conservative'' variables
$(\rho_*,\ttau,\tS_i,\tilde{B}^i)$ must satisfy certain inequalities. 
In this appendix, we derive the inequalities and provide a practical 
recipe to impose these inequalities approximately to reduce inversion
failures, which occur mainly in regions with 
very low density in the artificial ``atmosphere'' or inside the BH 
horizon where high accuracy is difficult to maintain but not crucial.
Even when applying this recipe,
inversion failures sometimes occur.  In that case, we employ an
alternative inversion scheme, described in Sec.~\ref{sec:fontfix},
that always works. Readers who are only interested in our recipe 
may jump directly to Sec.~\ref{mhd_hyd_ineq_algorithm} and skip 
the rest of this appendix outlining the derivation.

Since the inversion between $\tilde{B}^i$ and $B^i$ is trivial and does not 
involve other primitive variables, we will treat values of $B^i$ as given and 
only consider the inversion from $(\rho_*,\ttau,\tS_i)$ to $(\rho_0,P,v^i)$. 
We assume that the 
EOS $P=P(\rho_0,\epsilon)$ always gives $P \geq 0$ whenever $\rho_0 \geq 0$ and 
$\epsilon \geq 0$. We also assume that the 
metric, lapse and shift are in the physical range. In particular, we require 
$\alpha>0$ and 
$\gamma_{ij}k^i k^j >0$ for any real vector
$k^i$. The requirement $\alpha>0$ is always satisfied 
by our particular time slicing. However, $\gamma_{ij}$ 
may lose positive-definiteness due to numberical error during the evolution, 
especially in the region deep inside the BH, near the ``puncture''. Before performing 
the inversion, we check if $\gamma_{ij}$ is positive definite by finding its 
eigenvalues. 
If $\gamma_{ij}$ is not positive definite, we reset 
$\gamma_{ij} \rightarrow \psi^4 f_{ij}$ during the inversion, where 
$f_{ij}$ is the 3D flat metric tensor.

\subsection{Derivation of conservative variable inequalities: Pure hydrodynamic case} 
\label{app:taust_hydro}
In the absence of magnetic fields, the conservative variables are given by 
\beqn
  \rhos &=&  \sgam \rho_0 \gamma_v \\
  \tS_i &=& \rho_* h u_i \label{eq:Stilde_hydro} \\
  \ttau &=& h(w-\rhos) + (h-1)\rhos - \sgam\, P ,
\label{eq:tau_hydro}
\eeqn
where 
\beq
  \gamma_v = \alpha u^0 \ \ \ , \ \ \
  w = \gamma_v \rhos ,
\eeq
and $h = 1+\epsilon + P/\rho_0 \geq 1$.
It follows from $u_\mu u^\mu=-1$ that
\beq
  \gamma_v = \sqrt{1+\gamma^{ij} u_i u_j} .
\label{eq:alpha_u0}
\eeq
Since $\gamma^{ij} u_i u_j$ is positive definitive, $\gamma_v \geq 1$. 
We therefore conclude that 
\beq
  \rho_* \geq 0 
\eeq
for $\rho_0 \geq 0$ and $u_i \in (-\infty,\infty)$. 
In addition, $w= \gamma_v \rhos \geq \rho_*$, from which we obtain 
\beqn
  \ttau &\geq & (h-1)\rhos - \sgam\, P  \cr 
 &=& \sgam [ \gamma_v \rho_0 \epsilon + (\gamma_v-1)P ] \geq 0 .
\eeqn
Hence we conclude that 
\beq
  \ttau \geq 0 ,
\eeq
which is Eq.~(\ref{eq:taucon}). 

To derive the inequality (\ref{eq:tau_stilde_ineqs}), 
Eqs.~(\ref{eq:Stilde_hydro}) and (\ref{eq:alpha_u0}) are combined to
yield
\beq
  \tS^2 = \gamma^{ij} \tS_i \tS_j = (\rhos h)^2 (\gamma_v^2-1) =  h^2 (w^2-\rhos^2) .
\label{eq:tS2_hydro}
\eeq
A straightforward calculation yields
\beqn
  (\ttau + \rhos)^2 &=& \tS^2 + h^2 \rhos^2 - \sgam\, P (2hw-\sgam\, P) \cr
 &=& \tS^2 + (\gamma_v \sgam)^2 [\rho_0^2 (1+\epsilon)^2 - P^2]  \cr 
 && + (\sgam \, P)^2 \ . 
\label{eq:tau_plus_rhos2}
\eeqn
Since there is no upper limit on $\gamma_v$, the sum of the second and
third terms is always positive if and only if the dominant energy
condition -- $P^2 \leq \rho_0^2 (1+\epsilon)^2$ -- holds. Hence we conclude that
\beq
 \tS^2 \leq (\ttau + \rhos)^2,
\eeq
if the dominant energy condition is satisfied. The
inequality~(\ref{eq:tau_stilde_ineqs}) is more stringent, and hence
more useful. It can be derived using Eq.~(\ref{eq:tau_plus_rhos2}):
\[
  \ttau (\ttau + 2\rhos) 
 = \tS^2 + (\gamma_v \sgam)^2 (2\rho_0^2 \epsilon + \rho_0^2 \epsilon^2
- P^2) + (\sgam \, P)^2 . 
\]
In this case, the sum of the second and third terms is always positive
if and only if $P^2 \leq 2\rho_0^2 \epsilon + \rho_0^2 \epsilon^2$. Hence
we have derived the inequality (Eq.~\ref{eq:tau_stilde_ineqs}):
\beq
  \tS^2 \leq \ttau (\ttau + 2\rhos) \ \ \ \mbox{iff }
P^2 \leq 2\rho_0^2 \epsilon + \rho_0^2 \epsilon^2 .
\eeq
Whether the
condition $P^2 \leq 2\rho_0^2 \epsilon + \rho_0^2 \epsilon^2$ is
satisfied depends on the EOS. For the $\Gamma$-law
EOS $P=(\Gamma-1)\rho_0 \epsilon$, simple calulation gives
\beq
  P^2 - 2\rho_0^2 \epsilon - \rho_0^2 \epsilon^2 =
\rho_0^2 \epsilon [ \Gamma(\Gamma-2)\epsilon - 2 ].
\eeq
Hence the inequality $\tS^2 \leq \ttau (\ttau + 2\rhos)$ holds 
iff $\Gamma(\Gamma-2)\epsilon - 2 \leq 0$. Restricting to the parameter
space where the sound speed is subluminal, i.e.\ $c_s^2 = \Gamma
P/(\rho_0 h) < 1$, we have $\Gamma(\Gamma-2) \epsilon < 1$.
Therefore, $\tS^2 \leq \ttau (\ttau + 2\rhos)$ holds for the $\Gamma$-law
EOS in regions where the sound speed is subluminal. 
For the $\Gamma$-law EOS it can be shown that
$c_s^2 < \Gamma-1$ holds for any nonnegative $\rho_0$ and $P$. 
Thus, the sound speed will always be subluminal when $ 1 < \Gamma \leq 2$. 
Thus, $\tS^2 \leq \ttau (\ttau + 2\rhos)$  is satisfied for 
the $\Gamma$-law EOS, when $ 1 < \Gamma \leq 2$. 

We have just proved that 
the inequalities $\rho_*\geq 0$, $\ttau \geq 0$ and $\tS^2 \leq \ttau (\ttau + 2\rhos)$
are {\it necessary} conditions for the primitive inversion to yield a physical solution 
for $1 < \Gamma \leq 2$. We now want to prove that they are also the {\it sufficient} 
conditions for the $\Gamma$-law EOS with $\Gamma >1$. For the $\Gamma$-law EOS, the 
enthalpy is related to the pressure $P$ by 
\beq
  h = 1+\frac{\Gamma P}{(\Gamma-1)\rho_0} .
\eeq
Combining the above equation with Eq.~(\ref{eq:tau_hydro}) yields
\beq
  h = \frac{\Gamma w(\ttau+\rho_*)-(\Gamma-1)\rho_*^2}{\Gamma w^2-(\Gamma-1)\rho_*^2} .
\label{eq:hGamma-law}
\eeq
It is useful to define a variable $x\equiv(w-\rho_*)/\ttau$. It follows from 
Eqs.~(\ref{eq:hGamma-law}) and (\ref{eq:tS2_hydro}) that 
\beq
  h-1 = \frac{\Gamma \ttau (1-x)(x\ttau+\rho_*)}{\Gamma \ttau x (\ttau x+2\rho_*) + \rho_*^2}
\label{eq:hm1hydro}
\eeq
and 
\beq
  f(x) \equiv \ttau x(\ttau x+2\rho_*) - \frac{\tS^2}{h^2} = 0 .
\eeq
These two equations can be combined to yield a quartic equation in $x$. However, 
it is easier to analyze the equations in the present form. 
For any given $\rho_* \geq 0$, $\ttau \geq 0$ and $\tS^2 \leq \ttau (\ttau+2\rho_*)$, 
when $x=0$, $h=1+\Gamma \ttau/\rho_* \geq 1$ and $f(0)=-\tS^2/h^2\leq 0$; when $x=1$, 
$h=1$ and $f(1)=\ttau (\ttau + 2\rho_*) - \tS^2 \geq 0$. Hence the
intermediate value theorem implies there exists $x_0 \in [0,1]$ so
that $f(x_0)=0$. 
The primitive variables are recovered
from the following expressions
\beqn
  \gamma_v &=& \frac{w}{\rho_*} = 1+\frac{\ttau x_0}{\rho_*} , \\ 
  \rho_0 &=& \frac{\rho_*}{\gamma_v \sgam} , \\ 
  h-1 &=& \frac{\Gamma \ttau (1-x_0)(x_0\ttau+\rho_*)}{\Gamma \ttau x_0 (\ttau x_0+2\rho_*) + \rho_*^2} , \\ 
  P &=& \frac{\Gamma-1}{\Gamma} \rho_0 (h-1) , \\ 
  u_i &=& \frac{\tS_i}{\rho_* h} ,  \label{eq:u_i} \\ 
  v^i &=& \frac{\alpha}{\gamma_v} \gamma^{ij} u_j - \beta^i . \label{eq:vi}
\eeqn
Upon inspection, all the recovered primitive
variables lie in the physically acceptable range for $x_0 \in [0,1]$
and $\Gamma>1$.

We therefore conclude that 
the inequalities $\rho_*\geq 0$, $\ttau \geq 0$ and $\tS^2 \leq \ttau (\ttau + 2\rhos)$ 
are both necessary and sufficient conditions for the inversion to
yield physically-acceptable primitive variables for the $\Gamma$-law
EOS with $1< \Gamma \leq 2$. Since the $\Gamma$-law EOS with
$\Gamma=2$ is adopted in our simulations, we impose  
these inequalities in regions where there are no magnetic fields.

\subsection{Derivation of conservative variable inequalities: MHD case} 
\label{app:taust_mhd}
In the presence of magnetic fields, the conservative variables $\tS_i$ and 
$\ttau$ are given by 
\beqn
  \tS_i &=& \tS^{\rm fluid}_i + \sg\, (b^2 u^0 u_i - b^0 b_i) ,
\label{eq:mhdsti} \\ 
  \ttau &=& \ttau_{\rm fluid} + \sgam \, \left[ \gamma_v^2 b^2 - \frac{b^2}{2} 
- (\alpha b^0)^2 \right] \label{eq:mhdtau} .
\eeqn
Here $\tS^{\rm fluid}_i=\rho_* h u_i$ and 
$\ttau_{\rm fluid}=h(w-\rhos) + (h-1)\rhos - \sgam\, P$, which are the 
same expressions as Eqs.~(\ref{eq:Stilde_hydro}) and (\ref{eq:tau_hydro}). 
The variable $\rho_*=\gamma_v \sgam \, \rho_0$ remains unchanged and hence the 
inequality $\rho_* \geq 0$ still holds in the presence of magnetic fields.

It is convenient to introduce the following three quantities 
\beqn
  W &\equiv& \gamma_v^2 \rho_0 h = wh/\sgam = \frac{\sqrt{\tS_{\rm fluid}^2+(\rhos h)^2}}{\sgam} ,
\label{eq:W} \\
  V &\equiv& \sqrt{\gamma_v^2-1}/\gamma_v , \\
 \bB^i &\equiv& B^i/\sqrt{4\pi}  .
\eeqn
Following the algebra in Sec.~3.1 of~\cite{ngmz06}, one can
show that (c.f.\ Eqs.~(26), (27) and (29) of~\cite{ngmz06})
\beqn
  \bB^i \tS_i &=& \bB^i \tS^{\rm fluid}_i , \label{eq:BdotS} \\
  \tS^2 &=& (\sgam)^2 V^2 (W+\bB^2)^2 - \frac{(\bB^i \tS_i)^2 (\bB^2+2W)}{W^2} , \ \ 
\ \ \ \ \ \label{eq:tS2} \\ 
  \ttau &=& \ttau_{\rm fluid} + \frac{\sgam}{2}(1+V^2) \bB^2 
- \frac{(\bB^i \tS_i)^2}{2\sgam\, W^2} ,\label{eq:ttau}
\eeqn
where $\bB^2 = \gamma_{ij} \bB^i \bB^j$. It follows from Eq.~(\ref{eq:BdotS}) 
and the Cauchy-Schwartz inequality that 
\beqn
  (\bB^i \tS_i)^2 &=& (\gamma^{ij} \bB_i \tS^{\rm fluid}_j)^2 \cr 
&\leq & (\gamma^{ij} \bB_i \bB_j) (\gamma^{ij} \tS^{\rm fluid}_i \tS^{\rm fluid}_j) \cr 
&=& \bB^2 \tS^2_{\rm fluid} . \nonumber
\eeqn
Hence we obtain
\beq
  \tS^2_{\rm fluid} \geq (\hat{\bB}^i \tS_i)^2 ,
\label{eq:tS2flb}
\eeq
where $\hat{\bB}^i \equiv \bB^i/\bB$.

Using Eq.~(\ref{eq:tS2}), we can write 
\beq
  V^2 = \frac{1}{\gamma (W+\bB^2)^2} \left[ \tS^2 +
\frac{(\bB^i \tS_i)^2 (\bB^2+2W)}{W^2} \right] ,
\label{eq:V2}
\eeq 
and 
\beq
  \tS_{\rm fluid}^2 = \gamma V^2 W^2 = \frac{W^2 \tS^2 +
(\bB^i \tS_i)^2 (\bB^2+2W)}{(W+\bB^2)^2} .
\label{eq:tsfluid1}
\eeq
Given values of $\tS_i$ and $\bB^i$, the only independent variable in the above 
equation is $W$. Straightforward calculation yields 
\beq
  \frac{d \tS_{\rm fluid}^2}{dW} = \frac{2W[\bB^2 \tS^2 - (\bB^i \tS_i)^2]}
{(W+\bB^2)^3} \geq 0 ,
\label{eq:dS2fdW}
\eeq
where we have applied the Cauchy-Schwarz inequality 
\[
  (\bB^i \tS_i)^2 = (\gamma^{ij} \bB_i \tS_j)^2 \leq (\gamma^{ij} \bB_i \bB_j) 
(\gamma^{ij} \tS_i \tS_j) = \bB^2 \tS^2 .
\]
Hence the maximum value of $\tS_{\rm fluid}^2$ is achieved when $W\rightarrow \infty$, 
which gives $\tS_{\rm fluid}^2 \leq \tS^2$. The minimum value of $\tS_{\rm fluid}^2$ 
is achieved when $W=W_m$, where $W_m$ is the minimum value of $W$ for  
given values of $\rhos$, $\tS_i$ and $\bB^i$. Hence we obtain 
\beq
 \tS_m^2 \leq \tS_{\rm fluid}^2 \leq \tS^2 ,
\label{eq:intSf}
\eeq
where 
\beq
  \tS_m^2 = \frac{W_m^2 \tS^2 + (\bB^i \tS_i)^2 (\bB^2+2W_m)}{(W_m+\bB^2)^2} .
\label{eq:tSm2}
\eeq

The upper and lower bounds of $\ttau_{\rm fluid}$ can be derived by first combining 
Eqs.~(\ref{eq:ttau}) and (\ref{eq:V2}): 
\beq
  \ttau_{\rm fluid} = \ttau - \frac{\sgam}{2} \bB^2
 - \frac{ \bB^2 \tS^2 - (\bB^i \tS_i)^2}{2\sgam\, (W+\bB^2)^2} .
\eeq
For fixed values of $\ttau$, $\tS_i$ and $\bB^i$, $\ttau_{\rm fluid}$ 
increases with increasing $W$. Therefore, we conclude that 
\beq
  \ttau_m \leq \ttau_{\rm fluid} \leq \ttau - \frac{\sgam}{2}\bB^2 ,
\label{eq:intauf}
\eeq
where 
\beq
  \ttau_m = \ttau - \frac{\sgam}{2} \bB^2
 - \frac{ \bB^2 \tS^2 - (\bB^i \tS_i)^2}{2\sgam\, (W_m+\bB^2)^2} .
\label{eq:taum}
\eeq
To calculate $W_m$, we can combine the last equality in Eq.~(\ref{eq:W}) and 
Eq.~(\ref{eq:tsfluid1}) to obtain an implicit relation of $W=W(h)$. 
Using $V\leq 1$, it is then
straightforward to show that $dW/dh>0$, and hence $W=W_m$ 
when $h$ is minimized. Given that $h\geq 1$, we therefore have  
\beq
  W_m = \frac{\sqrt{\tS_m^2+\rhos^2}}{\sgam} .
\label{eq:Wm}
\eeq
This equation can be combined with Eq.~(\ref{eq:tSm2}), resulting in a 
quartic equation for $W_m$: 
\[
  (\gamma W_m^2-\rho_*^2)(W_m+\bB^2)^2-W_m^2\tS^2 
- (\bB^i \tS_i)^2(\bB^2+2W_m)=0 ,
\]
which may be solved analytically. 
Alternatively, Eqs.~(\ref{eq:tSm2}) 
and (\ref{eq:Wm}) may be solved using an iterative scheme. We start by
using Eq.~(\ref{eq:tS2flb}) and choose an initial guess $W_m =
[(\hat{\bB}^i \tS_i)^2 + \rhos^2]^{1/2} /\sgam$. Next we compute the
initial guess $\tS_m^2$ using Eq.~(\ref{eq:tSm2}). We then recompute
$W_m$ using Eq.~(\ref{eq:Wm}) and $\tS^2_m$ using
Eq.~(\ref{eq:tSm2}). We keep iterating until the values of $\tS_m^2$
and $W_m$ converge.

In the previous subsection, we proved that in the absence of
magnetic fields $\rho_* \geq 0$, $\ttau \geq 0$ and 
$\tS^2 \leq \ttau (\ttau+2\rho_*)$ are the necessary 
and sufficient conditions for the inversion to produce the primitive variables in the 
physical range for the $\Gamma$-law EOS with $1<\Gamma \leq 2$. In the presence of magnetic 
fields, the necessary and sufficient conditions 
for the $\Gamma$-law EOS with $1 < \Gamma \leq 2$ are replaced by 
the following inequalities:
\beqn 
  \rho_* & \geq & 0 , \label{eq:ttt1} \\ 
  \ttau_{\rm fluid} & \geq & 0 ,  \\ 
  \ttau_{\rm fluid} (\ttau_{\rm fluid}+2\rho_*) & \geq & \tS^2_{\rm fluid} .
\label{eq:ttt3}
\eeqn

Unfortunately, no simple, analytic expression for necessary and sufficient conditions between 
the conservatives $\rho_*$, $\tS_i$, $\ttau$ seems to exist in the presence of 
magnetic fields. However, necessary and sufficient conditions can be
derived separately by combining Eqs.~(\ref{eq:ttt1})--(\ref{eq:ttt3}), (\ref{eq:intSf}) and 
(\ref{eq:intauf}). The results are as follows. 

Necessary conditions for guaranteeing a physical solution:
\beqn 
  \rho_* & \geq & 0 , \label{eq:nece1}  \\ 
  \ttau & \geq & \frac{\sgam}{2} \bB^2 , \label{eq:nece2}  \\ 
  \tS^2_m & \leq &\left( \ttau - \frac{\sgam}{2} \bB^2 \right) 
 \left( \ttau - \frac{\sgam}{2} \bB^2 +2\rho_* \right) .
\label{eq:nece3}
\eeqn

Sufficient conditions for guaranteeing a physical solution: 
\beqn
  \rho_* & \geq & 0 , \label{eq:suff1} \\ 
  \ttau_m & \geq & 0 , \label{eq:suff2}  \\ 
  \ttau_m (\ttau_m+2\rho_*) & \geq & \tS^2 . 
\label{eq:suff3}
\eeqn

Both $\tS^2_m$ and $\ttau_m$ are nonlinear functions of $\rho_*$, $\tS_i$, 
and $\ttau$.  
Unlike the pure hydrodynamic case, these inequalities
are not trivial to impose strictly, so they are imposed approximately
as follows.  First, a parameter $\psi^6_{\rm thr}$ is introduced,
which determines whether 
the region under consideration is deep inside the BH horizon. For
regions deep inside the BH horizon, defined by $\sgam = \psi^6 \geq
\psi^6_{\rm thr}$, the primary goal is to keep the evolution stable and
prevent inaccurate data from leaking out of the BH horizon.  We find
that imposing the sufficient
conditions~(\ref{eq:suff1})--(\ref{eq:suff3}) approximately in this
region is adequate (detailed recipe below).  In regions where $\psi^6
\leq \psi^6_{\rm thr}$, the goal is to evolve the GRMHD equations as
accurately as possible, which means that imposing the sufficient
conditions is not appropriate.  We instead impose two of the necessary
conditions~(\ref{eq:nece1})  and (\ref{eq:nece2}) only. Since we do
not strictly impose all the inequalities, inversion failures sometimes
occur. Failures are fixed by replacing the $\ttau$ equation by the
equation $P=P_{\rm cold}(\rho_0)$. We will demonstrate in 
Sec.~\ref{sec:fontfix} that the set of equations $\rho_* = \gamma_v
\sgam\, \rho_0$, Eq.~(\ref{eq:mhdsti}) and $P=P_{\rm cold}(\rho_0)$ always
results in the primitive variables in the physical range.  Our
detailed recipe is described in the following subsection.

\subsection{Algorithm for Imposing MHD/Hydrodynamic Inequalities}
\label{mhd_hyd_ineq_algorithm}

\begin{enumerate}

\item In any region, if $\rho_* \leq 0$, set $\rho_0=\rho_{\rm atm}$,
  $P=P_{\rm atm}$, $u_i=0$ and recompute the conservative variables. 
  If $\ttau \leq \sgam \, \bB^2/2$, reset $\ttau = \ttau_{\rm atm}+\sgam\, \bB^2/2$.

\item In the region where $B^i=0$, if $\ttau \leq 0$, 
  reset $\ttau = \ttau_{\rm atm}$. If $\tS^2 > \ttau (\ttau + 2\rho_*)$, 
  replace 
\beq
  \tS_i \rightarrow \tS_i \sqrt{ \frac{\ttau (\ttau+2\rho_*)}{\tS^2}} .
\eeq

\item In the region where $\psi^6 \geq \psi^6_{\rm thr}$ (deep inside the BH horizon): 
First, estimate $W_m$ and $S_m^2$ as follows
\beqn
  W_{m0} &=& \psi^{-6} \left[ (\hat{\bar{B}}^i \tilde{S}_i)^2 + \rho_*^2\right]^{1/2} , \\ 
  \tilde{S}^2_{m0} &=& \frac{W_{m0}^2 \tilde{S}^2 + (\hat{\bar{B}}^i \tilde{S}_i)^2
(\bar{B}^2+2W_{m0})}{(W_{m0}+\bar{B}^2)^2}  , \\ 
  W_{m} &=& \psi^{-6} \left( \tilde{S}^2_{m0} + \rho_*^2\right)^{1/2} .
\eeqn
Next, calculate $S_m$ and $\ttau_m$ from
Eqs.~(\ref{eq:tSm2}) and (\ref{eq:taum}). Then check if $\ttau_m \geq
\ttau_{\rm atm}$.  If $\ttau_m<\tau_{\rm atm}$ reset $\ttau$ (without 
changing $W_m$ and $\tS_i$) according to 
\beq
  \ttau = \ttau_{\rm atm} + \frac{\sgam}{2} \bB^2 
+ \frac{ \bB^2 \tS^2 - (\bB^i \tS_i)^2}{2\sgam\, (W_m+\bB^2)^2} .
\eeq
Then check if $\tS^2 \leq \ttau_m (\ttau_m + 2\rho_*)$. If not, reset 
$\tS_i$ (without changing $\ttau_m$ and $\rho_*$) according to 
\beq
 \tS_i \rightarrow \tS_i \sqrt{\frac{\ttau_m (\ttau_m + 2\rho_*)}{\tS^2}} .
\eeq
These procedures do not strictly impose the sufficient
conditions~(\ref{eq:suff2}) and (\ref{eq:suff3}), but can
significantly reduce inversion failures.

\item If the inversion still fails after going through all the above steps, 
replace the $\ttau$ equation~(\ref{eq:mhdtau}) by $P=P_{\rm cold}(\rho_0)$ 
and perform the inversion. This procedure guarantees to produce the primitive 
variables in the physical range, as will be shown in the next subsection.
\end{enumerate}

\subsection{Inversion using $\rho_*$, $\tS_i$ and $P=P_{\rm cold}(\rho_0)$}
\label{sec:fontfix}

After imposing the conservative variables inequalities as described in
the previous subsection, sometimes the conservatives$\to$primitives
variable inversion still fails.  In this case, an alternative
inversion is imposed, solving for $\rho_0$ and $u_i$ from the
equations $\rho_* = \sgam\, \gamma_v \rho_0$, $P=P_{\rm cold}(\rho_0)$
and Eq.~(\ref{eq:mhdsti}). We will prove that this
inversion always results in the primitive variables in the physical
range.

First consider the case $B^i=0$ for the $\Gamma$-law EOS with
$1<\Gamma \leq 2$. In this case, simple, analytic expressions for
the necessary and sufficient conditions exist, are easy to implement (as
described in Secs.~\ref{app:taust_hydro} \&
\ref{mhd_hyd_ineq_algorithm}), and guarantee successful inversions.
However, since the analysis is much simpler than the general case, it
is instructive to study the alternative inversion scheme for this case
first.  It follows from $\rho_*=\sgam\, \gamma_v \rho_0$, 
$P=P_{\rm cold}(\rho_0)$ and Eq.~(\ref{eq:tS2_hydro}) that 
\beq
  \rho_0 = \frac{\rho_*}{\sgam} \left[ 1 + \frac{\tS^2}{\rho_*^2 h_{\rm cold}^2(\rho_0)}\right]^{-1/2} ,
\label{eq:fontrho0}
\eeq
where $h_{\rm cold} = 1+\epsilon_{\rm cold}(\rho_0) + P_{\rm cold}(\rho_0)/\rho_0$ is the specific 
enthalpy for the cold EOS. The above equation is an implicit equation for $\rho_0$. 
Define the function 
\beq
  f(\rho_0) = \rho_0 - \frac{\rho_*}{\sgam} \left[ 1 + \frac{\tS^2}{\rho_*^2 h_{\rm cold}^2(\rho_0)}\right]^{-1/2} .
\eeq
and introduce two variables
\beq
  \rho_1 \equiv \frac{\rho_*}{\sgam}\left(1+\frac{\tS^2}{\rho_*^2}\right)^{-1/2} \ \ \ , \ \ \ 
 \rho_2 \equiv \frac{\rho_*}{\sgam} . 
\label{eq:rho0bounds}
\eeq
Since $h_{\rm cold} \geq 1$, 
for given values of $\rho_* > 0$ and $\tS_i \in (-\infty,\infty)$, $f(\rho_1) \leq 0$ and $f(\rho_2) \geq 0$. 
Hence there exists $\rho_0 \in [\rho_1,\rho_2]$ that satisfies Eq.~(\ref{eq:fontrho0}) 
provided that $h_{\rm cold}(\rho_0)$ is continuous in $[\rho_1,\rho_2]$, which is true 
for the $\Gamma$-law EOS. 
This proves that 
the inversion always produces $\rho_0$, as well as $P$ in the physical range. The value of $\gamma_v$ is 
then given by $\gamma_v=\rho_*/(\sgam \, \rho_0)$. For $\rho_0 \in [\rho_1,\rho_2]$, we have 
$1 \leq \gamma_v \leq \sqrt{1+\tS^2/\rho_*^2}$, which is also in the physical range. Finally, the velocity 
$v^i$ is recovered from Eqs.~(\ref{eq:u_i}) and (\ref{eq:vi}).

Next consider the case $B^i \neq 0$. Equations~(\ref{eq:W}) and
(\ref{eq:tsfluid1}) yield 
\beqn
W &=& \frac{\sqrt{\tS_{\rm fluid}^2+(\rhos h_{\rm cold})^2}}{\sgam} , \label{eq:Wcold} \\ 
\tS_{\rm fluid}^2 &=& \frac{W^2 \tS^2 + (\bB^i \tS_i)^2 (\bB^2+2W)}{(W+\bB^2)^2} , \label{eq:tsfluid2} \\ 
\rho_0 &=& \frac{\rho_*}{\sgam} \left[ 1 + \frac{\tS^2_{\rm fluid}}{\rho_*^2 h_{\rm cold}^2(\rho_0)}\right]^{-1/2} ,
\label{eq:fontrho0mhd}
\eeqn
where $\tS^2_{\rm fluid}$ is regarded as an implicit function of
$\tS_i$, $\bB^i$, $\rho_*$ and $\rho_0$ through Eqs.~(\ref{eq:Wcold})
and (\ref{eq:tsfluid2}). Hence Eq.~(\ref{eq:fontrho0mhd}) is an
implicit equation for $\rho_0$.  Next define
\beq
  f(\rho_0) = \rho_0 
-  \frac{\rho_*}{\sgam} \left[ 1 + \frac{\tS^2_{\rm fluid}}{\rho_*^2 h_{\rm cold}^2(\rho_0)}\right]^{-1/2} ,
\eeq
where $\rho_1$ and $\rho_2$ are as in
Eq.~(\ref{eq:rho0bounds}). Section~\ref{mhd_hyd_ineq_algorithm} proved
that $(\hat{\bB}^i \tS_i)^2 \leq \tS^2_{\rm fluid} \leq \tS^2$
[Eqs.~(\ref{eq:tS2flb}) and (\ref{eq:intSf})].  It follows that
$f(\rho_1) \leq 0$ and $f(\rho_2) \geq 0$. Hence a solution exists for
$\rho_0 \in [\rho_1,\rho_2]$ as in the pure hydro case, provided  
that both $h_{\rm cold}(\rho_0)$ and $\tS^2_{\rm fluid}(\rho_0)$ are continuous in $[\rho_1,\rho_2]$. 
The inversion thus produces $\rho_0$ as well as $P$ and $\gamma_v$ in
the physical range.  Applying some algebraic manipulations to
Eqs.~(\ref{eq:mhdsti}) and (\ref{eq:bmu}), we recover the 4-velocity
using the formula (c.f.\ Eq.~(31) of~\cite{ngmz06})
\[
u_i = \left[ \tS_i + \frac{\sgam\, (\bB^j \tS_j)}{\gamma_v \rhos h_{\rm cold}(\rho_0)}\bB_i 
\right] \left[ \rhos h_{\rm cold}(\rho_0) + \frac{\sgam \, \bB^2}{\gamma_v} \right]^{-1} 
\]
and the 3-velocity $v^i$ from Eq.~(\ref{eq:vi}). 

In conclusion, when using the equations $\rho_* = \sgam\, \gamma_v \rho_0$, 
$P=P_{\rm cold}(\rho_0)$ and Eq.~(\ref{eq:mhdsti}), the inversion will
always produce the primitive variables in the physical range for any
$\rho_*>0$ and $\tS_i \in (-\infty,\infty)$.

\bibliography{paper}

\end{document}